\documentclass[preprint,3p,a4,onecolumn]{elsarticle}

\usepackage{epsfig}
\usepackage{times}
\usepackage{latexsym}
\usepackage{subfigure}
\usepackage{amsfonts,amsmath,amsthm,amssymb}
\usepackage{graphicx}
\usepackage{natbib}
\usepackage{lineno}

\usepackage[bookmarks=true,colorlinks=false,pdfborder={0 0 0 0}]{hyperref}

\newcommand{\EXP}[1]{\mathsf{E}\!\left(#1\right)}
\newcommand{\EXPSUB}[2]{\mathsf{E}_{#2}\!\left(#1\right)}

\newcommand{\ind}[1]{\mathbf{1}_{\left\{#1\right\}}}

\newcommand{\remove}[1]{}
\newcommand{\notes}[1]{}
\renewcommand{\qed}{\hfill \rule{2mm}{2mm}}

\newcommand{\bmath}[1]{\mbox{\boldmath$#1$}}
\newcommand{\first}[1]{$1^{\mathrm{st}}$}
\newcommand{\second}[1]{$2^{\mathrm{nd}}$}

\theoremstyle{definition}
\newtheorem{definition}{Definition}[section]
\newtheorem{approximation}{Approximation}[section]

\theoremstyle{remark}
\newtheorem{remarks}{Remarks}[section] 

\theoremstyle{plain}
\newtheorem{theorem}{Theorem}[section]

\newtheorem{lemma}{Lemma}[section]

\begin{document}

\begin{frontmatter}

\title{State Dependent Attempt Rate Modeling of Single Cell IEEE~802.11 WLANs \\ with Homogeneous Nodes and Poisson Packet Arrivals\tnoteref{whatsnew}} 

\tnotetext[whatsnew]{This paper is an extended and thoroughly revised version of our earlier work~\cite{wanet.manoj-anuragCOMSNETS09SDAR}. In this paper, we provide new simulation results in support of the State Dependent Attempt Rate (SDAR) model of contention, and also provide complete proofs and derivations.} 


\author[manojanurag]{Manoj K. Panda\corref{manojece}}
\ead{manoj.manojpanda@gmail.com}

\author[manojanurag]{Anurag Kumar}
\ead{anurag@ece.iisc.ernet.in}

\address[manojanurag]{Department of Electrical Communication Engineering
  \\ Indian Institute of Science, Bangalore -- 560012.} 


\cortext[manojece]{Corresponding author}



\begin{abstract}
Analytical models for IEEE 802.11-based WLANs are invariably based on approximations, such as the well-known \textit{decoupling approximation} proposed by Bianchi for modeling single cell WLANs consisting of saturated nodes. In this paper, we provide a new approach to model the situation when the nodes are not saturated. We study a State Dependent Attempt Rate (SDAR) approximation to model $M$ queues (one queue per node) served by the CSMA/CA protocol as standardized in the IEEE 802.11 DCF MAC protocol. The approximation is that, when $n$ of the $M$ queues are non-empty, the transmission attempt probability of the $n$ non-empty nodes is given by the long-term transmission attempt probability of $n$ ``saturated'' nodes as provided by Bianchi's model. The SDAR approximation reduces a single cell WLAN with non-saturated nodes to a ``coupled queue system''. When packets arrive to the $M$ queues according to independent Poisson processes, we provide a Markov model for the coupled queue system with SDAR service. \textit{The main contribution of this paper is to provide an analysis of the coupled queue process by studying a lower dimensional process, and by introducing a certain conditional independence approximation}. We show that the SDAR model of contention provides an accurate model for the DCF MAC protocol in single cells, and report the simulation speed-ups thus obtained by our \textit{model-based simulation}. 
\end{abstract} 



\end{frontmatter}


\section{Introduction} 
\label{sec:introduction} 

The IEEE 802.11 standard~\cite{wanet.IEEE802dot11standard2007} has been widely adopted as the \textit{de facto} standard for accessing shared wireless media in Wireless Local Area Networks (WLANs). The Distributed Coordination Function (DCF) Medium Access Control (MAC) protocol, which is a particular version of the \textit{Carrier Sense Multiple Access with Collision Avoidance} (CSMA/CA) protocols, provides the fundamental access method in 802.11 WLANs~\cite{wanet.IEEE802dot11standard2007}. The Hybrid Coordination Function (HCF) which provides service differentiation and Quality of Service (QoS), and the optional Point Coordination Function (PCF) are also built on top of the DCF~\cite{wanet.IEEE802dot11standard2007}. Hence, understanding the behavior of the DCF is the key to understand the performance of 802.11 based WLANs. 

\begin{figure}[tb]
  \centering \
  \begin{minipage}{6cm}
    \begin{center}
      \includegraphics[scale=0.4]{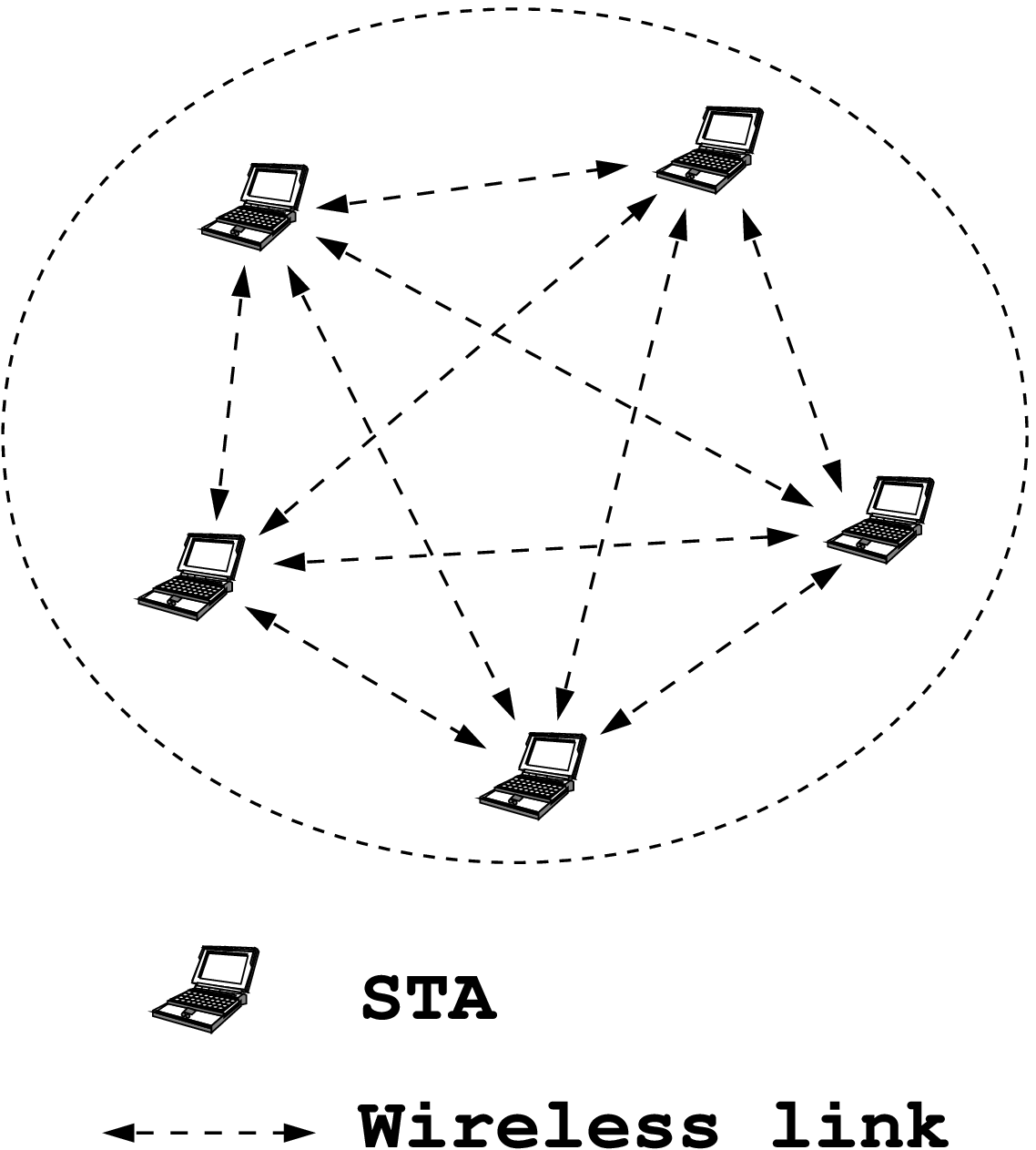}
      \caption{A single cell in the \textit{ad hoc} mode: In this
        case, users or client stations (STAs) can directly communicate
        among themselves. \label{fig:singleCellAdHoc}} 
    \end{center}
  \end{minipage}
  \hfill
  \begin{minipage}{9cm}
    \begin{center}
      \includegraphics[scale=0.4]{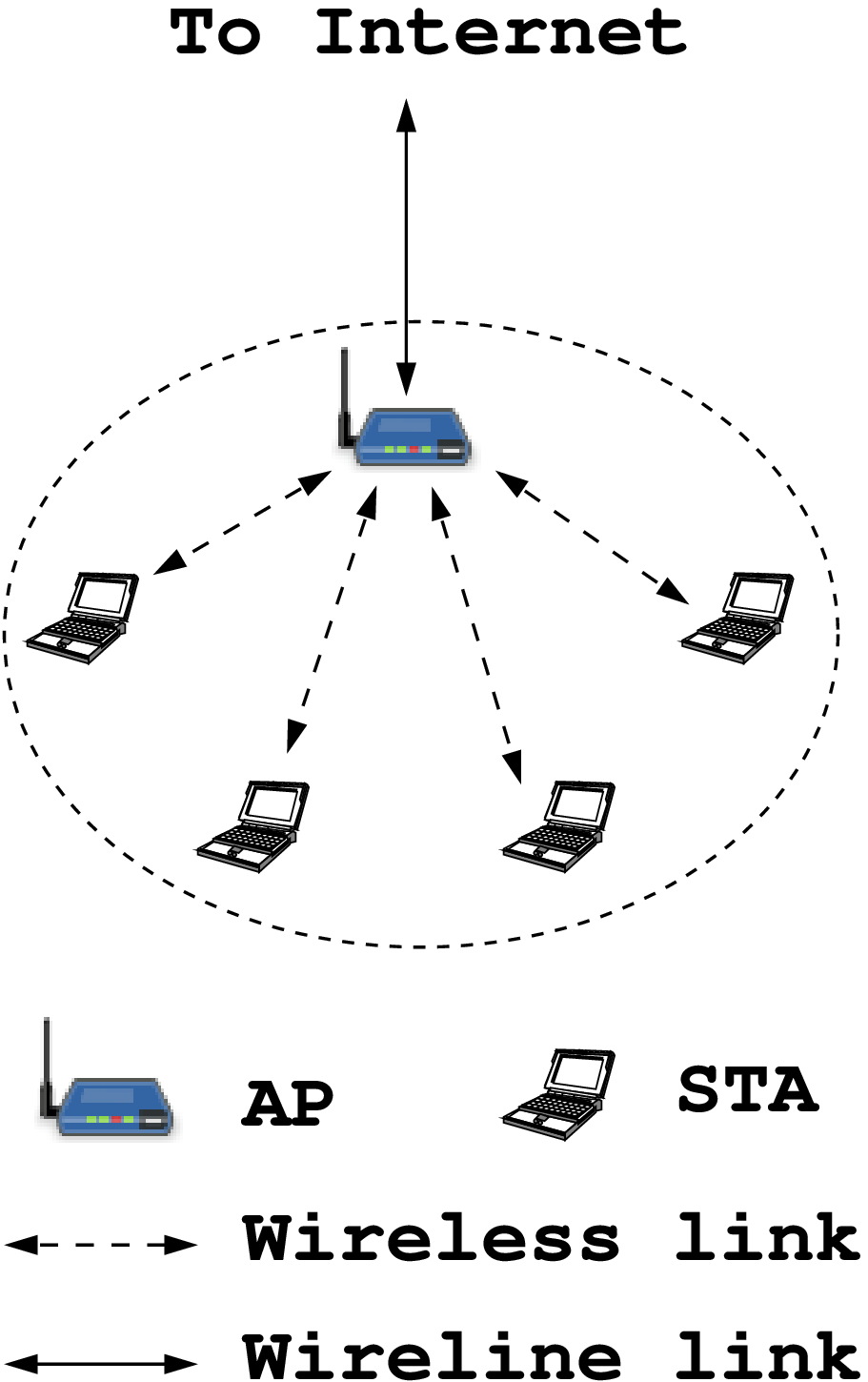}
  \caption{A single cell in the \textit{infrastructure} mode: In this
        case, STAs communicate among themselves and access the
        Internet through an Access Point (AP). \label{fig:singleCellInfra}}  
    \end{center}
  \end{minipage}
\end{figure}

This paper is concerned with analytical modeling of DCF-based WLANs in the \textit{single cell} scenario. As in~\cite{wanet.kumar_etal07new_insights}, we define a single cell to be a set of closely located 802.11 nodes such that: (i) every node can sense the transmissions by every other node, and (ii) every node can decode the transmissions by every other node in absence of interference. Figure~\ref{fig:singleCellAdHoc} depicts an example of a single cell WLAN which operates in the \textit{ad hoc} mode. In this example, users or client stations (STAs) can directly communicate among themselves. Figure~\ref{fig:singleCellInfra} depicts another example of a single cell WLAN which operates in the \textit{infrastructure} mode. In this latter example, STAs are primarily interested in accessing the Internet through an Access Point (AP). The analytical model in this paper applies to single cells that operate either in the ad hoc mode or in the infrastructure mode. To develop our analytical model, we do not need to distinguish between the AP and the STAs, and we call an AP or a STA, a ``node''. We assume that the readers are familiar with the DCF MAC protocol, the details of which can be found in the standard document~\cite{wanet.IEEE802dot11standard2007}.


Analytical modeling of 802.11 WLANs has been a topic of great interest ever since the publication of the original version of the standard~\cite{wanet.IEEE8021199standard}. In his seminal work, Bianchi~\cite{wanet.bianchi00performance} proposed an accurate analytical model for DCF-based single cell WLANs consisting of \textit{saturated} nodes. A node is said to be saturated if it always has a packet to transmit, i.e., if its transmission queue never becomes empty. The saturation model is often useful in providing a sufficient condition for stability~\cite{queueing.baccelli-foss95saturation,wanet.kumar-patil97cdma-aloha}.\footnote{It is shown in~\cite{queueing.baccelli-foss95saturation} that, if the arrival point process is a stationary and ergodic \textit{marked point process} of finite intensity, then under certain constraints (which are satisfied by several classical open queueing networks), the system is stable if the intensity of the arrival process is strictly smaller than the ``intensity of the departure process obtained by saturating the queues.'' By ``stability'' here we mean that the (joint) queue length process has a \textit{proper} stationary distribution.} In fact, the saturation assumption adopted by Bianchi~\cite{wanet.bianchi00performance} was partly motivated by the requirement to determine the maximum traffic intensity that the system can sustain under stable conditions. The saturation assumption, of course, simplifies the modeling problem since the node queues never become empty and queueing dynamics can be ignored. However, the saturation assumption is not valid, in general. In many real applications, e.g., web browsing and packetized voice telephony, a single cell WLAN might operate well below saturation.  

When a node is not saturated, its (transmission) queue can be empty for a positive fraction of time during which it does not contend for accessing the medium. Thus, queueing dynamics must be studied in order to determine the fraction of time for which a node contends. Due to the contention for medium access, however, the queue length processes of the nodes are \textit{coupled}. Hence, modeling a single cell WLAN with non-saturated nodes involves analyzing a \textit{``coupled queue system''} where the queues are served according to the CSMA/CA protocol as standardized in IEEE 802.11 DCF~\cite{wanet.IEEE802dot11standard2007}. In this paper, we apply a \textbf{State Dependent Attempt Rate} (SDAR) approximation (see Approximation~\ref{approx:SDAR} in Section~\ref{subsec:model}) to study such a coupled queue system.

\subsection{Literature Survey}
\label{subsec:literature}

Bianchi~\cite{wanet.bianchi00performance} developed the seminal analytical model for DCF-based single cell WLANs consisting of saturated nodes. The key approximation in Bianchi's model is the following: 

\begin{approximation}[Bianchi's Approximation] 
\label{approx:bianchi-decoupling}
\textit{Every transmission attempt by a node collides with a constant probability $\gamma$ regardless of the history of collisions.} 
\hfill \qed 
\end{approximation} 

Approximation~\ref{approx:bianchi-decoupling} is popularly known as the \textit{decoupling approximation} since it decouples the backoff processes of the nodes through a constant parameter $\gamma$. In reality, the backoff processes of the nodes are coupled precisely due to collisions. Applying the decoupling approximation, Approximation~\ref{approx:bianchi-decoupling}, Bianchi considered a tagged node, and developed a two-dimensional Markov chain to model the evolution of the backoff process of the tagged node. Since the \textit{backoff parameters} (i.e., $CW_{min}$, $CW_{max}$, and the retransmit limit) of the nodes are identical in the DCF, any node can be taken as the tagged node. The two-dimensional Markov chain keeps track of the \textit{backoff stage} (i.e., the number of times the same packet has already collided) and the \textit{backoff count} (i.e., the remaining number of backoff slots for the next transmission attempt to begin) of the tagged node. From the stationary distribution of the two-dimensional Markov chain, Bianchi obtained ``the probability $\beta$ that a node attempts a transmission in a randomly chosen slot'' as a function of the collision probability $\gamma$. Let \[\beta = G(\gamma)\] denote this function. With $n$ saturated nodes in the single cell, Bianchi obtained the collision probability $\gamma$ as a function of the attempt probability $\beta$ by \[\gamma = \Gamma(\beta) := 1 - (1 - \beta)^{n-1},\] since every node attempts a transmission in any randomly chosen slot with probability $\beta$. We emphasize that, $\gamma$ is \textit{not} the probability that a collision occurs in a randomly chosen slot; $\gamma$ is the probability that an attempted transmission collides. The above two equations yield a fixed point equation $\gamma = \Gamma(G(\gamma))$. The transmission attempt probability $\beta$ and the system throughput could be obtained using the solution of the fixed point equation. 

Cali et al.~\cite{wanet.cali-etal00throughput-limit,wanet.cali-etal00adaptive-backoff} proposed and analyzed a $p$-persistent version of IEEE 802.11. Considering saturated nodes, they obtained a theoretical upper limit, and also proposed an adaptive backoff mechanism to achieve the theoretical throughput limit. Kumar et al.~\cite{wanet.kumar_etal07new_insights} generalized the Bianchi model to arbitrary distributions of backoff times, backoff multipliers and retransmit limits. 

Motivated by the need to understand the performance of WLANs under realistic traffic conditions, modeling of the non-saturated case has attracted much attention in the recent past. Several models for the traffic arrival processes have been considered in the literature. Under the ``Poisson traffic'' model, packets of fixed size are assumed to arrive into the node queues according to independent Poisson processes of given rates~\cite{wanet.tickoo-sikdar04finite-load-queueing-model-802.11-MAC,wanet.Cantieni_etal05finite-load-and-multirate,wanet.malone_etal07finite-load-heterogeneous,wanet.ganesh-duffy-07-nonsat-commletter,wanet.choi_etal08matrix-analytic,wanet.huang-duffy09buffering-hypothesis,wanet.garetto-chiasserini05802.11-MAC-sporadic-traffic,wanet.foh-zukerman02SDSR}. Models for packetized voice telephony have been considered in~\cite{wanet.hegde_etal05voice-capacity,wanet.harsha07WiNet}. In~\cite{wanet.winands_etal04finite-source-feedback}, the authors consider ON-OFF traffic sources with exponentially distributed OFF periods, geometrically distributed number of packets during ON periods, and exponentially distributed packet payload sizes. The Poisson, voice, and ON-OFF traffic models belong to the so-called \textit{open-loop} type where the traffic sources \textit{do not} adjust their sending rates depending on the level of congestion in the network. Arrivals according to the \textit{closed-loop} control of TCP have been considered for \textit{long-lived} flows in~\cite{wanet.harsha07WiNet,wanet.bruno-etal06tcp-over-dot11,wanet.miorandi_etal06http_over_wlans}, and for \textit{short-lived} flows in~\cite{wanet.miorandi_etal06http_over_wlans,wanet.litjens_etalITC03integrated_packet_flow}.


As pointed out earlier, modeling the non-saturated case requires analyzing a coupled queue system. To that end, several simplifying assumptions have been made in the literature, either explicitly or implicitly. A detailed account of the veracity of common modeling hypotheses, in the context of 802.11 WLANs, can be found in~\cite{wanet.huang_etal08MAC-modeling-hypotheses}. We follow the precise terminology of~\cite{wanet.huang_etal08MAC-modeling-hypotheses} to comment on the relevant assumptions. For any tagged node, define $C_k := 1$ if the $k^{th}$ transmission attempt by the node results in a collision, and define $C_k := 0$ otherwise. Also, for any tagged node, define $Q_k := 1$ if there is at least one packet awaiting transmission after the $k^{th}$ successful transmission from the node, and define $Q_k := 0$ otherwise. The simplifying assumptions in~\cite{wanet.tickoo-sikdar04finite-load-queueing-model-802.11-MAC}-\cite{wanet.ganesh-duffy-07-nonsat-commletter} are then equivalent to the following assumptions regarding the sequences $\{C_k\}$ and $\{Q_k\}$:

\begin{itemize}

\item [(A1)] The sequence $\{C_k\}$ consists of independent random variables. 

\vspace{-1mm} 

\item [(A2)] The sequence $\{C_k\}$ consists of identically distributed random variables. 

\vspace{-1mm} 

\item [(A3)] The sequence $\{Q_k\}$ consists of independent random variables. 

\vspace{-1mm} 

\item [(A4)] The sequence $\{Q_k\}$ consists of identically distributed random variables. 

\end{itemize}

In particular, the authors in~\cite{wanet.tickoo-sikdar04finite-load-queueing-model-802.11-MAC}-\cite{wanet.ganesh-duffy-07-nonsat-commletter} assume that: (i) in a randomly chosen slot, each node attempts a transmission with constant probability $\beta$, (ii) every (transmission) attempt collides with constant probability $\gamma$, and (iii) each queue is not empty after a departure with constant probability $q$.\footnote{In general, if the nodes are not identically parametrized, e.g., in 802.11e, then Node-$i$ is associated with the probabilities $\beta_i$, $\gamma_i$, and $q_i$. However, for a given node $i$, the probabilities $\beta_i$, $\gamma_i$, and $q_i$ are assumed to be constants independent of the current state of the system.} They analyze the evolution of the backoff process and the queue length process of each node \textit{in isolation} and obtain fixed point equations relating $\beta$, $\gamma$, and $q$. The probabilities $\beta$, $\gamma$, and $q$ are obtained by solving the fixed point equations. The collision probabilities, the throughputs and the mean packet delays are obtained using the solution of the fixed point equations. 

Note that, Assumptions (A1)-(A4), essentially \textit{``decouple''} the transmission attempt processes, the backoff processes and the queue length processes of the nodes through the (unknown) constant probabilities $\beta$, $\gamma$, and $q$, which are obtained by solving certain fixed point equations. Assumptions (A1) and (A2) regarding the collision sequence $\{C_k\}$ constitute the decoupling approximation introduced by Bianchi~\cite{wanet.bianchi00performance} which can be called a \textit{collision-decoupling approximation}. Assumptions (A3) and (A4) regarding the queue-status sequence $\{Q_k\}$ are specific to the non-saturated case and, together, they can be called a \textit{queue-decoupling approximation}. 

In~\cite{wanet.choi_etal08matrix-analytic}, the authors assume only (A1) and (A2) and apply a matrix-geometric analytic method. It was shown in~\cite{wanet.huang-duffy09buffering-hypothesis} that Assumption (A4) leads to inaccurate predictions for throughputs when the arrival rates into the queues differ significantly from each other. In~\cite{wanet.garetto-chiasserini05802.11-MAC-sporadic-traffic} it is argued, by providing results from NS-2 simulations~\cite{wanet.ns2}, that the collision-decoupling approximation of Bianchi works well in the saturated case, but leads to inaccurate results in the non-saturated case. A detailed discussion of the validity of Assumptions (A1)-(A4) can be found in~\cite{wanet.huang_etal08MAC-modeling-hypotheses} where the authors provide evidence from NS-2 simulations to conclude that:

\begin{enumerate}

\item Assumption (A1) is valid regardless of whether the nodes are saturated or not. 


\item Assumption (A2) is valid when the nodes are saturated but is not valid, in general, when the nodes are not saturated. 


\item Assumption (A3) and (A4) are not valid. 

\end{enumerate}

In~\cite{wanet.garetto-chiasserini05802.11-MAC-sporadic-traffic}, the authors propose to model the transmission attempt probability of the nodes at any instant $t$ as a function of the number of non-empty nodes in the system at $t$. Clearly, in~\cite{wanet.garetto-chiasserini05802.11-MAC-sporadic-traffic}, the attempt probabilities of the nodes are \textit{state-dependent}. The approach of state-dependent attempt probabilities has also been adopted in~\cite{wanet.harsha07WiNet,wanet.bruno-etal06tcp-over-dot11}. In~\cite{wanet.foh-zukerman02SDSR,wanet.miorandi_etal06http_over_wlans,wanet.litjens_etalITC03integrated_packet_flow} a \textit{state-dependent service rate} approach is adopted. 


\subsection{Preview of Contributions}
\label{subsec:contributions} 

We develop a new approach to model single cells with non-saturated nodes under Poisson packet arrivals. Guided by the reported inaccuracy of the state-independent approach, we adopt a state-dependent attempt probability approach as in~\cite{wanet.garetto-chiasserini05802.11-MAC-sporadic-traffic,wanet.harsha07WiNet,wanet.bruno-etal06tcp-over-dot11}. The state-dependent attempt probabilities in~\cite{wanet.garetto-chiasserini05802.11-MAC-sporadic-traffic} are obtained by an iterative method which requires computations involving a three-dimensional Markov chain. We, however, apply an approximation proposed in~\cite{wanet.harsha07WiNet} to obtain the state-dependent attempt probabilities (see Approximation~\ref{approx:SDAR} in Section~\ref{subsec:model}). As explained in Section~\ref{subsec:complexity}, our model is computationally less expensive than that in~\cite{wanet.garetto-chiasserini05802.11-MAC-sporadic-traffic}. In particular, our model requires computations involving a two-dimensional Markov chain. We emphasize that, even though we apply the approximation proposed in~\cite{wanet.harsha07WiNet}, the problem we address in this paper is completely different from that in~\cite{wanet.harsha07WiNet}. The problem setting in~\cite{wanet.harsha07WiNet} is such that analysis of queueing dynamics is not required whereas we analyze a coupled queue system. 

Our contributions in this paper are the following:

\begin{itemize}

\item We develop a Markov model with Poisson packet arrivals. Our Markov model reduces a single cell WLAN with non-saturated nodes to a coupled queue system with SDAR service discipline (Section~\ref{sec:model}). We provide a sufficient condition under which the joint queue length Markov chain is positive recurrent (Theorem~\ref{thm:positive-recurrence-Qt}). 

\vspace{-1mm} 

\item For the case when the arrival rates into the queues are equal, we propose a technique to reduce the state space of the coupled queue system (Section~\ref{sec:reduction-state-space}). For the case when the buffer sizes of the queues are finite and equal, we propose an iterative method to obtain the stationary distribution of the reduced state process (Section~\ref{subsec:iterative-method}). Our iterative method is computationally less expensive than that in~\cite{wanet.garetto-chiasserini05802.11-MAC-sporadic-traffic} (Section~\ref{subsec:complexity}), and yet, it provides accurate predictions for important performance measures (Section~\ref{sec:results}). 

\vspace{-1mm} 

\item We applied the SDAR approximation to modify the MAC layer of NS-2, keeping all other layers unchanged. Originally, our objective in doing so was to validate the SDAR approximation itself by comparing the results obtained from the unmodified and modified NS-2 simulations. However, by doing so, we could also demonstrate the possibility of improving the speed of simulations by \textit{model-based simulation} at the MAC layer. We show that the SDAR model of contention provides an accurate model for the CSMA/CA protocol in single cells and, at the same time, achieves speed-ups (w.r.t.~MAC layer   operations) up to 1.55 to 5.4 depending on the arrival rates and the number of nodes in the single cell WLAN. 

\end{itemize}

\subsection{Outline of the Paper}
\label{subsec:outline}

We summarize our network model and assumptions in Section~\ref{sec:network-model-assumptions}. In Section~\ref{sec:model} we introduce the SDAR approximation and develop a Markov model with Poisson packet arrivals and infinite buffers. In Section~\ref{sec:reduction-state-space}, assuming equal arrival rates, we reduce the state space of the coupled queue system and obtain the transition probability matrix of the reduced state process. To demonstrate the predictive capability of our model, we restrict to the case of finite and equal buffers in Section~\ref{sec:finite-buffer-case}. In Section~\ref{subsec:iterative-method} we propose an iterative method to obtain the stationary distribution of the reduced state process (for equal arrival rates, finite and equal buffers), using which, we obtain predictions for important performance measures in Section~\ref{subsec:performance-measures}. In Section~\ref{sec:SDAR-simulation-technique} we report how the SDAR heuristic technique could be applied to improve the speed of simulations. In Section~\ref{sec:results} we validate our coupled queue model and our iterative method by comparing with NS-2 simulations where we also discuss some simulation results for the case of unequal arrival rates. Section~\ref{sec:conclusion} concludes the paper. Proofs and derivations have been provided in the appendices. 


\section{Network Model and Assumptions}
\label{sec:network-model-assumptions}

We consider a IEEE 802.11 DCF-based single cell WLAN consisting of $M$ nodes. Our network model and assumptions are given in the following: 

\begin{itemize}

\item The nodes are \textit{homogeneous}. This means that the nodes use identical backoff parameters, i.e., they use identical $CW_{min}$, $CW_{max}$, and ``retransmit limit'' (which is indeed true for the DCF MAC protocol). 

\vspace{-1mm}

\item The arrival processes bring packets of fixed size $L$ into the node queues according to independent Poisson processes. The arrival rate in packets/sec into Node-$i$'s queue is denoted by $\lambda_i$. 

\vspace{-1mm}

\item The nodes use equal Physical layer (PHY) rates to transmit their packets. 

\vspace{-1mm}

\item The wireless channel is \textit{error-free} which implies that \textit{single transmissions are always successful}. 

\vspace{-1mm}

\item There is no \textit{packet capture}. Simultaneous transmissions (i.e., collisions) always result in the failure of all the involved transmissions. Thus, \textit{there can be at most one successful transmission at any point of time}. 

\end{itemize}

\section{A Coupled Queue Model}
\label{sec:model}

Since there can be at most one successful transmission at any point of time, the system can be viewed as a single server serving multiple queues. The $M$ queues corresponding to the $M$ nodes are coupled essentially due to MAC contention and our immediate objective is to model the evolution of the joint queue length process. We proceed by embedding at the so-called ``channel slot boundaries'' described in the following.

\subsection{Channel Slots}
\label{subsec:channel-slot}

Let $\sigma$ denote the duration of a backoff slot in seconds. The duration $\sigma$ is a PHY parameter~\cite{wanet.IEEE802dot11standard2007}. We call a time unit equal to the duration $\sigma$ of a backoff slot, a \textit{system slot}. We call a period of time during which all the nodes in the system have empty queues, a \textit{system-empty period}. For analytical convenience, we make the following approximations: 

\begin{enumerate}

\item [($a_1$)] Nodes always sample non-zero backoffs. Consequently, when an activity period ends (i.e., after a successful transmission or a collision), subsequent transmission attempts can occur only after at least one backoff slot. 

\item [($a_2$)] System-empty periods are integer multiples of system slots. 

\end{enumerate}


\begin{figure}[t]
\centering \
  \begin{minipage}{15cm}
  \begin{center}
\includegraphics[scale=0.4]{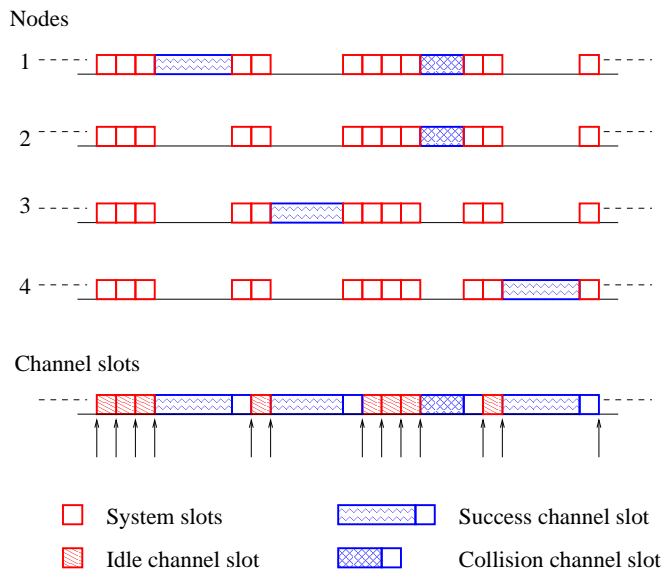}
     \caption{Description of channel slots: The channel slot
       boundaries indicated by arrows are the only possible attempt
       instants. \label{fig:channel_slot}} 
  \end{center}
  \end{minipage}
\end{figure}

As per the rules of the DCF, in reality, nodes do sample 0 backoffs with some positive probability. Approximation ($a_2$) is also not true, in general, since arrivals occur in real time. However, it will be clear from the accuracy of our analytical model that the errors due to Approximations ($a_1$) and ($a_2$) are negligible. Owing to Approximation~($a_1$), transmission attempts in the system can possibly occur only immediately after the end of a backoff slot. These possible attempt instants have been indicated by arrows in Figure~\ref{fig:channel_slot}, which depicts the backoffs and the activities in a single cell. The channel slots that occur on the common medium have also been shown. We call the time interval between any two such possible attempt instants a \textit{channel slot}, and observe that the channel activity evolves over cycles of channel slots. Note that, \textit{channel slot boundaries are the only possible attempt instants}. 

When the system is empty, an idle channel slot of duration $\sigma$ occurs (see Approximation~($a_2$)). A succession of idle channel slots occur until arrivals make some of the nodes non-empty. Non-empty nodes sample non-zero backoffs (see Approximation~($a_1$)) and attempt transmissions when their backoff counters become 0. Depending on whether there are no attempts, only one attempt, or more than one attempt made in the system, an idle, a success, or a collision channel slot occurs. The duration of an idle channel slot when the system is non-empty is equal to the duration $\sigma$ of a backoff slot. By Approximation~($a_2$), when an activity period ends, subsequent transmission attempts can only occur after a backoff slot. Hence, we combine the time duration of $\sigma$ seconds, which immediately follows an activity period, with the activity period itself to form success or collision channel slots. The attempt process resumes at the end of channel slots, thereby creating more channel slots and the process repeats. 

Let $T_s$ (resp.~$T_c$) denote the duration of a successful transmission (resp.~a collision). The success time $T_s$ and the collision time $T_c$ depend on the PHY rate of transmission. Since the nodes use equal PHY rates and the packets are of fixed size $L$, the durations of $T_s$ and $T_c$ are fixed and equal for all the nodes. The duration $T_s$ corresponds to ``DATA-SIFS-ACK-DIFS'' in the Basic Access mode, or ``RTS-SIFS-CTS-SIFS-DATA-SIFS-ACK-DIFS'' in the RTS/CTS mode. The duration $T_c$ corresponds to ``DATA-DIFS'' in the Basic Access mode, or ``RTS-DIFS'' in the RTS/CTS mode. Note that $T_s$ and $T_c$ need not be integer multiples of system slots. The duration of a success channel slot is equal to $T_s+\sigma$ and that of a collision channel slot is equal to $T_c+\sigma$.

\subsection{Coupled Queue Formulation}
\label{subsec:model}

\begin{figure}[tb]
\centering \
  \begin{minipage}{15cm}
  \begin{center}
\includegraphics[scale=0.6]{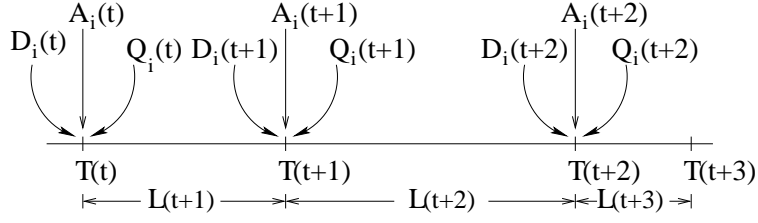}
     \caption{Evolution of Node-$i$'s queue. \label{fig:evolution}} 
  \end{center}
  \end{minipage}
\end{figure}

We model the evolution of the system over discrete time instants embedded at the channel slot boundaries. Figure~\ref{fig:evolution} depicts the evolution of the queue length process of a typical node in the system. Let $T(t), t = 0,1,2,3,\ldots$, with $T(0) = 0$, denote the channel slot boundaries. The $t^{th}$ channel slot $(t \geq 1)$ is precisely the time interval $\left[T(t-1),T(t) \right)$. The duration of the $t^{th}$ channel slot is denoted by $L(t)$. We analyze the system in discrete time, where \textit{the discrete time index $t$ corresponds to the actual (i.e., continuous) time instant $T(t)$}. 

Let $Q_i(t), t \geq 0, i = 1, 2, \ldots, M$, denote the number (of packets) in the $i^{th}$ node's queue at time $t$. Let $A_i(t)$ (resp.~$D_i(t)$), $t \geq 1, i = 1, 2, \ldots, M$, denote the number of arrivals into (resp.~departures from) the $i^{th}$ node's queue in the $t^{th}$ channel slot. Notice the embedding of $Q_i(t)$, $A_i(t)$ and $D_i(t)$ in Figure~\ref{fig:evolution}. Departures from the queues occur at the end of channel slots, since a packet is removed from a transmission queue only when an ACK is successfully received or a timeout occurs. Arrivals that occur during a channel slot are taken into account only at the next channel slot boundary, since, if a node is empty in the beginning of a channel slot, it can attempt only at the next channel slot boundary provided that packets arrive into its queue during the channel slot. The queue lengths are updated after the arrivals and the departures in the previous channel slot have been taken into account. 

In this section, we assume that \textit{each node has infinite buffer space}.\footnote{The infinite buffer assumption will be dropped in Section~\ref{sec:finite-buffer-case} and we will analyze the finite buffer case as well.} Thus, $\forall t \geq 0, i = 1, 2, \ldots, M$, we have $Q_i(t) \in \mathbb{N}$ where $\mathbb{N} := \{0, 1, 2, \ldots\}$. Also, $\forall t \geq 1, i = 1, 2, \ldots, M$, we have $A_i(t) \in \mathbb{N}$, and $D_i(t) \in \{0,1\}$. The last constraint follows from the fact that there can be at most one successful transmission, and thus, at most one departure, in a channel slot. Clearly, the ``number in the queue'' processes $\{Q_i(t), t \geq 0\}$, $1 \leq i \leq M$, evolve as: \begin{equation}
\label{eqn:queue-evolution}
Q_i(t+1) = Q_i(t) - D_i(t+1) + A_i(t+1).
\end{equation} Evidently, it must hold that $\left(Q_i(t) = 0\right) \Rightarrow \left(D_i(t+1) = 0\right)$, since there cannot be a departure from an empty queue. 

Due to the ``Poisson arrivals'' assumption, the distribution of the number of arrivals in a channel slot depends only on the duration of the channel slot. The duration of a channel slot is known if the channel slot type (i.e., whether it is an idle, a success or a collision channel slot) is known (see Section~\ref{subsec:channel-slot}). Let $L_{idle}$, $L_{succ}$ and $L_{coll}$ denote the duration in seconds of an idle, a success and a collision channel slot, respectively. Then, we have $L_{idle} = \sigma$, $L_{succ} = T_s + \sigma$ and $L_{coll} = T_c + \sigma$. For a given PHY, the duration $\sigma$ is known. Also, given the packet payload size $L$ and the PHY layer (transmission) rate, the durations $T_s$ and $T_c$ can be computed~\cite{wanet.bianchi00performance,wanet.kumar_etal07new_insights}. With slight abuse of notation, we indicate the occurrence of an idle, a success and a collision channel slot by $L(t) = L_{idle}$, $L(t) = L_{succ}$ and $L(t) = L_{coll}$, respectively, and define, $\forall t \geq 0, 1 \leq i \leq M$, the following probabilities: \begin{eqnarray}
\label{eqn:definition-dj}
d_i(j) &:=& P\big(A_i(t+1) = j \; \big | \; L(t+1) = L_{idle}\big)
\nonumber \\ 
&=& \left\{ \begin{array}{ll} e^{\displaystyle -\lambda_i \sigma} \;
  \displaystyle \frac{\left(\lambda_i \sigma \right)^{j}}{j!} &
  \forall j \geq 0, \\ 0 & \forall j < 0. \end{array} \right. 
\end{eqnarray} \begin{eqnarray}
\label{eqn:definition-sj}
s_i(j) &:=& P\big(A_i(t+1) = j \; \big | \; L(t+1) = L_{succ}\big)
\nonumber \\ 
&=& \left\{ \begin{array}{ll} e^{\displaystyle -\lambda_i \left(T_s +
    \sigma\right)} \; \displaystyle \frac{\big(\lambda_i \left(T_s +
    \sigma\right) \big)^{j}}{j!} & \forall j \geq 0, \\ 0 & \forall j
  < 0. \end{array} \right. 
\end{eqnarray} \begin{eqnarray}
\label{eqn:definition-cj}
c_i(j) &:=& P\big( A_i(t+1) = j \; \big | \; L(t+1) = L_{coll}\big)
\nonumber \\ 
&=& \left\{ \begin{array}{ll} e^{\displaystyle -\lambda_i \left(T_c +
    \sigma\right)} \; \displaystyle \frac{\big(\lambda_i \left(T_c +
    \sigma\right) \big)^{j}}{j!} & \forall j \geq 0, \\ 0 & \forall j
  < 0. \end{array} \right. 
\end{eqnarray}

As described in Section~\ref{subsec:channel-slot}, nodes can attempt only at the channel slot boundaries. Only those nodes that are non-empty at $t$ can attempt a transmission at $t$. Let $N(t)$ denote the number of non-empty nodes in the system at $t$. Then, by definition, we have \begin{equation}
\label{eqn:definition-Nt}
N(t) = \sum_{i=1}^M \ind{Q_i(t) > 0},
\end{equation} where $\ind{\cdot}$ denotes the indicator function. We now introduce an important approximation regarding the attempt processes of the nodes which was first proposed in~\cite{wanet.harsha07WiNet}. 

\begin{approximation}[State Dependent Attempt Rate] 
\label{approx:SDAR}
At any channel slot boundary $t$, $t \geq 0$, every non-empty node attempts a transmission with probability $\beta_{N(t)}$ where $\beta_{n}$ is the attempt probability of the nodes in a single cell consisting of $n$ \textbf{homogeneous} and \textbf{saturated} nodes. \hfill \qed 
\end{approximation} 

We call Approximation~\ref{approx:SDAR} the \textbf{State Dependent Attempt Rate (SDAR)} approximation. The $\beta_{n}$'s in the SDAR approximation can be obtained by a saturation analysis as in~\cite{wanet.bianchi00performance,wanet.kumar_etal07new_insights}. As a consequence of the SDAR approximation, given $N(t) = n$, the number of transmission attempts made in the system at the channel slot boundary $t$ is binomially distributed with parameters $n$ and $\beta_n$. Hence, the probability that the $(t+1)^{th}$ channel slot is an idle, a success or a collision channel slot can be obtained as follows: \begin{eqnarray}
\label{eqn:channel-slot-probabilities}
\left. \begin{array}{rcl} p_{idle,n} &:=& P\big(L(t+1) = L_{idle} \;
  \big| \; N(t) = n\big) = (1-\beta_n)^{n} \\ p_{succ,n} &:=&
  P\big(L(t+1) = L_{succ} \; \big| \; N(t) = n\big) = n \beta_n
  (1-\beta_n)^{n-1} \\ p_{coll,n} &:=& P\big(L(t+1) = L_{coll} \;
  \big| \; N(t) = n\big) = 1 - p_{idle,n} - p_{succ,n} \end{array}
\right\} 
\end{eqnarray} Furthermore, in case of a success channel slot, the packet departure can occur from any of the non-empty queues with equal probability. Thus, for all $t \geq 0$, $1 \leq i \leq M$, the number of departures $D_i(t+1)$ from the $i^{th}$ queue in the $(t+1)^{th}$ channel slot satisfy \begin{equation}
\label{eqn:departure-equallylikely}
P\left(D_i(t+1) = 1 \; \big| \; N(t) = n, L(t+1) = L_{succ}, Q_i(t) >
0 \right) = \frac{1}{n} \; .
\end{equation} Note that, if $L(t+1) \neq L_{succ}$ or $Q_i(t) = 0$, then $D_i(t+1) = 0$. 

Owing to the channel slot structure imposed by Approximations ($a_1$) and ($a_2$), the ``Poisson arrivals'' assumption, and the SDAR approximation, the joint queue length process $\{\bmath{Q}(t), t \geq 0\}$, where \[\bmath{Q}(t) := \big( Q_1(t), Q_2(t), \ldots, Q_M(t) \big),\] is an $M$-dimensional Discrete Time Markov Chain (DTMC) embedded at the channel slot boundaries.

\begin{theorem}
\label{thm:positive-recurrence-Qt}

The DTMC $\{\bmath{Q}(t), t \geq 0\}$ is positive recurrent if, $\forall i, 1 \leq i \leq M$, we have $\lambda_i > 0$, and \[\left(\sum_{i=1}^M \lambda_i \right) < \min_{1 \leq n \leq M} \Theta_{sat,n},\] where $\Theta_{sat,n}$ is the aggregate throughput in packets/sec in a single cell consisting of $n$ homogeneous and saturated nodes. 

\end{theorem}

\textit{Proof:} See Appendix~\ref{app:positive-recurrence-Qt}. \hfill \qed

\begin{remarks}
The variation of $\Theta_{sat,n}$ with $n$, in general, depends on the backoff parameters. However, for the default backoff parameters as prescribed in the 802.11 DCF, we observe that, for $M$ large enough ($M \geq 5$ suffices), we have, $\min_{1 \leq n \leq M} \Theta_{sat,n} = \Theta_{sat,M}$. Then, the DTMC $\{\bmath{Q}(t), t \geq 0\}$ is positive recurrent if the aggregate arrival rate $\sum_{i=1}^M \lambda_i$ is strictly less than the aggregate throughput $\Theta_{sat,M}$ of $M$ saturated nodes. Thus, Theorem~\ref{thm:positive-recurrence-Qt} exhibits the connection between the saturation throughput and the maximum stable throughput of the system. \hfill \qed 
\end{remarks}

By Theorem~\ref{thm:positive-recurrence-Qt}, there exist $\lambda_i > 0$, $1 \leq i \leq M$, such that the DTMC $\{\bmath{Q}(t), t \geq 0\}$ is positive recurrent. We denote the stationary distribution of the DTMC $\{\bmath{Q}(t), t \geq 0\}$ by $\nu(\bmath{k})$, $\bmath{k} = (k_1, k_2, \ldots, k_M) \in \mathbb{N}^M$. In principle, important performance measures such as collision probability, throughput and mean packet delay can be obtained once the stationary distribution of the DTMC $\{\bmath{Q}(t), t \geq 0\}$ is known. However, the DTMC $\{\bmath{Q}(t), t \geq 0\}$ has a state space $\mathbb{N}^M$, which is difficult to handle for $M > 1$. The transition structure of the DTMC $\{\bmath{Q}(t), t \geq 0\}$ is such that, for $M > 1$, it is difficult to obtain closed-form expressions for the $\nu(\bmath{k})$'s.\footnote{The $M = 1$ case can be easily analyzed, for example, by the ``generating function approach.''} Furthermore, it is not possible to numerically solve the infinite number of balance equations of the DTMC $\{\bmath{Q}(t), t \geq 0\}$. 

Thus, to validate the SDAR approximation, we replace ``the detailed implementation of the IEEE 802.11 DCF in NS-2 MAC layer'' with ``the SDAR model of contention'' keeping all other layers unchanged. The modifications are summarized in Section~\ref{sec:SDAR-simulation-technique}. In Section~\ref{sec:results} we compare the simulation results obtained from (a) the unmodified NS-2 with (b) the SDAR approximation in NS-2, and show that the simulation results obtained from (a) and (b) match extremely well. This validates the SDAR approximation for infinite or finite buffers, and equal or unequal arrival rates. In Section~\ref{subsec:performance-measures} we obtain numerical predictions for the performance measures such as collision probability, throughput and mean packet delay for the case of equal arrival rates, equal PHY rates, and finite and equal buffers. To proceed towards that goal, we first propose a state space reduction technique in Section~\ref{sec:reduction-state-space}.

\section{Reduction of the State Space}
\label{sec:reduction-state-space}

Suppose that the arrival rates into the queues are equal, i.e., let $\lambda_i = \lambda$, $\forall i, 1 \leq i \leq M$. (We still retain the ``infinite buffer'' assumption in this section.) Thus, $\forall i, 1 \leq i \leq M$, $\forall j, j = 0, 1, \ldots$, we have $d_i(j) = d(j)$, $s_i(j) = s(j)$, and $c_i(j) = c(j)$ where $d(j)$, $s(j)$, and $c(j)$ can be obtained by substituting $\lambda_i = \lambda$ in Equations~\eqref{eqn:definition-dj}-\eqref{eqn:definition-cj}.

\begin{definition}[Exchangeability~\cite{theory.feller71volII}]
\label{defn:exchangeability}
The random variables $X_1, X_2, \ldots, X_n$ are said to be \textbf{exchangeable} if all the $n!$ permutations $\left(X_{i_1}, X_{i_2}, \ldots, X_{i_n} \right)$ of $\left(X_1, X_2, \ldots, X_n \right)$, where $(i_1, i_2, \ldots, i_n)$ denotes a permutaion of $(1, 2, \ldots, n)$, have the same $n$-dimensional distribution. \hfill \qed 
\end{definition}

Since the nodes are homogeneous and they use \textit{equal} PHY rates, and since packets arrive into the node queues according to \textit{equal} rate Poisson processes, it follows that, $\forall t \geq 1$, the random variables $Q_1(t), Q_2(t), \ldots, Q_M(t)$, are exchangeable provided that the random variables $Q_1(0), Q_2(0), \ldots, Q_M(0)$, are exchangeable. We assume throughout this paper that the random variables $Q_1(0), Q_2(0), \ldots, Q_M(0)$, are exchangeable. We remark that, with unequal arrival rates into the node queues or with unequal PHY rate of transmission, the desired symmetry will be lost and the random variables $Q_1(t),Q_2(t), . . . ,Q_M(t)$, will not be exchangeable. 

Since, $\forall t \geq 0$, the random variables $Q_1(t), Q_2(t), \ldots, Q_M(t)$, are exchangeable, we consider the following alternative description of the system. We define the state of the system at a channel slot boundary $t$ by \[\mathcal{X}(t) := \big(Q_1(t),\mathcal{M}(t)\big)\] where $Q_1(t)$ denotes the number (of packets) in the queue of a tagged node\footnote{Any node can be taken as the tagged node due to symmetry.} at the channel slot boundary $t$ and $\mathcal{M}(t)$ denotes the number of nodes, other than the tagged node, that are non-empty at $t$. Thus, by definition, we have \begin{equation} 
\label{eqn:definition-Mt}
\mathcal{M}(t) = \displaystyle \sum_{i = 2}^M \ind{Q_i(t) > 0}.
\end{equation} Note that, $\forall t \geq 0$, we have, $0 \leq \mathcal{M}(t) \leq M-1$, and \begin{equation}
\label{eqn:relation-Nt-Mt}
N(t) = \ind{Q_1(t) > 0} + \mathcal{M}(t). 
\end{equation} The state space has now reduced from $\mathbb{N}^M$ to $\mathbb{N} \times \{0,1,\ldots,M-1\}$.

\subsection{Approximating the Process $\{\mathcal{X}(t)\}$ by a Process with a Few Unknown Parameters} 
\label{subsec:markov-approximation-of-Xt}

In Appendix~\ref{app:balance-eqn-Xt} we derive the one-step transition probabilities of the process $\{\mathcal{X}(t)\}$ where we show that the transition probabilities are functions of some unknown quantities which appear because of the following reason. To model the change in $\mathcal{M}(t)$ over one step, i.e., over one channel slot, one requires the probability that ``a departure from a non-tagged node leaves the queue empty.'' Clearly, a departure leaves the queue empty if the following two events occur together:

\begin{itemize}

\item [(E1)] In the beginning of the success channel slot, the queue contains exactly one packet (which departs at the end of the success channel slot), and 

\item [(E2)] The queue does not receive any packets in the success channel slot. 

\end{itemize}

Event (E2) occurs with probability $s(0)$ (see Equation~\eqref{eqn:definition-sj} and recall that the arrival rates are equal). However, with the state description $\mathcal{X}(t)$, the probability that event (E1) occurs cannot be determined for the non-tagged queues, since $\mathcal{X}(t)$ does not keep track of the number of packets in the non-tagged queues. As shown in Appendix~\ref{app:balance-eqn-Xt}, the unknown quantities are precisely the probabilities $q(i,n,t)$, $t \geq 0$, $i = 0, 1, 2, \ldots$, $0 \leq n \leq M-1$, given by \begin{equation*}
q(i,n,t) := P\big(Q_l(t) = 1 \; \big| \; Q_1(t) = i, \mathcal{M}(t) =
n, L(t+1) = L_{succ}, D_l(t+1) = 1, Q_l(t) > 0\big) 
\end{equation*} where $l$, $2 \leq l \leq M$, denotes the index of a non-tagged node. Since the nodes are homogeneous, and the arrival rates and the PHY rates are equal across the node indices, the index $l$ does not really matter, and hence, does not appear in the notation $q(i,n,t)$. The condition ${L(t+1)} = L_{succ}$ indicates that there is a departure in the $(t+1)^{th}$ channel slot. The condition $D_l(t+1) = 1$ indicates that the departure occured from the $l^{th}$ queue. The condition $Q_l(t) > 0$ must accompany the condition $D_l(t+1) = 1$, since otherwise $\left(Q_l(t) = 0\right) \Rightarrow \left(D_l(t+1) = 0\right)$, which is a contradiction. Clearly, \textit{$q(i,n,t)$ is the probability that the non-tagged queue from which a departure occurs in the $(t+1)^{th}$ channel slot contains exactly one packet at $t$, given the state description $\left(Q_1(t) = i, \mathcal{M}(t) = n\right)$}. 

The $q(i,n,t)$'s are not known precisely because the number of packets in the non-tagged queues are not kept in the state description $\mathcal{X}(t)$. Moreover, the $q(i,n,t)$'s cannot be obtained from the known quantities, namely, the arrival rate $\lambda$ and the state-dependent attempt probabilities $\beta_n$'s. All the other probabilities, namely, $p_{idle,n}$, $p_{coll,n}$, $p_{succ,n}$, $0 \leq n \leq M$, $d(j)$, $c(j)$, $s(j)$, $j = 0, 1, 2, \ldots$, that appear in the transition probabilities of the process $\{\mathcal{X}(t)\}$ (see Equations~\eqref{eqn:pi-jk-tplus1-first-part-final} and~\eqref{eqn:pi-jk-tplus1-second-part-final}) can be obtained from $\lambda$ and the $\beta_n$'s. 

The conditions $L(t+1) = L_{succ}$ and $D_l(t+1) = 1$ can be eliminated from the definition of the $q(i,n,t)$'s by the following lemma. 

\begin{lemma}
\label{lem:conditional-probability-simplification-exact}

$\forall j \geq 1$, $\forall l, 2 \leq l \leq M$, $\forall i, i = 0, 1, 2, \ldots$, $\forall n, 0 \leq n \leq M-1$, the following is true: \begin{eqnarray*} 
\label{eqn:conditional-probability-simplification-exact}
\lefteqn{P\big(Q_l(t) = j \; \big| \; Q_1(t) = i, \mathcal{M}(t) = n,
  L(t+1) = L_{succ}, D_l(t+1) = 1, Q_l(t) > 0 \big)} \nonumber \\ 
&=& P\big(Q_l(t) = j \; \big| \; Q_1(t) = i, \mathcal{M}(t) = n,
Q_l(t) > 0\big). \; \; \; \; \; \; \; \; \; \; \; \; \; \; \; \; \; \;
\; \; \; \; \; \; \; \; \; \; \; \; \; \; \; \; \; \; \; \; \; \; \;
\end{eqnarray*} 

\end{lemma}

\textit{Proof:} See  Appendix~\ref{app:proof-lemma-conditional-probability-simplification-exact}. \hfill \qed

Lemma~\ref{lem:conditional-probability-simplification-exact} says that, the probability that a non-tagged queue contains exactly $j$ packets ($j \geq 1$), given that it is non-empty and given the state description, does not depend on whether a departure occurs from that queue. Thus, applying Lemma~\ref{lem:conditional-probability-simplification-exact}, we redefine, $\forall t \geq 0$, $\forall i, i = 0, 1, 2, \ldots$, $\forall n, 0 \leq n \leq M-1$, $\forall l, 2 \leq l \leq M$, \begin{equation} 
\label{eqn:qint-definition-two}
q(i,n,t) := P\big(Q_l(t) = 1 \; \big| \; Q_1(t) = i, \mathcal{M}(t) = n, Q_l(t) > 0\big). 
\end{equation} We now apply an approximation first introduced in~\cite{wanet.sykasetal86buffanal} in the context of ALOHA networks, and later, also applied in~\cite{wanet.garetto-chiasserini05802.11-MAC-sporadic-traffic} in the context of 802.11 WLANs.

\begin{approximation}[Conditional Independence]
\label{approx:independence-exact-number}

$\forall t \geq 0$, $\forall i, i = 0, 1, 2, \ldots$, $\forall n, 0 \leq n \leq M-1$, $\forall l, 2 \leq l \leq M$, we impose the following: \begin{eqnarray}
\label{eqn:independence-exact-number}
\lefteqn{P\big(Q_l(t) = 1 \; \big| \; Q_1(t) = i,\mathcal{M}(t) = n,
  Q_l(t) > 0\big)} \nonumber \\ 
&=& P\big(Q_l(t) = 1 \; \big| \; Q_1(t) = i, N(t) = \ind{i > 0}+n,
Q_l(t) > 0\big) \nonumber \\ 
&=& P\big(Q_l(t) = 1 \; \big| \; N(t) = \ind{i > 0}+n, Q_l(t) >
0\big). \; \; \; \; \; \; \; \; 
\end{eqnarray}

\vspace{-2mm}

\hfill \qed 

\end{approximation}

Approximation~\ref{approx:independence-exact-number} pertains to the second step in Equation~\eqref{eqn:independence-exact-number}, which amounts to saying that, the probability that a non-tagged queue contains exactly one packet, given that it is non-empty and given the number of non-empty nodes in the system, is independent of the \textbf{exact number of packets} in the tagged queue. 

We define, $\forall t \geq 0$, $\forall n, 1 \leq n \leq M$, $\forall l, 2 \leq l \leq M$, \[q(n,t) := P\big(Q_l(t) = 1 \; \big| \; N(t) = n, Q_l(t) > 0\big).\] Then, Approximation~\ref{approx:independence-exact-number} says that, \begin{eqnarray} 
\label{eqn:relation-qint-qnt}
q(i,n,t) = \left\{ \begin{array}{ll} q(n,t) & \mbox{if} \; i = 0,
  \\ q(n+1,t) & \mbox{if} \; i > 0. \end{array} \right.
\end{eqnarray} 

We approximate the process $\{\mathcal{X}(t), t \geq 0\}$ by a process $\{\tilde{\mathcal{X}}(t), t \geq 0\}$ as follows. The process $\{\tilde{\mathcal{X}}(t), t \geq 0\}$ has the same state description as that of $\{\mathcal{X}(t), t \geq 0\}$. However, the transition probabilities of the process $\{\tilde{\mathcal{X}}(t), t \geq 0\}$ are obtained from the transition probabilities of the process $\{\mathcal{X}(t), t \geq 0\}$ by first applying Lemma~\ref{lem:conditional-probability-simplification-exact} and then applying Approximation~\ref{approx:independence-exact-number}, i.e., the transition probabilities of the process $\{\tilde{\mathcal{X}}(t), t \geq 0\}$ now involve the unknowns $q(n,t)$, $1 \leq n \leq M$. We further impose the conditions that, $\forall n, 1 \leq n \leq M$, $\forall t \geq 0$, \[q(n,t) = \tilde{q}(n),\] where the $\tilde{q}(n)$'s are unknown constants independent of $t$. Thus, the process $\{\tilde{\mathcal{X}}(t), t \geq 0\}$ models the event (E1) through constant time-independent probabilities $\tilde{q}(n)$'s. 

With the $\tilde{q}(n)$'s thus defined, the probability that ``a departure from a non-tagged queue leaves the queue empty given that it is non-empty in the beginning of the success channel slot and that there are $n$ non-empty nodes in the system in the beginning of the success channel slot'' is given by $\tilde{q}(n)s(0)$ which is the joint probability of the events (E1) and (E2). Thus, \textit{we can regard the process $\{\tilde{\mathcal{X}}(t), t \geq 0\}$ as a DTMC (embedded at the channel slot boundaries) whose transition probabilities are functions of the unknown parameters $\tilde{q}(n)$'s yet to be determined}. 

In the remainder of this paper, a random variable $X$ (resp.~a quantity $x$) defined for the process $\{\mathcal{X}(t), t \geq 0\}$ will have an analogous random variable $\tilde{X}$ (resp.~a quantity $\tilde{x}$) defined for the process $\{\tilde{\mathcal{X}}(t), t \geq 0\}$, and vice versa. Also, random variables and probabilities without the time argument would correspond to the stationary regime assuming that the stationary regime exists.

\subsection{Transition Probability Matrix of the DTMC
  $\{\tilde{\mathcal{X}}(t)\}$} 
\label{subsec:transition-probabilities}

We define the transition probabilities of the DTMC $\{\tilde{\mathcal{X}}(t), t \geq 0\}$ as follows: 

\vspace{1mm}

$A_j(n,k) :=$ \parbox[t] {14 cm} {transition probability from the state $(0,n)$ to the state $(j,k)$, $j = 0, 1, 2, \ldots$, $0 \leq n,k \leq M-1$.} 

$B_j(n,k) :=$ \parbox[t] {14 cm} {transition probability from the state $(i,n)$ to the state $(i+j,k)$, $i = 1, 2, \ldots$, $j = -1, 0, 1, \ldots$, $0 \leq n,k \leq M-1$.} 

\vspace{1mm}

The transition probabilities of the process $\{\tilde{\mathcal{X}}(t), t \geq 0\}$ have been derived in  ~\ref{app:balance-eqn-tilde-Xt} (see Equations~\eqref{eqn:pi-jk-tplus1-first-part-final-tilde} and~\eqref{eqn:pi-jk-tplus1-second-part-final-tilde}). It follows from Equations~\eqref{eqn:pi-jk-tplus1-first-part-final-tilde} and~\eqref{eqn:pi-jk-tplus1-second-part-final-tilde} that each of the $A_j(n,k)$'s and the $B_j(n,k)$'s can be separated into two parts, a part that contains an unknown parameter $\tilde{q}(\cdot)$ and a part that does not, and we define \begin{eqnarray} 
\label{eqn:Aj0nk-Bjnk-qn}
A_j(n,k) &:=& A_j^{(0)}(n,k) + \tilde{q}(n) A_j^{(1)}(n,k), \nonumber \\
B_j(n,k) &:=& B_j^{(0)}(n,k) + \tilde{q}(n+1) B_j^{(1)}(n,k),
\end{eqnarray} where $A_j^{(0)}(n,k)$, $A_j^{(1)}(n,k)$, $B_j^{(0)}(n,k)$ and $B_j^{(1)}(n,k)$ can be obtained from Equations~\eqref{eqn:pi-jk-tplus1-first-part-final-tilde} and~\eqref{eqn:pi-jk-tplus1-second-part-final-tilde}, and are given by \begin{eqnarray} 
\label{eqn:Ajnk-Bjnk-1}
A_j^{(0)}(n,k) &=& {{M-n-1 \choose k-n}} \left( p_{idle,n} \; d(j) \;
\left(1-d(0)\right)^{k-n} d(0)^{M-k-1} \right. \nonumber \\ 
&& \; \; \; \; \; \; \; \; \; \; \; \; \; \; \; \; \; \; \; \; \; \;
\; \; \; \; \; \; \; \; + p_{coll,n} \; c(j) \;
\left(1-c(0)\right)^{k-n} c(0)^{M-k-1} \nonumber \\ 
&& \left. \; \; \; \; \; \; \; \; \; \; \; \; \; \; \; \; \; \; \; \; \; \;
\; \; \; \; \; \; \; \; \; \; \; \; \; \; + p_{succ,n} \; s(j) \;
\left(1-s(0)\right)^{k-n} s(0)^{M-k-1} \right), \; \; \; \; \; \; \; \;
\; \; \; \; \; \; \; \; \; \; \; \; \; \; \; \; \; \; \; \; \; \; \;
\; \; \; \; \; \; \;  
\end{eqnarray} \begin{eqnarray}
\label{eqn:Ajnk-Bjnk-2}
A_j^{(1)}(n,k) &=& p_{succ,n} \; s(j) \left(1-s(0)\right)^{k-n}
s(0)^{M-k-1} \left( {{M-n \choose k-n+1}} \left(1-s(0)\right) -
{{M-n-1 \choose k-n}} \right), \; \; \; \; \; \; \; \; \; \; \; 
\end{eqnarray} \begin{eqnarray}
\label{eqn:Ajnk-Bjnk-3}
B_j^{(0)}(n,k) &=& {{M-n-1 \choose k-n}} \left( p_{idle,n+1} \; d(j)
\; \left(1-d(0)\right)^{k-n} d(0)^{M-k-1} \right. \nonumber \\
&& \; \; \; \; \; \; \; \; \; \; \; \; \; \; \; \; \; \; \; \; \; \;
\; \; \; \; \; \; \; \; + p_{coll,n+1} \; c(j) \;
\left(1-c(0)\right)^{k-n} c(0)^{M-k-1} \nonumber \\ 
&& \left. \; \; \; \; \; \; \; \; \; \; \; \; \; \; \; \; \; \; \; \; \; \;
\; \; \; \; \; \; \; \; \; \; \; \; \; \; \; + p_{succ,n+1} \; s(j) \;
\left(1-s(0)\right)^{k-n} s(0)^{M-k-1} \right) \nonumber \\
&& + \; {{M-n-1 \choose k-n}} \frac{p_{succ,n+1}}{n+1}
\left(1-s(0)\right)^{k-n} s(0)^{M-k-1} \left(s(j+1) - s(j)\right), \;
\; \; \; \; \; \; \; \; \; \; \; \; \; \; \; \; \; \; \; \; \; \; \;
\; \; \; \; \; 
\end{eqnarray} \begin{eqnarray}
\label{eqn:Ajnk-Bjnk-4}
B_j^{(1)}(n,k) &=& \left( \frac{n}{n+1} \right) p_{succ,n+1} \; s(j) \;
\left( 1-s(0)\right)^{k-n} s(0)^{M-k-1} \nonumber \\
&& \; \; \; \; \; \; \; \; \; \; \; \; \; \; \; \; \; \; \; \; \; \;
\; \cdot \; \left( {{M-n \choose k-n+1}} \left(1-s(0)\right) - {{M-n-1
    \choose k-n}} \right). \; \; \; \; \; \; \; \; \; \; \; \; \; \; \;
\; \; \; \; \; \; \; \; \; \; \; \; \; \; \; \; \; \; \; \; \; \; \;
\; \; \; \; 
\end{eqnarray} 

Let $\bmath{A_j}^{(0)}$, $\bmath{A_j}^{(1)}$,  $\bmath{B_j}^{(0)}$ and $\bmath{B_j}^{(1)}$ denote the $M \times M$  matrices with their $(n,k)^{th}$ entries given by $A_j^{(0)}(n,k)$, $A_j^{(1)}(n,k)$, $B_j^{(0)}(n,k)$ and $B_j^{(1)}(n,k)$, respectively. Using this matrix notation, Equation~\eqref{eqn:Aj0nk-Bjnk-qn} can be rewritten as \begin{eqnarray}
\label{eqn:Aj0nk-Bjnk-Deltaq}
\bmath{A_j} := \bmath{A_j}^{(0)} + \bmath{\Delta_{\tilde{q},A}}
\bmath{A_j}^{(1)} \; \; , \; \; \bmath{B_j} := \bmath{B_j}^{(0)} +
\bmath{\Delta_{\tilde{q},B}} \bmath{B_j}^{(1)},
\end{eqnarray} where $\bmath{\Delta_{\tilde{q},A}} = diag \left( 0, \tilde{q}(1), \ldots, \tilde{q}(M-1)\right)$ and $\bmath{\Delta_{\tilde{q},B}} = diag \left( \tilde{q}(1), \tilde{q}(2), \ldots, \tilde{q}(M)\right)$ are $M \times M$ diagonal matrices. Let $\tilde{\pi}(j,k)$, $j \geq 0$, $0 \leq k \leq M - 1$, denote the stationary distribution of the DTMC $\{\tilde{\mathcal{X}}(t), t \geq 0\}$ (assuming that the stationary distribution exists). We define, $\forall j, j \geq 0$, \[\bmath{\tilde{\pi}_{j}} := \big(\tilde{\pi}(j,0), \tilde{\pi}(j,1), \ldots, \tilde{\pi}(j,M-1)\big),\] and \[\bmath{\tilde{\pi}} := \big(\bmath{\tilde{\pi}_{0}}, \bmath{\tilde{\pi}_{1}}, \bmath{\tilde{\pi}_{2}}, \ldots\big).\] Using the above notation, the balance equations for the $\tilde{\pi}(j,k)$'s can be written as \begin{eqnarray}
\label{eqn:2D-balance-eqn-matrix-form}
\bmath{\tilde{\pi}} = \bmath{\tilde{\pi}} \bmath{P},
\end{eqnarray} where the transition probability matrix $\bmath{P}$ has the following \textit{``$M/G/1$ type''}~\cite{theory.neuts89MG1} structure \begin{eqnarray}
\label{eqn:definition-tpm}
\bmath{P} = \left[ \begin{array}{ccccc} \bmath{A_0} & \bmath{A_1} &
    \bmath{A_2} & \bmath{A_3} & \cdots \\ \bmath{B_{-1}} & \bmath{B_0}
    & \bmath{B_1} & \bmath{B_2} & \cdots \\ \bmath{0} & \bmath{B_{-1}}
    & \bmath{B_0} & \bmath{B_1} & \cdots \\ \bmath{0} & \bmath{0} &
    \bmath{B_{-1}} & \bmath{B_0} & \cdots \\ \bmath{\vdots} &
    \bmath{\vdots} & \bmath{\vdots} & \bmath{\vdots} & \bmath{\ddots} 
\end{array} \right].
\end{eqnarray}


\section{The Finite Buffer Case}
\label{sec:finite-buffer-case} 

Let us now turn to the question of determining the unknown parameters $\tilde{q}(n)$'s. Applying exchangeability, we can write \begin{eqnarray} 
\label{eqn:exchangeability-application}
q(n,t) &:=& P\big(Q_l(t) = 1 \; \big| \; N(t) = n, Q_l(t) > 0\big) \;
\; \; \; \; \; \; \; \; \; \; \; \; \; \; \; (2 \leq l \leq M)
\nonumber \\ 
&=& P\big(Q_1(t) = 1 \; \big| \; N(t) = n, Q_1(t) > 0\big) \; \; \; \;
\; \; \; \; \; \; \; \mbox{(exchangeability).} 
\end{eqnarray} 

Equation~\eqref{eqn:exchangeability-application} provides the clue to estimate the $\tilde{q}(n)$'s from the stationary distribution of the DTMC $\{\tilde{\mathcal{X}}(t), t \geq 0\}$ as follows: ( recall that random variables and probabilities without a time argument indicate the stationary values) \begin{eqnarray}
\label{eqn:approx-q-tilde-n}
\tilde{q}(n) &\approx& P\big(\tilde{Q}_1 = 1 \; \big| \; \tilde{N} =
n, \tilde{Q}_1 > 0\big) \nonumber \\
&=& \frac{P\big(\tilde{Q}_1 = 1, \tilde{N} = n\big)}{P\big(\tilde{Q}_1
  > 0, \tilde{N} = n\big)} = \frac{P\big(\tilde{Q}_1 = 1,
  \tilde{\mathcal{M}} = n-1\big)}{P\big(\tilde{Q}_1 >0,
  \tilde{\mathcal{M}} = n-1\big)} \nonumber \\ 
&=& \frac{\tilde{\pi}(1,n-1)}{\sum_{j=1}^{\infty} \tilde{\pi}(j,n-1)},
\end{eqnarray} where  recall that $\tilde{\pi}(j,k)$, $j = 0, 1, 2, \ldots$, $0 \leq k \leq M-1$, denote the stationary distribution of the DTMC $\{\tilde{\mathcal{X}}(t), t \geq 0\}$, assuming that it exists. This suggests an iterative method of solution as follows. Given the arrival rate $\lambda$ and the state-dependent attempt probabilities $\beta_n$'s, one can begin with some \textit{``guess''} values for the $\tilde{q}(n)$'s and obtain the transition probabilities. Then, the stationary state probabilities can be computed and new estimates for the $\tilde{q}(n)$'s can be computed from the stationary distribution by applying Equation~\eqref{eqn:approx-q-tilde-n}. This procedure can be repeated several times until the solutions converge within some specified tolerance limit. \textit{However, with infinite buffers, the DTMC $\{\tilde{\mathcal{X}}(t), t \geq 0\}$ would have infinite number of states and the stationary distribution cannot be computed numerically. Thus, we will apply the above idea of estimating the $\tilde{q}(n)$'s only for the case when the nodes have finite and equal buffers so that the finite number of balance equations can be solved numerically.} 

Let the buffer size of each queue be $K$ packets (one for the packet in service, if any, and $K-1$ waiting for service in the queue). We denote the finite buffer version of the process $\{\tilde{\mathcal{X}}(t), t \geq 0\}$ by $\{\tilde{\mathcal{X}}^{(K)}(t), t \geq 0\}$. (Processes, random variables and probabilities pertaining to the finite buffer case will be denoted by adding a superscript $^{(K)}$.) Recall that we interpreted the process $\{\tilde{\mathcal{X}}(t), t \geq 0\}$ as a DTMC embedded at the channel slot boundaries whose transition probabilities are functions of the unknown parameters $\tilde{q}(n)$'s. Similarly, we interpret the process $\{\tilde{\mathcal{X}}^{(K)}(t), t \geq 0\}$ as a DTMC embedded at the channel slot boundaries whose transition probabilities are functions of the unknown parameters $\tilde{q}^{(K)}(n)$'s. Let $\tilde{\pi}^{(K)}(j,k)$, $0 \leq j \leq K$, $0 \leq k \leq M - 1$, denote the stationary distribution of the DTMC $\{\tilde{\mathcal{X}}^{(K)}(t), t \geq 0\}$ (assuming that the stationary distribution exists). 

We define, $\forall j, 0 \leq j \leq K$, \[\bmath{\tilde{\pi}^{(K)}_{j}} := \big(\tilde{\pi}^{(K)}(j,0), \tilde{\pi}^{(K)}(j,1), \ldots, \tilde{\pi}^{(K)}(j,M-1) \big),\] and \[\bmath{\tilde{\pi}^{(K)}} := \big(\bmath{\tilde{\pi}^{(K)}_{0}}, \bmath{\tilde{\pi}^{(K)}_{1}}, \bmath{\tilde{\pi}^{(K)}_{2}}, \ldots \bmath{\tilde{\pi}^{(K)}_{K}}\big).\] Then the balance equations for the finite buffer case can be written as \begin{eqnarray} 
\label{eqn:2D-balance-eqn-matrix-form-K}
\bmath{\tilde{\pi}^{(K)}} = \bmath{\tilde{\pi}^{(K)}} \bmath{P^{(K)}},
\end{eqnarray} where the transition probability matrix $\bmath{P^{(K)}}$ has the following $M/G/1/K$ structure \begin{eqnarray}
\label{eqn:definition-tpm-K}
\bmath{P^{(K)}} = \left[ \begin{array}{cccccc} \bmath{A_0} &
    \bmath{A_1} & \bmath{A_2} & \cdots & \bmath{A_{K-1}} & \sum_{j =
      K}^{\infty} \bmath{A_{j}} \\ \bmath{B_{-1}} & \bmath{B_0} &
    \bmath{B_1} & \cdots & \bmath{B_{K-2}} & \sum_{j = K-1}^{\infty}
    \bmath{B_{j}} \\ \bmath{0} & \bmath{B_{-1}} & \bmath{B_0} & \cdots
    & \bmath{B_{K-3}} & \sum_{j = K-2}^{\infty} \bmath{B_{j}} \\
    \bmath{0} & \bmath{0} & \bmath{B_{-1}} & \cdots & \bmath{B_{K-4}}
    & \sum_{j = K-3}^{\infty} \bmath{B_{j}} \\
    \bmath{\vdots} &
    \bmath{\vdots} & \bmath{\vdots} & \bmath{\ddots} & \bmath{\vdots}
    & \bmath{\vdots} \\ \bmath{0} & \bmath{0} & \bmath{0} & \cdots &
    \bmath{B_{-1}} & \sum_{j = 0}^{\infty} \bmath{B_{j}} 
\end{array} \right].
\end{eqnarray} 

The $\bmath{A_{j}}$'s and the $\bmath{B_{j}}$'s in $\bmath{P^{(K)}}$ are given by Equation~\eqref{eqn:Aj0nk-Bjnk-Deltaq}, except that $\bmath{\Delta_{\tilde{q},A}}$ and $\bmath{\Delta_{\tilde{q},B}}$ need to be replaced by $\bmath{\Delta_{\tilde{q}^{(K)},A}}$ and $\bmath{\Delta_{\tilde{q}^{(K)},B}}$, respectively, where $\bmath{\Delta_{\tilde{q}^{(K)},A}} = diag \left(0, \tilde{q}^{(K)}(1), \ldots, \tilde{q}^{(K)}(M-1)\right)$ and $\bmath{\Delta_{\tilde{q}^{(K)},B}} = diag \left(\tilde{q}^{(K)}(1), \tilde{q}^{(K)}(2), \ldots, \tilde{q}^{(K)}(M)\right)$ are $M \times M$ diagonal matrices, and $\bmath{A_j}^{(0)}$, $\bmath{A_j}^{(1)}$, $\bmath{B_j}^{(0)}$ and $\bmath{B_j}^{(1)}$ are as defined for the infinite buffer case. We emphasize that, $\bmath{P^{(K)}}$ denotes the transition probability matrix for the finite buffer case, and that $\bmath{P^{(K)}}$ is \textbf{not} the $K$-step transition probability matrix associated with the one-step transition probability matrix $\bmath{P}$ for the infinite buffer case. We also emphasize that computing the infinite sums in the last column of $\bmath{P^{(K)}}$ only requires summing up probabilities from Poisson distributions which can be simplified by observing that $\sum_{j = k}^{\infty} d(j) = 1 - \sum_{j = 0}^{k-1} d(j)$, and so on.

\subsection{An Iterative Method}
\label{subsec:iterative-method}

In this subsection we propose an iterative method which can be applied in the ``equal arrival rates, and finite and equal buffers'' case to obtain the $\tilde{q}^{(K)}(n)$'s and the $\tilde{\pi}^{(K)}(j,k)$'s. Given the $\tilde{\pi}^{(K)}(j,k)$'s, we can ``estimate'' the $\tilde{q}^{(K)}(n)$'s as follows: ( recall that random variables and probabilities without a time argument indicate the stationary values) \begin{eqnarray}
\label{eqn:qn-in-pijk}
\tilde{q}^{(K)}(n) &\approx& P\big(\tilde{Q}_1^{(K)} = 1 \; \big| \;
\tilde{Q}_1^{(K)} > 0, \tilde{N}^{(K)} = n\big) \nonumber \\ 
&=& \displaystyle \frac{\tilde{\pi}^{(K)}(1,n-1)}{\displaystyle
  \sum_{j=1}^{K} \tilde{\pi}^{(K)}(j,n-1)}. \; \; \; \; \; \; \; \; \;
\; \; \; \; \; \; \; \; \; \; \; \; \; \; \; \; \; \; \; \; \; \; \;
\end{eqnarray} For a given $\lambda$, assuming some values for the $\tilde{q}^{(K)}(n)$'s, $\tilde{q}^{(K)}(n) \in [0,1], 1 \leq n \leq M$, the finite number of balance equations given by Equation~\eqref{eqn:2D-balance-eqn-matrix-form-K} can be numerically solved along with the normalization equation \begin{equation}
\label{eqn:normalization}
\sum_{j=0}^{K} \sum_{k=0}^{M-1} \tilde{\pi}_K(j,k) = 1 
\end{equation} to obtain \textit{unique positive} solutions for the $\tilde{\pi}^{(K)}(j,k)$'s (provided that $\{\tilde{\mathcal{X}}^{(K)}(t), t \geq 0\}$ is positive recurrent for the values assumed for the $\tilde{q}^{(K)}(n)$'s). Equation~\eqref{eqn:qn-in-pijk} then provides estimates for the $\tilde{q}^{(K)}(n)$'s. The balance equations given by Equation~\eqref{eqn:2D-balance-eqn-matrix-form-K} and the normalization equation, Equation~\eqref{eqn:normalization}, can be solved with the new set of $\tilde{q}^{(K)}(n)$'s yielding new estimates for the $\tilde{\pi}^{(K)}(j,k)$'s. This iterative procedure should continue until the solution converges within some specified tolerance limit. 

Clearly, for $\lambda >0$, the finite-state DTMC $\{\tilde{\mathcal{X}}^{(K)}(t), t \geq 0\}$ is irreducible if and only if the $\tilde{q}^{(K)}(n)$'s are \textit{positive}, since $\tilde{\mathcal{M}}^{(K)}(t)$ can decrease with positive probability if and only if the $\tilde{q}^{(K)}(n)$'s are positive. Thus, for $\lambda >0$, if the $\tilde{q}^{(K)}(n)$'s are positive, then the DTMC $\{\tilde{\mathcal{X}}^{(K)}(t), t \geq 0\}$ is irreducible, and hence, it is positive recurrent since it has finite number of states. This implies that, if the iterations began with positive $\tilde{q}^{(K)}(n)$'s, then, Equations~\eqref{eqn:2D-balance-eqn-matrix-form-K} and~\eqref{eqn:normalization} would yield unique positive solutions for the $\tilde{\pi}^{(K)}(j,k)$'s. Equation~\eqref{eqn:qn-in-pijk} would then provide new estimates for the $\tilde{q}^{(K)}(n)$'s which would also be positive and this procedure would continue. 

Recall that for a given $\lambda$, the $d(j)$'s, $c(j)$'s and $s(j)$'s are known for all $j \geq 0$ (see Equations~\eqref{eqn:definition-dj}-\eqref{eqn:definition-cj}) and the $\beta_n$'s are obtainable from a saturation analysis~\cite{wanet.bianchi00performance,wanet.kumar_etal07new_insights}. Thus, for a given $\lambda$, the matrices $\bmath{A_j}^{(0)}$, $\bmath{A_j}^{(1)}$, $j \geq 0$, and $\bmath{B_j}^{(0)}$, $\bmath{B_j}^{(1)}$, $j \geq -1$, need to be computed only once. Each iteration gives a new set of $\tilde{q}^{(K)}(n)$'s, and the matrices $\bmath{\Delta_{\tilde{q}^{(K)},A}}$ and $\bmath{\Delta_{\tilde{q}^{(K)},B}}$ get updated in each iteration resulting in updated $\bmath{A_j}$'s and the $\bmath{B_j}$'s (see Equation~\eqref{eqn:Aj0nk-Bjnk-Deltaq}). This, in turn, updates $\bmath{P^{(K)}}$ resulting in a new set of balance equations which yields a new set of $\tilde{q}^{(K)}(n)$'s. Important performance measures can be obtained from the converged values of the $\tilde{\pi}^{(K)}(j,k)$'s (Section~\ref{subsec:performance-measures}).

\subsection{Complexity of our Iterative Method}
\label{subsec:complexity}

The SDAR approximation has the following advantages: 

\begin{enumerate}

\item It enables computation of the $\beta_n$'s using the saturation analysis of~\cite{wanet.bianchi00performance}  or~\cite{wanet.kumar_etal07new_insights}. Thus, in contrast to the model in~\cite{wanet.garetto-chiasserini05802.11-MAC-sporadic-traffic}, the $\beta_n$'s in our model are independent of the arrival rates and of the average queue occupancies. They are computed only once in the beginning of the overall computation by a separate procedure and then used as given parameters in the iterative method described in Section~\ref{subsec:iterative-method}. 

\item The effects of the backoff parameters are effectively captured in the pre-computed $\beta_n$'s (which will be evident from the accuracy of our model as demonstrated in Section~\ref{sec:results}). Hence, they are not considered when analyzing the queueing dynamics. This enables us to eliminate the first dimension, namely, ``the backoff stage of the tagged node'' from the three-dimensional Markov chain of~\cite{wanet.garetto-chiasserini05802.11-MAC-sporadic-traffic} and our model requires computations involving a two-dimensional Markov chain. 

\item Since the $\beta_n$'s are computed independent of the arrival rate of the Poisson arrival processes, they need not be computed for each arrival rate when studying the effect of arrival rate on performance measures. 

\end{enumerate}

Thus, the SDAR approximation makes our iterative method computationally less expensive than that in~\cite{wanet.garetto-chiasserini05802.11-MAC-sporadic-traffic} as follows. The complexity of the model in~\cite{wanet.garetto-chiasserini05802.11-MAC-sporadic-traffic} is $O(RKM)$ where $R$ denotes the retransmit limit, $K$ denotes the buffer size and $M$ denotes the number of nodes (see Section III-C of~\cite{wanet.garetto-chiasserini05802.11-MAC-sporadic-traffic}). The complexity of obtaining $\beta_n$, $1 \leq n \leq M$, by a separate procedure is $O(RM)$ and the complexity of our finite buffer model, given the $\beta_n$'s, is $O(KM)$. Thus, the overall complexity of our finite buffer model is $O(KM) + O(RM)$ which is less than the complexity $O(RKM)$ of the finite buffer model of~\cite{wanet.garetto-chiasserini05802.11-MAC-sporadic-traffic}. If one needs to solve for, say, $l$ different arrival rates, to examine how the protocol behaves with the variation of traffic intensity, then the complexity of our model is $O(lKM) + O(RM)$ since we compute the $\beta_n$'s only once and use them for all the $l$ arrival rates whereas the complexity of the model in~\cite{wanet.garetto-chiasserini05802.11-MAC-sporadic-traffic} for $l$ different arrival rates is $O(lRKM)$. Thus, when studying the effect of arrival rates on the performance measures, our model is far superior to that in~\cite{wanet.garetto-chiasserini05802.11-MAC-sporadic-traffic}. This reduction in complexity is achieved precisely due to the SDAR approximation.

\subsection{Prediction of Performance Measures}
\label{subsec:performance-measures}

The processes $\{\bmath{Q}(t), t \geq 0\}$ and its finite buffer version $\{\bmath{Q}^{(K)}(t), t \geq 0\}$ are DTMCs embedded at the channel slot boundaries. It is easy to see that $\{(\bmath{Q}(t),T(t)), t \geq 0\}$ and $\{(\bmath{Q}^{(K)}(t),T(t)), t \geq 0\}$ are Markov renewal sequences~\cite{theory.kulkarni95modeling-analysis-stochastic-systems}. We apply Markov regenerative analysis to predict performance measures such as collision probability and throughput. Also, we obtain the mean packet delay by applying the $M/G/1/K$ queue analysis developed in~\cite{theory.kobayashi78modeling} (see pages 201-205 of~\cite{theory.kobayashi78modeling}). 

Let $\{N^{(K)}(t), t \geq 0\}$ denote the finite buffer version of the process $\{N(t), t \geq 0\}$. Let $\{\tilde{N}^{(K)}(t), t \geq 0\}$ be derived from $\{\tilde{\mathcal{X}}^{(K)}(t), t \geq 0\}$ by applying $\tilde{N}^{(K)}(t) := \tilde{Q}_1^{(K)}(t) + \tilde{\mathcal{M}}^{(K)}(t)$. We denote the stationary distribution of the process $\{N^{(K)}(t), t \geq 0\}$ by $p^{(K)}(n)$, $0 \leq n \leq M$, and that of the process $\{\tilde{N}^{(K)}(t), t \geq 0\}$ by $\tilde{p}^{(K)}(n)$, $0 \leq n \leq M$. The $\tilde{p}^{(K)}(n)$'s provide ``approximations'' for the $p^{(K)}(n)$'s and can be obtained from the $\tilde{\pi}^{(K)}(j,k)$'s as follows: \begin{eqnarray}
\label{eqn:pn-definition}
\tilde{p}^{(K)}(n) &:=& P\big(\tilde{N}^{(K)} = n\big) = P\big(
\tilde{Q}_1^{(K)} = 0, \tilde{\mathcal{M}}^{(K)} = n\big) +
P\big(\tilde{Q}_1^{(K)} > 0, \tilde{\mathcal{M}}^{(K)} = n - 1 \big)
\nonumber \\ 
&=& \tilde{\pi}^{(K)}(0,n) + \sum_{j=1}^{K} \tilde{\pi}^{(K)}(j,n-1). 
\end{eqnarray} The $\tilde{\pi}^{(K)}(j,k)$'s, in turn, can be obtained by the iterative method discussed in Section~\ref{subsec:iterative-method}. Once the $p^{(K)}(n)$'s are approximated by the $\tilde{p}^{(K)}(n)$'s, collision probability and throughput can be obtained as follows.

\textbf{Collision Probability:} Let $\mathcal{A}^{(K)}(t)$ and $\mathcal{C}^{(K)}(t)$ denote the total number of attempts and collisions, respectively, up to time $t$, where we recall that $t$ is the discrete time index. The (conditional) collision probability $\gamma$ is given by \begin{equation}
\label{eqn:gamma-finite-buffer-SDAR}
\gamma := \lim_{t \rightarrow \infty}
\frac{\mathcal{C}^{(K)}(t)}{\mathcal{A}^{(K)}(t)} \;
\displaystyle \stackrel{a.s.}{=} \; \frac{\displaystyle \sum_{n =
    0}^{M} p^{(K)}(n) \EXPSUB{C}{n}}{\displaystyle \sum_{n = 0}^{M}
  p^{(K)}(n) \EXPSUB{A}{n}} \; \approx \; \frac{\displaystyle \sum_{n
    = 0}^{M} \tilde{p}^{(K)}(n) \EXPSUB{C}{n}}{\displaystyle \sum_{n =
    0}^{M}\tilde{p}^{(K)}(n) \EXPSUB{A}{n}}, 
\end{equation} where $\EXPSUB{A}{n}$ and $\EXPSUB{C}{n}$ respectively denote the mean number of attempts and collisions \textit{per channel slot} given that $n$ nodes are non-empty in the beginning of the channel slot. It is easy to see that \[\EXPSUB{A}{n} = n \beta_n, \; \; \mbox{and} \; \; \EXPSUB{C}{n} = n \beta_n \left( 1 - (1-\beta_n)^{n-1}\right).\]

\textbf{Throughput:} Let $\mathcal{S}^{(K)}(t)$ denote the total number of successful transmissions up to time $t$. The aggregate system throughput $\Theta$ in packets/sec is given by \begin{equation}
\label{eqn:theta-finite-buffer-SDAR}
\Theta := \lim_{t \rightarrow \infty} \frac{\mathcal{S}^{(K)}(t)}{t}
\; \displaystyle \stackrel{a.s.}{=} \; \frac{\displaystyle \sum_{n =
    0}^{M} p^{(K)}(n) \EXPSUB{S}{n}}{\displaystyle \sum_{n = 0}^{M}
  p^{(K)}(n) \EXPSUB{L}{n}} \; \approx \; \frac{\displaystyle \sum_{n
    = 0}^{M} \tilde{p}^{(K)}(n) \EXPSUB{S}{n}}{\displaystyle \sum_{n =
    0}^{M} \tilde{p}^{(K)}(n) \EXPSUB{L}{n}},
\end{equation} where $\EXPSUB{S}{n}$ denotes the mean number of successful transmissions per channel slot and $\EXPSUB{L}{n}$ denotes the mean duration of a channel slot in seconds given that $n$ nodes are non-empty in the beginning of the channel slot. It is easy to see that \[\EXPSUB{S}{n} = n \beta_n (1-\beta_n)^{n-1},\] and \[\EXPSUB{L}{n} = \big(p_{idle,n}\sigma + p_{coll,n}(\sigma+T_c) + p_{succ,n}(\sigma+T_s)\big) = (\sigma + p_{coll,n}T_c+p_{succ,n}T_s).\] Due to symmetry, the per-node throughput $\theta$ is given by \[\theta = \displaystyle \frac{\Theta}{M}.\]

\textbf{Mean Packet Delay:} Consider the process $\{\bmath{Q}^{(K)}(t), t \geq 0\}$. Clearly, for $\lambda > 0$, $\{\bmath{Q}^{(K)}(t), t \geq 0\}$ is an irreducible and aperiodic DTMC with finite state space (see  Appendix~\ref{app:positive-recurrence-Qt} where we prove irreducibility and aperiodicity of $\{\bmath{Q}(t), t \geq 0\}$). Hence, for $\lambda > 0$, the DTMC $\{\bmath{Q}^{(K)}(t), t \geq 0\}$ is positive recurrent, stationary and ergodic. For any tagged queue, define the following:

\vspace{2mm}

$\alpha(j) :=$ \parbox[t] {13.5 cm} {fraction of (real) time that the tagged queue contains $j$ packets, $0 \leq j \leq K$.} 

\vspace{2mm}

$p^{(d)}(j) :=$ \parbox[t] {13 cm} {probability that a departure from the tagged queue leaves $j$ packets behind, $0 \leq j \leq K-1$. (Note that the number of packets left by a departure cannot be more than $K-1$)} 

\vspace{2mm}

$p^{(a)}(j) :=$ \parbox[t] {13 cm} {probability that a packet accepted into the tagged queue finds $j$ packets, $0 \leq j \leq K-1$. (Note that the number of packets as seen by an accepted packet cannot be more than $K-1$)} 

\vspace{2mm}

In  Appendix~\ref{app:derivation-eqn-pidj0toK-2}, we obtain the probabilities $p^{(d)}(j)$, $0 \leq j \leq K-2$, as follows: \begin{eqnarray}
\label{eqn:pidj0toK-2}
p^{(d)}(j) &=& \displaystyle \frac{\displaystyle \sum_{i=1}^{j+1}
  \displaystyle \sum_{n=0}^{M-1} \pi^{(K)}(i,n) \displaystyle
  \left(\frac{p_{succ,n+1}}{n+1}\right) s(j-i+1)}{\displaystyle
  \sum_{i=1}^{K} \displaystyle \sum_{n=0}^{M-1} \pi^{(K)}(i,n)
  \displaystyle \left(\frac{p_{succ,n+1}}{n+1}\right)}, \; \; \; \; \;
\; 
\end{eqnarray} where $\pi^{(K)}(i,n)$, $0 \leq i \leq K$, $0 \leq n \leq M-1$, denotes the stationary probability that the tagged queue contains $i$ packets and $n$ of the non-tagged queues are non-empty. The $s(j-i+1)$'s in Equation~\eqref{eqn:pidj0toK-2} can be obtained using Equation~\eqref{eqn:definition-sj} and the $\pi^{(K)}(i,n)$'s can be approximated by the $\tilde{\pi}^{(K)}(i,n)$'s. The $\tilde{\pi}^{(K)}(i,n)$'s, in turn, can be obtained by the iterative method discussed in Section~\ref{subsec:iterative-method}. The probability $p^{(d)}(K-1)$ is given by \begin{equation}
\label{eqn:pid-K-1}
p^{(d)}(K-1) = 1 - \displaystyle \sum_{j=0}^{K-2} p^{(d)}(j).
\end{equation} Since arrivals and departures occur one at a time, a \textit{level-crossing} analysis gives, $\forall j, 0 \leq j \leq K-1$, \begin{equation}
\label{eqn:relation-piaj-pidj}
p^{(a)}(j) = p^{(d)}(j).
\end{equation} Note that the probability that an arrival is blocked is given by $\alpha(K)$. The mean rate at which packets are accepted into the queue must be equal to the mean rate at which packets depart from the queue. Hence, $\alpha(K)$ can be obtained by solving \begin{equation}
\label{eqn:alphaK-from-theta}
\lambda (1 - \alpha(K)) = \theta = \frac{\Theta}{M}, 
\end{equation} where $\Theta$ can be computed applying Equation~\eqref{eqn:theta-finite-buffer-SDAR}. According to~\cite{theory.kobayashi78modeling}, we have, $\forall j, 0 \leq j \leq K-1$, \begin{equation}
\label{eqn:relation-alphaj-piaj-kobayashi}
\alpha(j) = p^{(a)}(j) (1 - \alpha(K)).
\end{equation} The mean queue length $\bar{Q}$ is then given by \begin{equation}
\label{eqn:mean-queue-length}
\bar{Q} = \displaystyle \sum_{j=0}^{K} j \alpha(j)
\end{equation} from which the mean sojourn time or the mean packet delay
$\bar{W}$ can be obtained as \begin{equation}
\label{eqn:mean-packet-delay}
\bar{W} = \frac{\bar{Q}}{\theta}. 
\end{equation}

\section{Applying the SDAR Approximation in the NS-2 Simulator}
\label{sec:SDAR-simulation-technique} 

Wireless network simulators invariably employ simple propagation models at the PHY layer to keep the simulations reasonably fast. In this section, we describe a model-based simulation technique at the MAC layer which is based on the SDAR approximation. We apply the SDAR approximation in NS-2 as follows. 

We modify the original implementation of NS-2 MAC layer to keep track of the number of non-empty nodes. Arrivals that occur during activity periods (i.e., during successful transmissions or collisions) are not taken into account to update the number of non-empty nodes until the activity finishes. Hence, the number of non-empty nodes does not change during the activity periods. Whenever an activity finishes or an arrival occurs during a channel idle period, the number of non-empty nodes is updated. Whenever the number of non-empty nodes is updated, all previously scheduled transmissions (if any) are canceled and random backoffs are sampled for each non-empty node using independent geometric random variables each having a mean $1 / \beta_n$ where $n$ denotes the current value of the number of non-empty nodes. Note that the geometric backoff durations with mean $1 / \beta_n$ are equivalent to Bernoulli attempt processes with probabilities $\beta_n$. The $\beta_n$'s are pre-computed using the saturation analysis in~\cite{wanet.kumar_etal07new_insights} and stored in a look-up table. 

From the sampled backoffs, it is easy to determine which node(s) sample the minimum backoff. If only one node samples the minimum backoff, the next event is a successful transmission. If two or more nodes sample the same minimum backoff, the next event is a collision. The appropriate event is then scheduled. If arrivals occur to empty queues before the ``beginning of the scheduled event'' epoch to increase the number of non-empty nodes, then the scheduled event is canceled. Else, the scheduler clock is moved to the ``end of the scheduled event'' epoch. In case the next event is a successful transmission, the DATA frame is handed over to the destination's MAC layer which then generates the corresponding ACK frame.

\begin{table*}[t]
\begin{center}
\caption{\label{table:SpeedUpI} Speed-up for MACHINE-I (Pentium dual
  core, 2.80 GHz, 1024 KB cache)} 
\vspace{5mm}
\begin{footnotesize}
\begin{tabular}{||c|c|c|c|c||} 
\hline
 & $\lambda$ & MAC time & MAC time & \\
 $M$ & (pkts/sec) & NS-2 (sec) & SDAR (sec) & Speed-up \\
\hline
\hline
50 & 5 & 45.045 & 9.405 & 4.79 \\
\hline
50 & 10 & 103.43 & 19.16 & 5.4 \\
\hline 
50 & 15 & 158.44 & 39.40 & 4.02 \\
\hline
\hline
30 & 5 & 10.8 & 3.25 & 3.08 \\
\hline
30 & 10 & 22.88 & 6.89 & 3.32 \\
\hline
30 & 15 & 32.49 & 10.24 & 3.17 \\
\hline
30 & 20 & 60.49 & 22.84 & 2.65 \\
\hline
\hline
10 & 10 & 2.233 & 0.943 & 2.37 \\
\hline
10 & 20 & 4.314 & 1.864 & 2.31 \\
\hline
10 & 30 & 7.256 & 3.114 & 2.33 \\
\hline
10 & 40 & 8.351 & 3.641 & 2.30 \\
\hline
10 & 50 & 10.777 & 4.767 & 2.26 \\
\hline
10 & 60 & 13.269 & 5.789 & 2.30 \\
\hline
10 & 70 & 19.922 & 10.312 & 1.93 \\
\hline
\end{tabular}
\end{footnotesize}
\vspace{5mm}
\caption{\label{table:SpeedUpII} Speed-up for MACHINE-II (Pentium,  1500 MHz, 256 KB cache)} 
\vspace{5mm}
\begin{footnotesize}
\begin{tabular}{||c|c|c|c|c||} 
\hline
 & $\lambda$ & MAC time & MAC time & \\
 $M$ & (pkts/sec) & NS-2 (sec) & SDAR (sec) & Speed-up \\
\hline
\hline
50 & 5 & 102.06 & 31.99 & 3.19 \\
\hline
50 & 10 & 268.23 & 66.86 & 4.0 \\
\hline 
50 & 15 & 369.13 & 97.77 & 3.78 \\
\hline
\hline
30 & 5 & 23.9 & 9.59 & 2.49 \\
\hline
30 & 10 & 48.16 & 19.06 & 2.53 \\
\hline
30 & 15 & 75.93 & 29.79 & 2.55 \\
\hline
30 & 20 & 136.82 & 54.11 & 2.53 \\
\hline
\hline
10 & 10 & 3.66 & 2.13 & 1.72 \\
\hline
10 & 20 & 7.31 & 4.25 & 1.72 \\
\hline
10 & 30 & 11.38 & 6.62 & 1.72 \\
\hline
10 & 40 & 15.52 & 9.00 & 1.72 \\
\hline
10 & 50 & 20.25 & 11.28 & 1.795 \\
\hline
10 & 60 & 25.87 & 13.99 & 1.85 \\
\hline
10 & 70 & 31.07 & 20.02 & 1.55 \\
\hline
\end{tabular}
\end{footnotesize}
\end{center}
\end{table*}

The above modifications have enabled us to achieve speed-ups up to 5.4. These speed-ups have been achieved with respect to the MAC layer operations and are summarized in Tables~\ref{table:SpeedUpI} and~\ref{table:SpeedUpII} for two different machines. The observed speed-ups are obtained due to the following reasons. In NS-2, one transmission event per non-empty node remains pending in the ``scheduler queue'' and $n$ backoff timers are kept running for $n$ non-empty nodes. When a timer expires resulting in a transmission, all the other $n-1$ timers remain ``paused'' until the transmission finishes. When the timers resume, the remaining $n-1$ transmission events have to be rescheduled. Similar pausing and rescheduling occurs in case of collisions when multiple timers expire simultaneously. Note that rescheduling of events requires searching the appropriate events from the scheduler's event queue which consumes the largest fraction of MAC simulation time. In our modifications, due to the memoryless property of the SDAR attempt model, we do not have to keep the backoffs sampled by the nodes. Moreover, at any point of time, only one event remains pending at the MAC layer which is the next event to occur on the common channel. This single pending event is interrupted and is rescheduled only if an arrival increases the number of non-empty nodes. Hence, the speed-up increases with the number of nodes $M$. The speed-up also increases with arrival rate $\lambda$ since the average number of non-empty nodes increases with increase in $\lambda$. However, above a certain $\lambda$, the rate of cancellation of already scheduled events dominates and speed-up actually decreases with $\lambda$. The speed-up w.r.t.~MAC layer operations remains constant in saturation conditions even though the simulation time consumed in higher layer operations increases with $\lambda$ due to increasing number of arrivals. 

These observations are supported by the data in Tables~\ref{table:SpeedUpI} and~\ref{table:SpeedUpII} where the last row for each $M$ corresponds to saturation. The increase in speed-up with $M$ is particularly desirable since NS-2 is found to become worse with increase in $M$ than $\lambda$. Also, note in Tables~\ref{table:SpeedUpI} and~\ref{table:SpeedUpII} that the speed-ups are more for the faster machine, i.e., MACHINE-I. This indicates that the speed-ups are not due to the incapability of the machines.

\section{Numerical and Simulation Results}
\label{sec:results}

\begin{figure*}[!p]
  \centering 
  \begin{minipage}{7.5cm}
    \begin{center}
      \includegraphics[height=6cm,width=7.5cm]{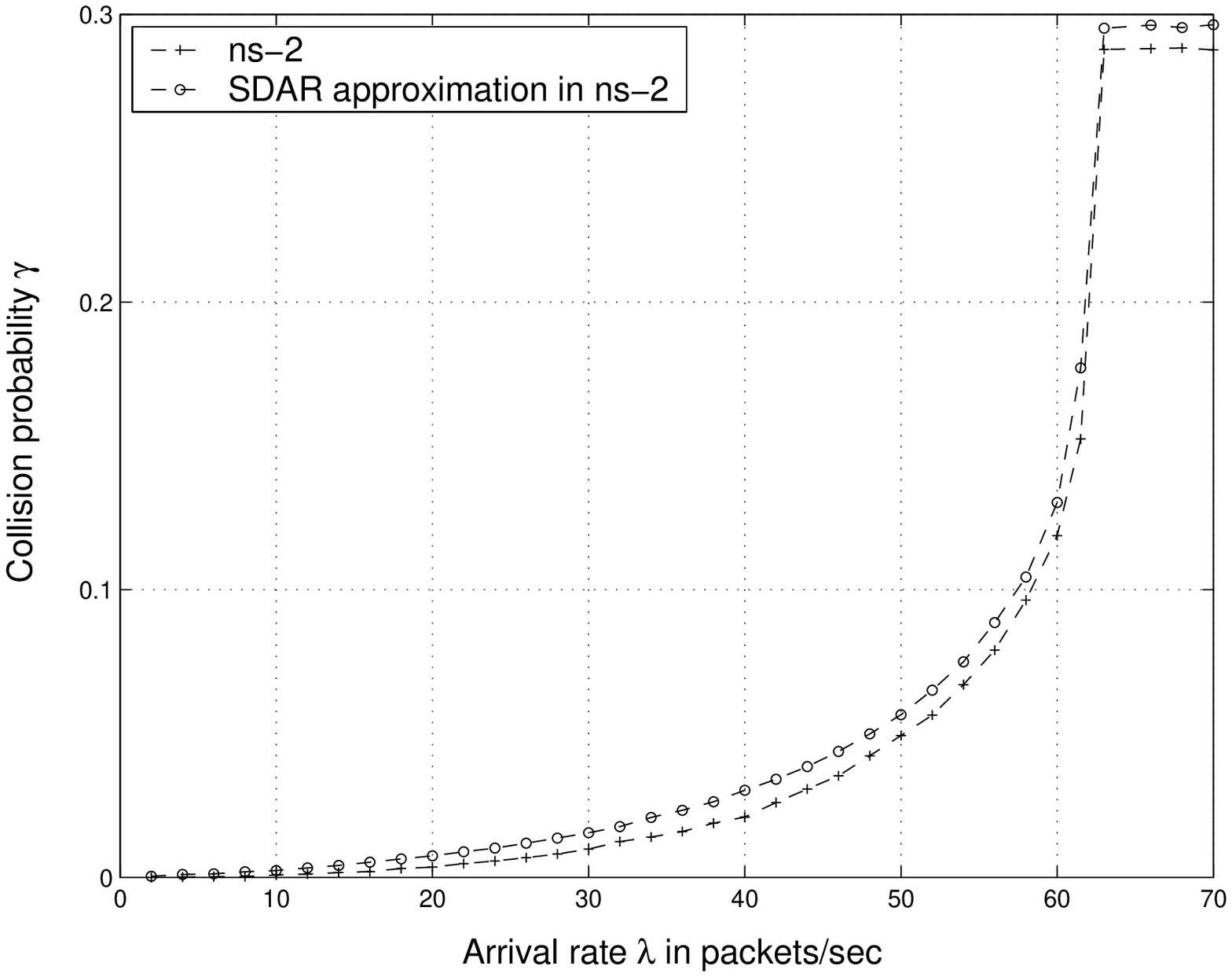}
      \caption{Comparison of collision probability $\gamma$ with $M =
        10$ nodes, infinite buffer size, and equal arrival
        rates. \label{fig:comparegamma}} 
      \vspace{3mm}
    \end{center}
  \end{minipage}
\hfill
  \begin{minipage}{7.5cm}
    \begin{center}
      \includegraphics[height=6cm,width=7.5cm]{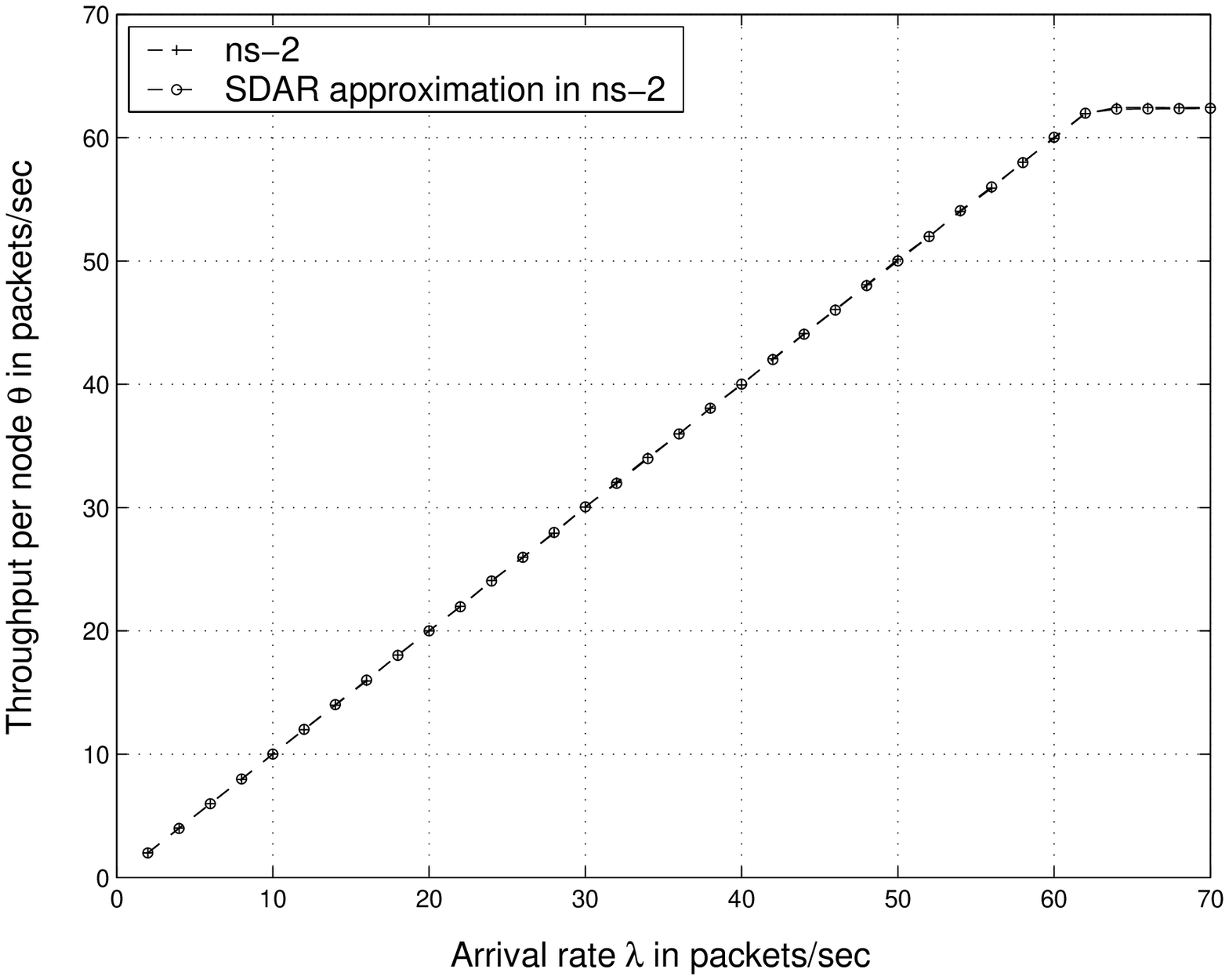}
      \caption{Comparison of throughput per node $\theta$ with $M = 10$
        nodes, infinite buffer size, and equal arrival
        rates. \label{fig:comparetheta}} 
      \vspace{3mm}
    \end{center}
  \end{minipage}
\hfill
  \begin{minipage}{7.5cm}
    \begin{center}
      \includegraphics[height=6cm,width=7.5cm]{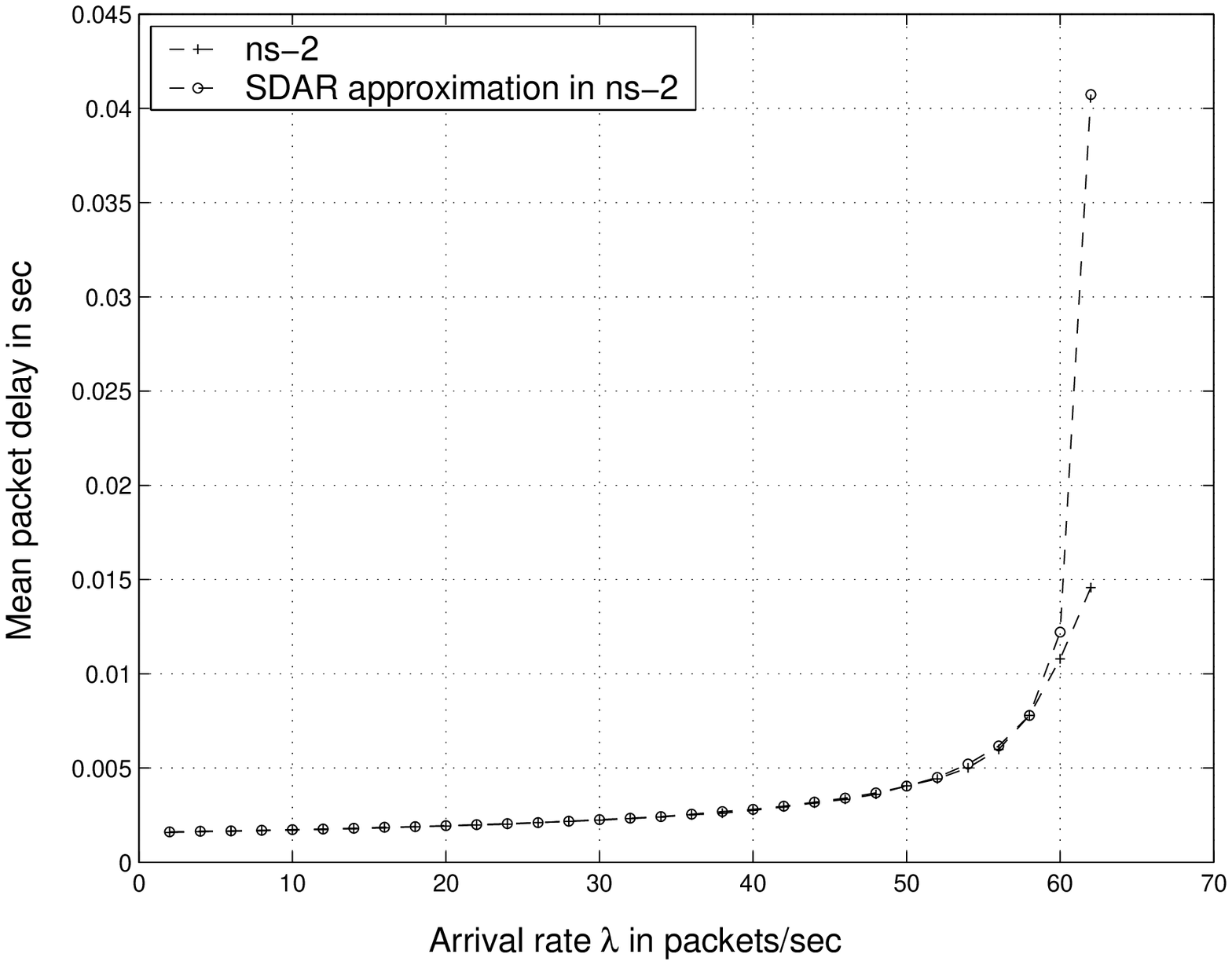}
      \caption{Comparison of mean packet delay $\bar{W}$ with $M = 10$
        nodes, infinite buffer size, and equal arrival
        rates. \label{fig:compareDelay}} 
      \vspace{3mm}
    \end{center}
  \end{minipage}
\hfill
  \begin{minipage}{7.5cm}
    \begin{center}
      \includegraphics[height=6cm,width=7.5cm]{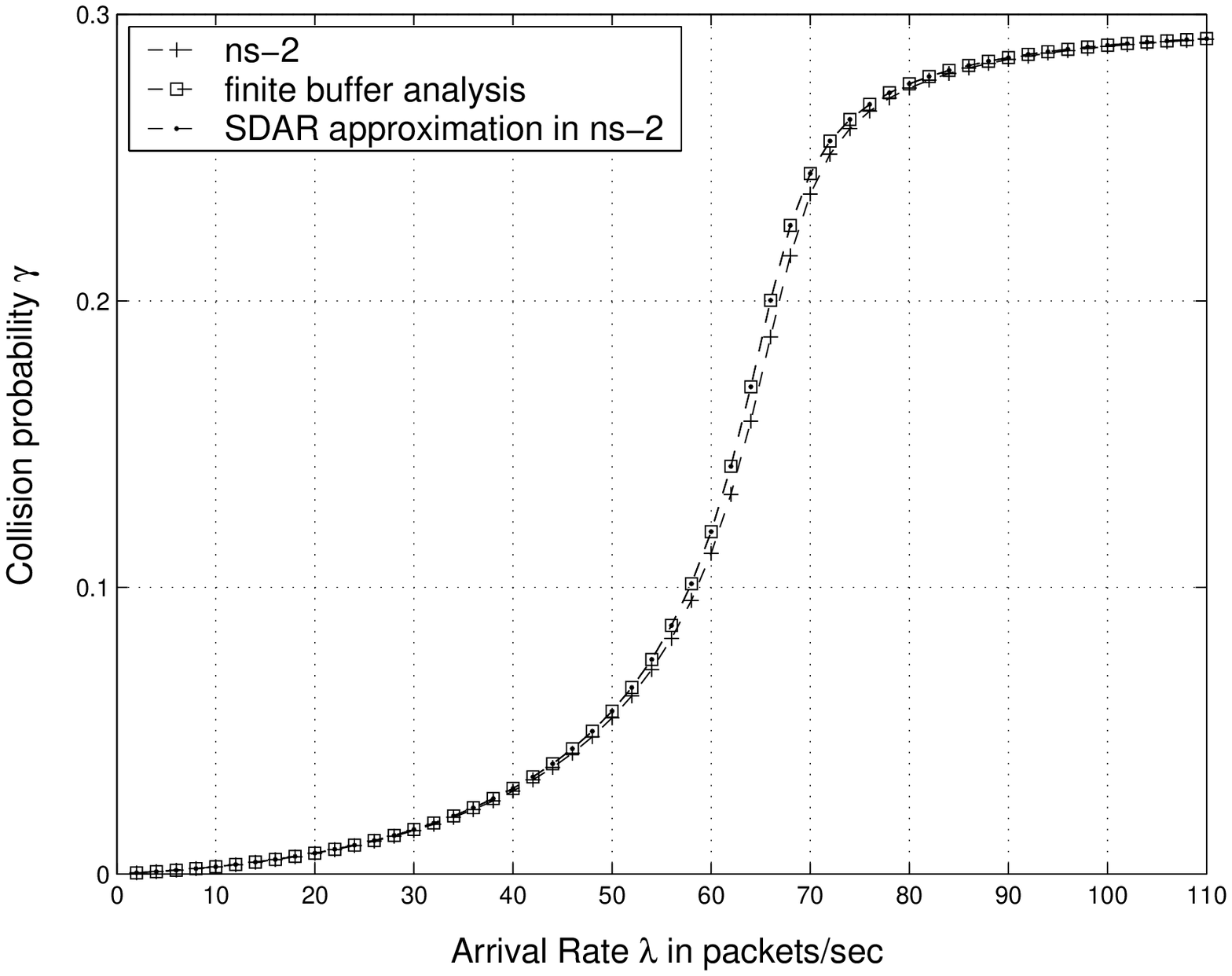}
      \caption{Comparison of collision probability $\gamma$ with $M =
        10$ nodes, finite buffer size $K = 5$, and equal arrival
        rates. \label{fig:comparegammaK5}} 
      \vspace{3mm}
    \end{center}
  \end{minipage}
\hfill
  \begin{minipage}{7.5cm}
    \begin{center}
      \includegraphics[height=6cm,width=7.5cm]{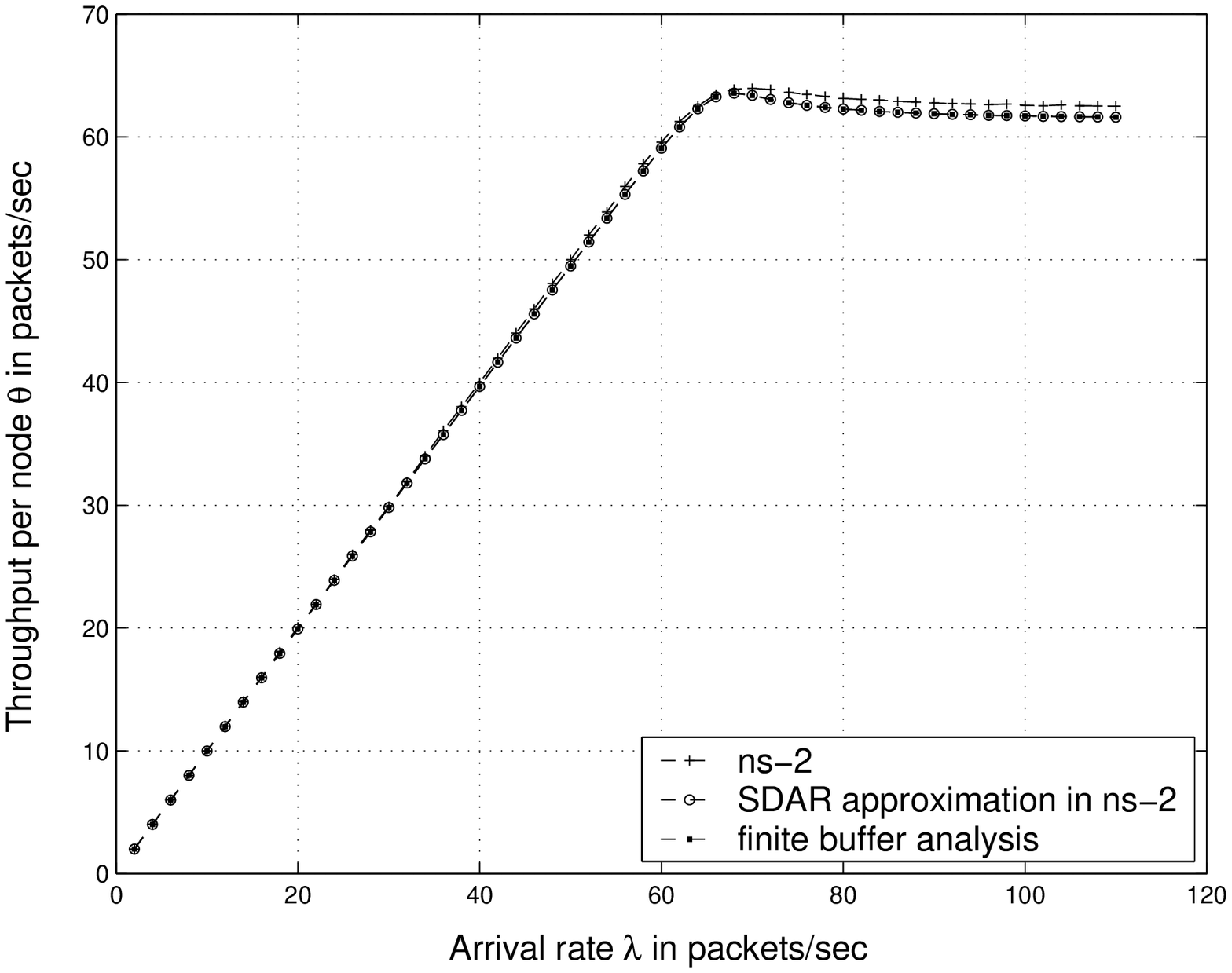}
     \caption{Comparison of throughput per node $\theta$ with $M = 10$
       nodes, finite buffer size $K = 5$, and equal arrival
       rates. \label{fig:comparethetaK5}} 
      \vspace{3mm}
    \end{center}
  \end{minipage}
\hfill
  \begin{minipage}{7.5cm}
    \begin{center}
      \includegraphics[height=6cm,width=7.5cm]{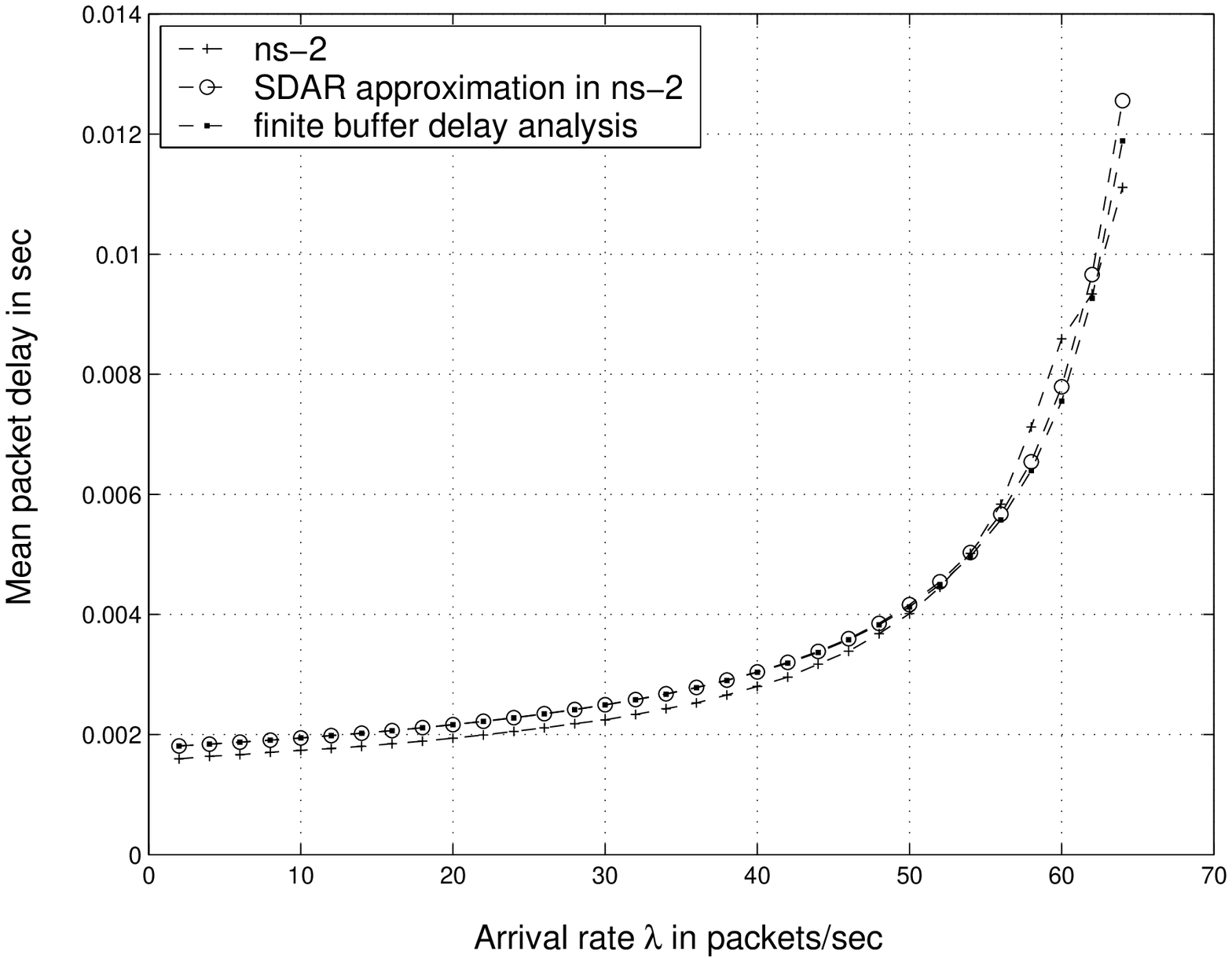}
      \caption{Comparison of mean packet delay $\bar{W}$ with $M = 10$
        nodes, finite buffer size $K = 5$, and equal arrival
        rates. \label{fig:comparedelayK5}} 
    \end{center}
  \end{minipage}
\end{figure*}

In this section we provide numerical and simulation results to validate our coupled queue model and our iterative method. The values of the backoff parameters were taken as per the 802.11b standard. We took Basic Rate = 2 Mbps, Data Rate = 11 Mbps, Packet Payload Size = 1000 Bytes. We provide results for the Basic Access mode. The iterative method was implemented using MATLAB. All other protocol parameters are set as per the standard setting in 802.11b. Figures~\ref{fig:comparegamma}, \ref{fig:comparetheta}, and \ref{fig:compareDelay} compare the collision probability $\gamma$, the per-node throughput $\theta$, and the mean packet delay $\bar{W}$ with $M=10$ nodes, infinite buffer size, and equal arrival rates. The buffer sizes were set to very large values to simulate infinite buffers. It can be seen that the simulation results obtained from the unmodified NS-2 and the SDAR approximation in NS-2 match extremely well. The mismatch in the collision probabilities which leads to visible mismatch in the mean packet delays near saturation is mainly due to the over-estimation of collision probability by the saturation analysis~\cite{wanet.bianchi00performance,wanet.kumar_etal07new_insights}, which clearly appears as a $\approx 5\%$ mismatch of collision probability beyond the saturation threshold (which is about $\lambda = 62.5$ packets/sec for $M=10$ nodes, (see Figure~\ref{fig:comparegamma})).

Notice, however, in Figure~\ref{fig:comparetheta} that the throughputs obtained from NS-2 simulations and from the SDAR approximation in NS-2 are indistinguishable. Let $\Theta_{sat}$ denote the aggregate system throughput under saturated conditions. For the default backoff parameters as prescribed in the 802.11b standard, $\Theta_{sat}$ is not sensitive to the change in γ. Hence, a less than $5\%$ difference in $\gamma$ does not lead to noticeable change in $\Theta_{sat}$ , and $\Theta_{sat}$  in the original NS-2 is almost equal to $\Theta_{sat}$ in NS-2 with SDAR. Since the number of nodes, $M$, remains constant, $\theta_{sat} := \Theta_{sat}/M$, in the original NS-2 is also equal to $\theta_{sat}$ in NS-2 with SDAR. This explains the good match of the flat portion of the curves in saturation. Since there are no packet drops due to buffer overflows (infinite buffer) and there are no packet discards, one can expect that, for $\lambda \leq \theta_{sat}$, the rate of departure is equal to the rate of arrival, i.e., $\theta = \lambda$, since the queues are stable. This explains the ``$y = x$'' behavior below saturation load.

Figures~\ref{fig:comparegammaK5}, \ref{fig:comparethetaK5} and \ref{fig:comparedelayK5} compare the collision probability $\gamma$, the throughput per node $\theta$ and the mean packet delay $\bar{W}$ with $M=10$ nodes, a buffer size of $K=5$ packets per queue, and equal arrival rates. It can be seen that the SDAR approximation in NS-2 and our iterative method of analysis both match extremely well with the unmodified NS-2. Furthermore, the results from the SDAR approximation in NS-2 and that from our iterative method of analysis are extremely close. This validates the conditional independence approximation, Approximation~\ref{approx:independence-exact-number} (at least from the point of view of predicting the performance measures), and also validates our iterative method. 

Notice the peak of throughput in Figure~\ref{fig:comparethetaK5} which occurs primarily because of the finite buffer size, and thus, not visible in the infinite buffer case. Note that, in the finite buffer case, the system cannot saturate for finite arrival rates. This can be understood by considering the finite state DTMC $\{\bmath{Q}^{(K)}(t), t \geq 0\}$, which is positive recurrent for all finite arrival rates, $0 < \lambda < \infty$. Hence, every queue must be empty for a positive fraction of time. However, as the arrival rate $\lambda$ increases, the fraction of time for which a node remains empty decreases. For very large arrival rates one can expect the throughput to become close to the saturation throughput. 

The peak throughput in Figure~\ref{fig:comparethetaK5} is slightly higher than the saturation throughput because of the following reason. Consider an arrival rate $\lambda$ higher than the per-node saturation throughput. Had the buffer sizes of the nodes been infinite, each node would have been saturated, the number of contending nodes would have been $M$, and the aggregate system throughput would have been $\Theta_{sat,M}$. However, due to finite buffers, the \textit{effective} number of contending nodes $n_{eff}(\lambda)$ corresponding to $\lambda$, is smaller than $M$. Since $n_{eff}(\lambda) < M$, we have $\Theta_{sat,n_{eff}(\lambda)} > \Theta_{sat,M}$, and the aggregate system throughput is larger than the saturation throughput. With increasing $\lambda$, $n_{eff}(\lambda)$ would approach $M$ and the aggregate system throughput would approach $\Theta_{sat,M}$. 

\begin{figure*}[t]
  \centering 
  \begin{minipage}{8cm}
    \begin{center}
      \includegraphics[height=7.5cm,width=8cm]{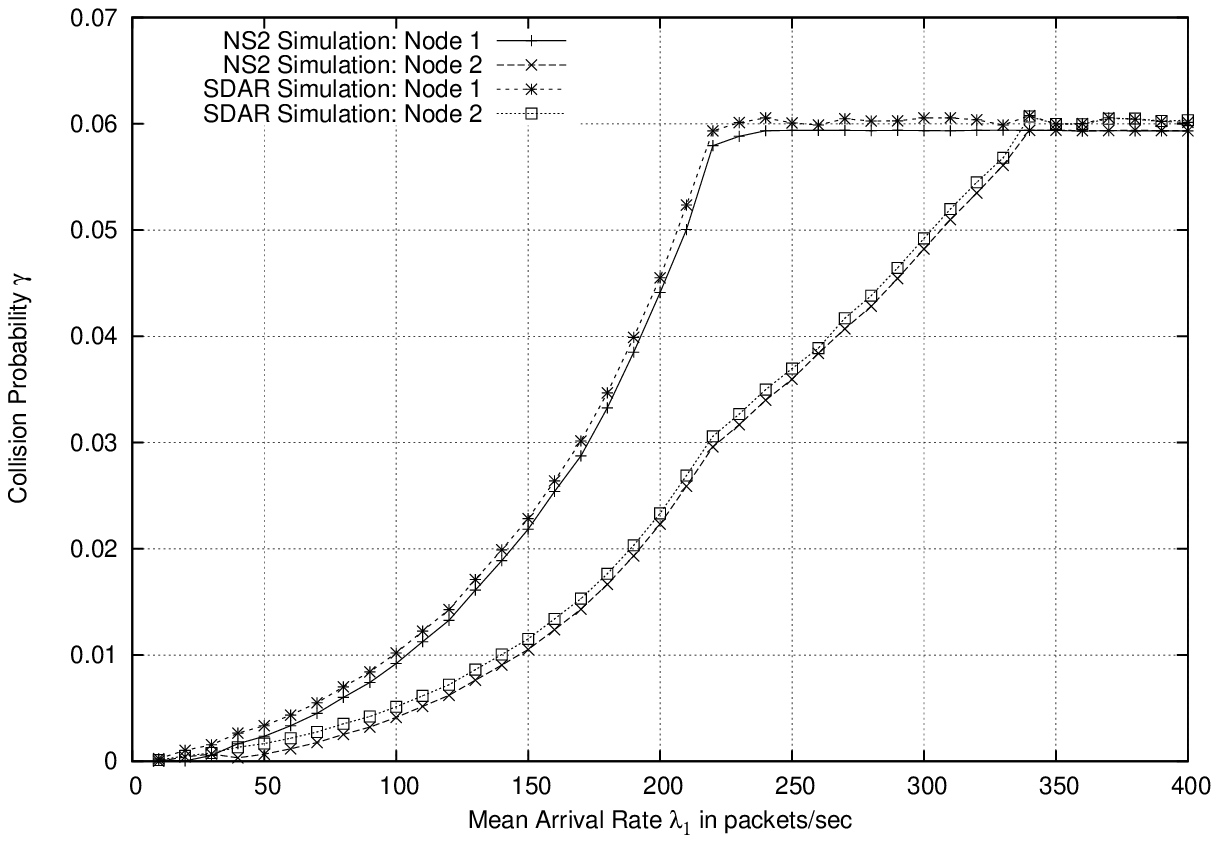}
      \caption{Comparison of collision probability $\gamma$ with $M = 2$
        nodes, infinite buffer size and unequal arrival
        rates. \label{fig:comparegammaUnequalTwoNodes}} 
      \vspace{3mm}
    \end{center}
  \end{minipage}
\hfill
  \begin{minipage}{8cm}
    \begin{center}
      \includegraphics[height=7.5cm,width=8cm]{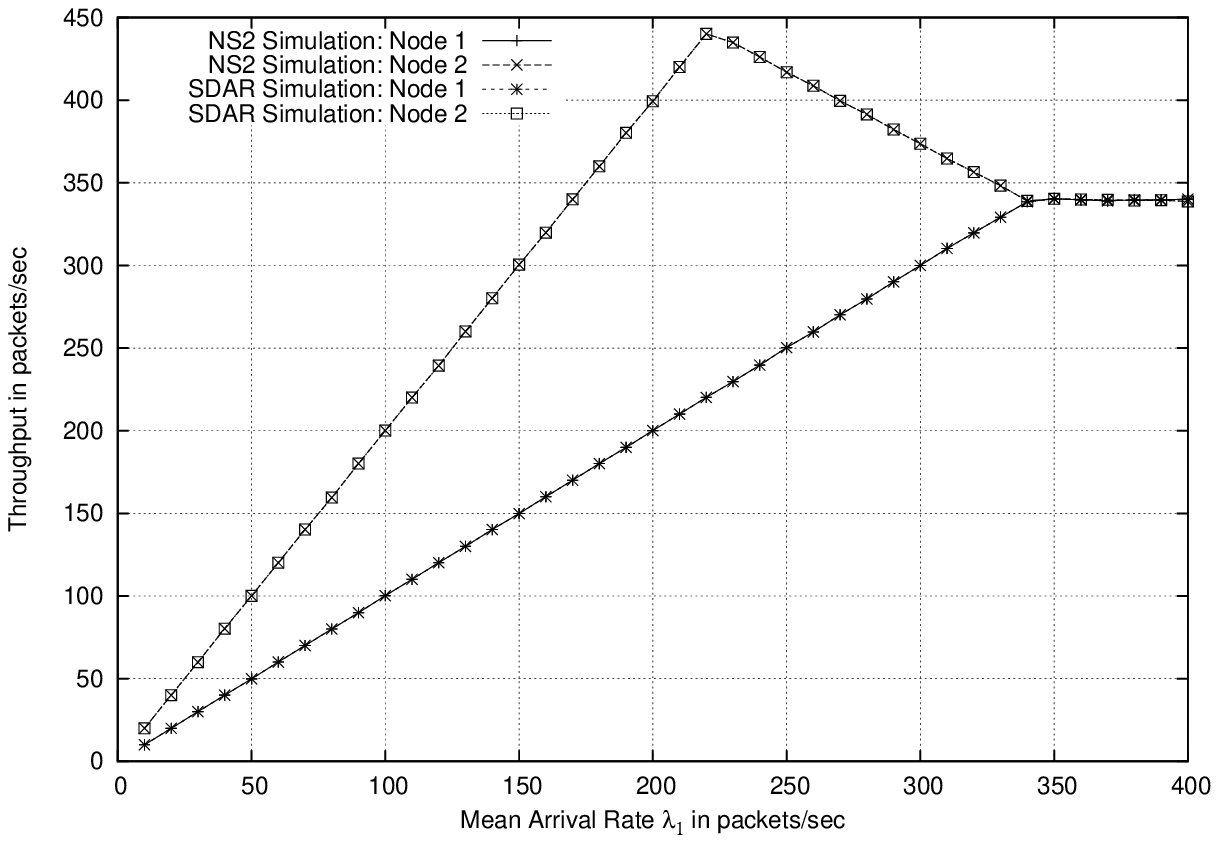}
      \caption{Comparison of throughput per node $\theta$ with $M = 2$
        nodes, infinite buffer size and unequal arrival
        rates. \label{fig:comparethetaUnequalTwoNodes}} 
      \vspace{3mm}
    \end{center}
  \end{minipage}
\hfill
  \begin{minipage}{8cm}
    \begin{center}
      \includegraphics[height=7.5cm,width=8cm]{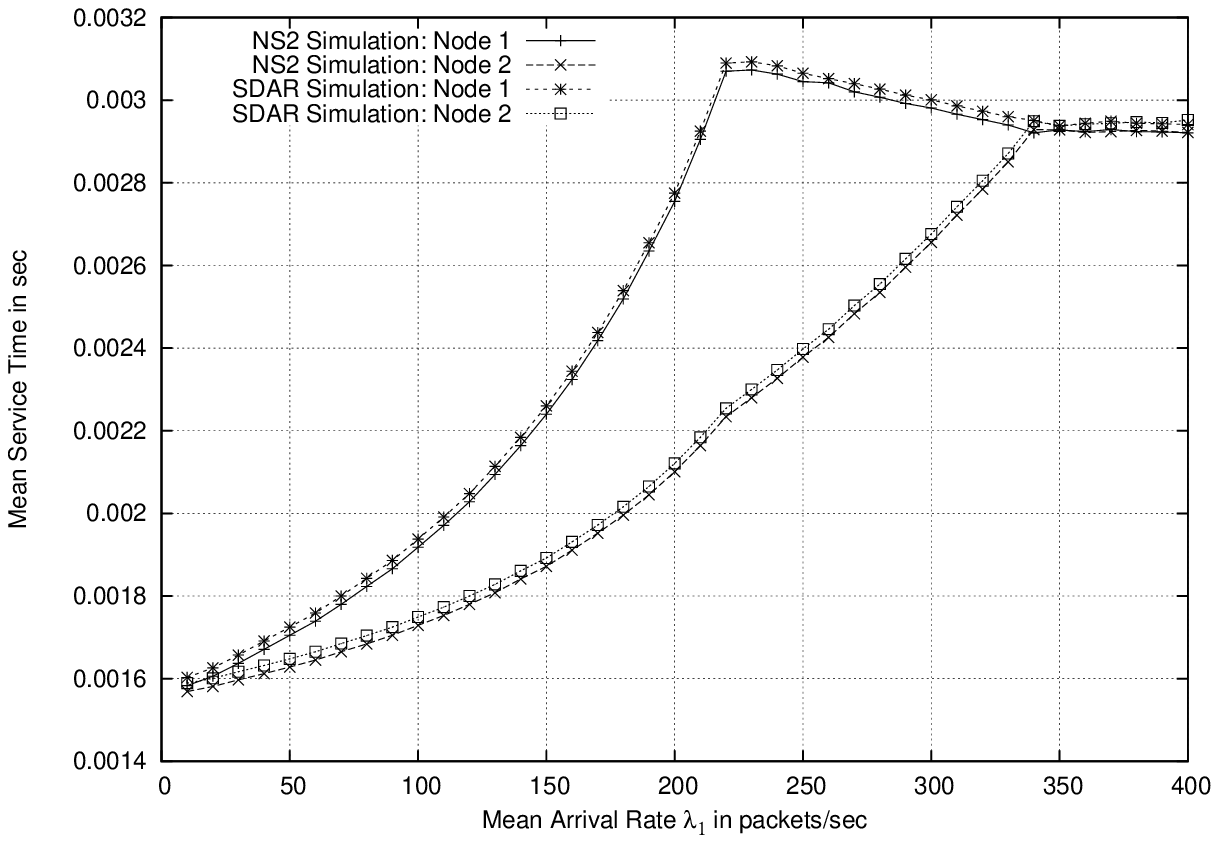}
     \caption{Comparison of mean service time with $M = 2$ nodes,
       infinite buffer size and unequal arrival
       rates. \label{fig:comparemuUnequalTwoNodes}} 
    \end{center}
  \end{minipage}
\hfill
  \begin{minipage}{8cm}
    \begin{center}
      \includegraphics[height=7.5cm,width=8cm]{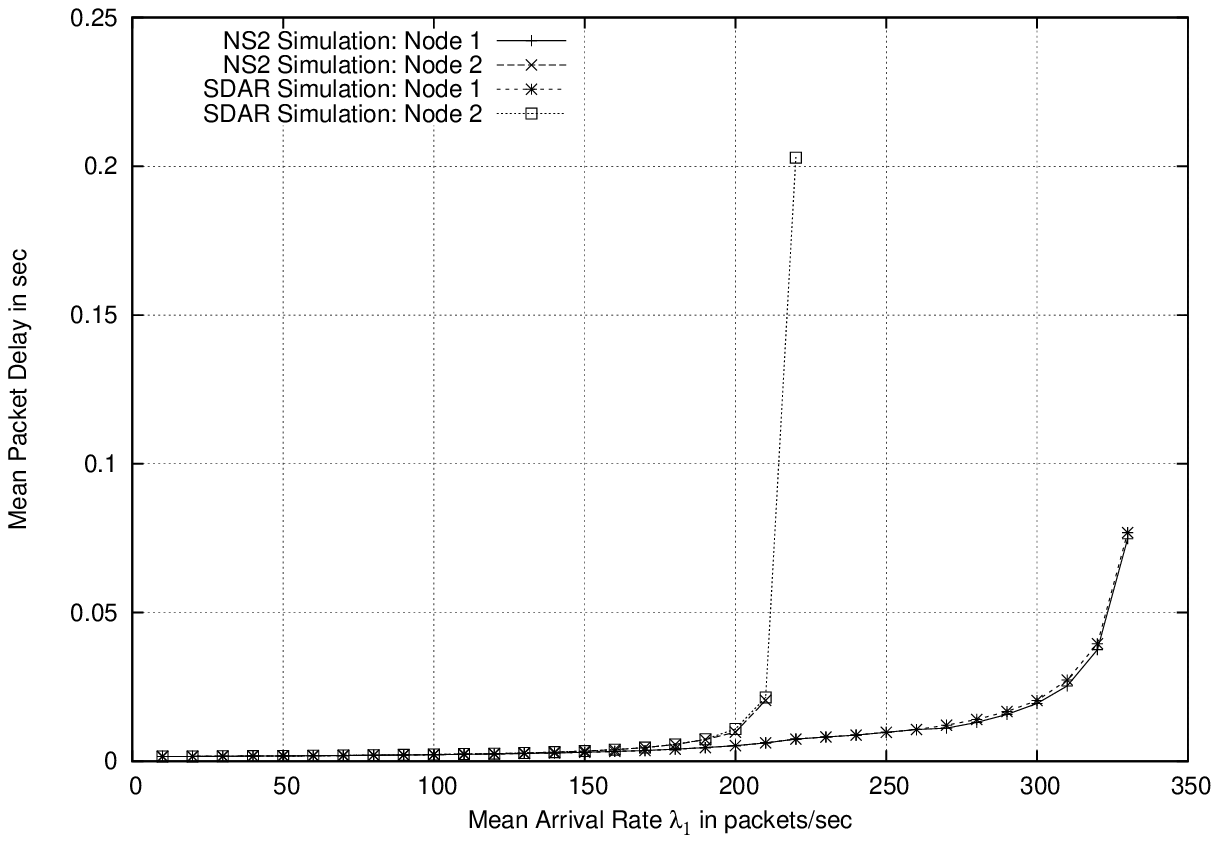}
     \caption{Comparison of mean packet delay $\bar{W}$ with $M = 2$
       nodes, infinite buffer size and unequal arrival
       rates. \label{fig:comparedelayUnequalTwoNodes}} 
    \end{center}
  \end{minipage}
\end{figure*}
\begin{figure*}[t]
  \centering 
  \begin{minipage}{8cm}
    \begin{center}
      \includegraphics[height=7.5cm,width=8cm]{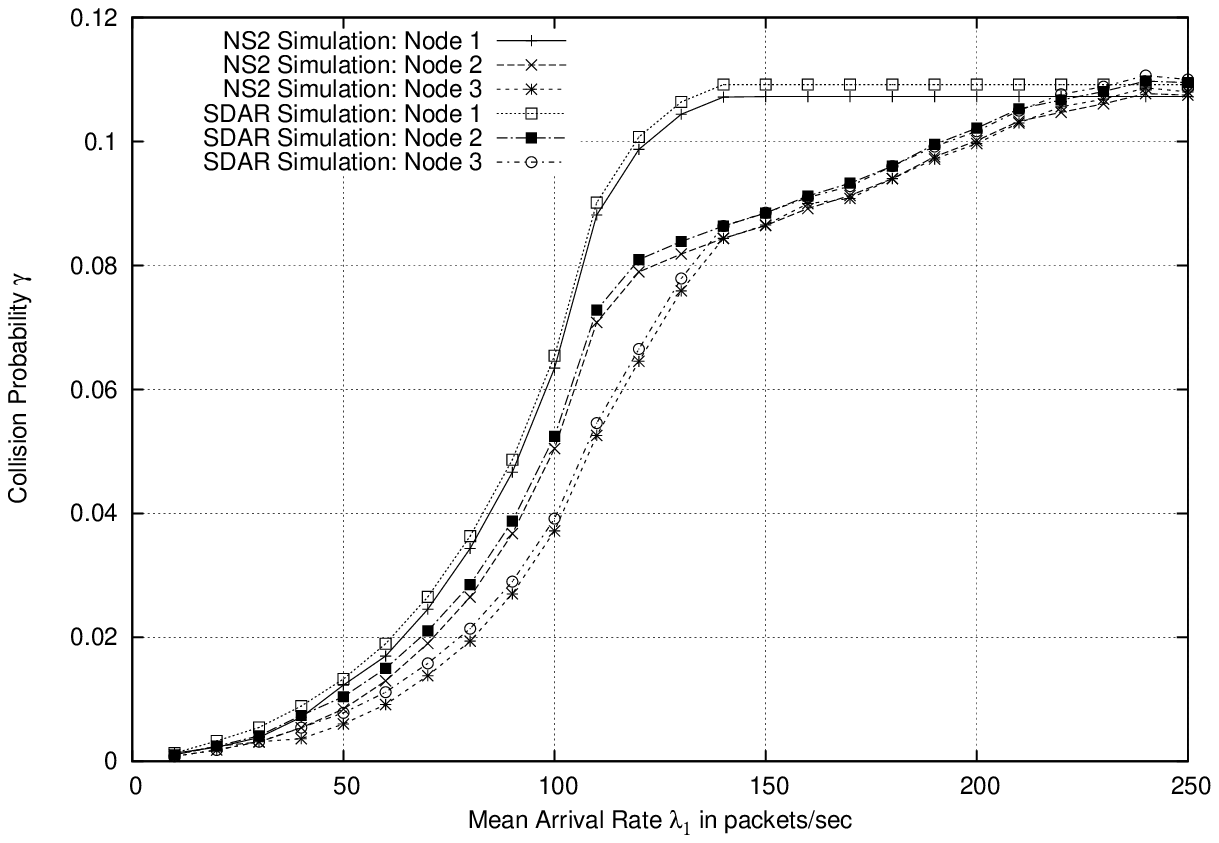}
      \caption{Comparison of collision probability $\gamma$ with $M = 3$
        nodes, infinite buffer size and unequal arrival
        rates. \label{fig:comparegammaUnequalThreeNodes}} 
      \vspace{3mm}
    \end{center}
  \end{minipage}
\hfill
  \begin{minipage}{8cm}
    \begin{center}
      \includegraphics[height=7.5cm,width=8cm]{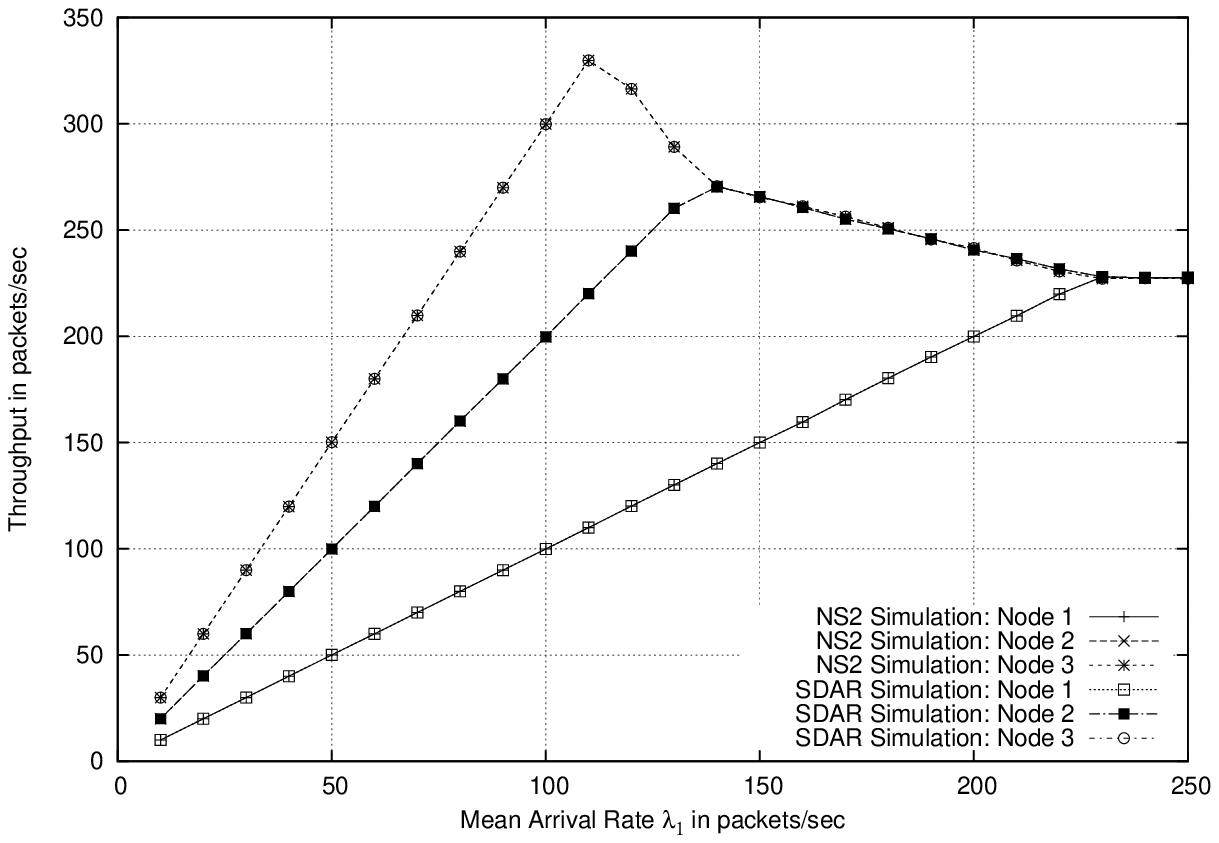}
      \caption{Comparison of throughput per node $\theta$ with $M = 3$
        nodes, infinite buffer size and unequal arrival
        rates. \label{fig:comparethetaUnequalThreeNodes}} 
      \vspace{3mm}
    \end{center}
  \end{minipage}
\hfill
  \begin{minipage}{8cm}
    \begin{center}
      \includegraphics[height=7.5cm,width=8cm]{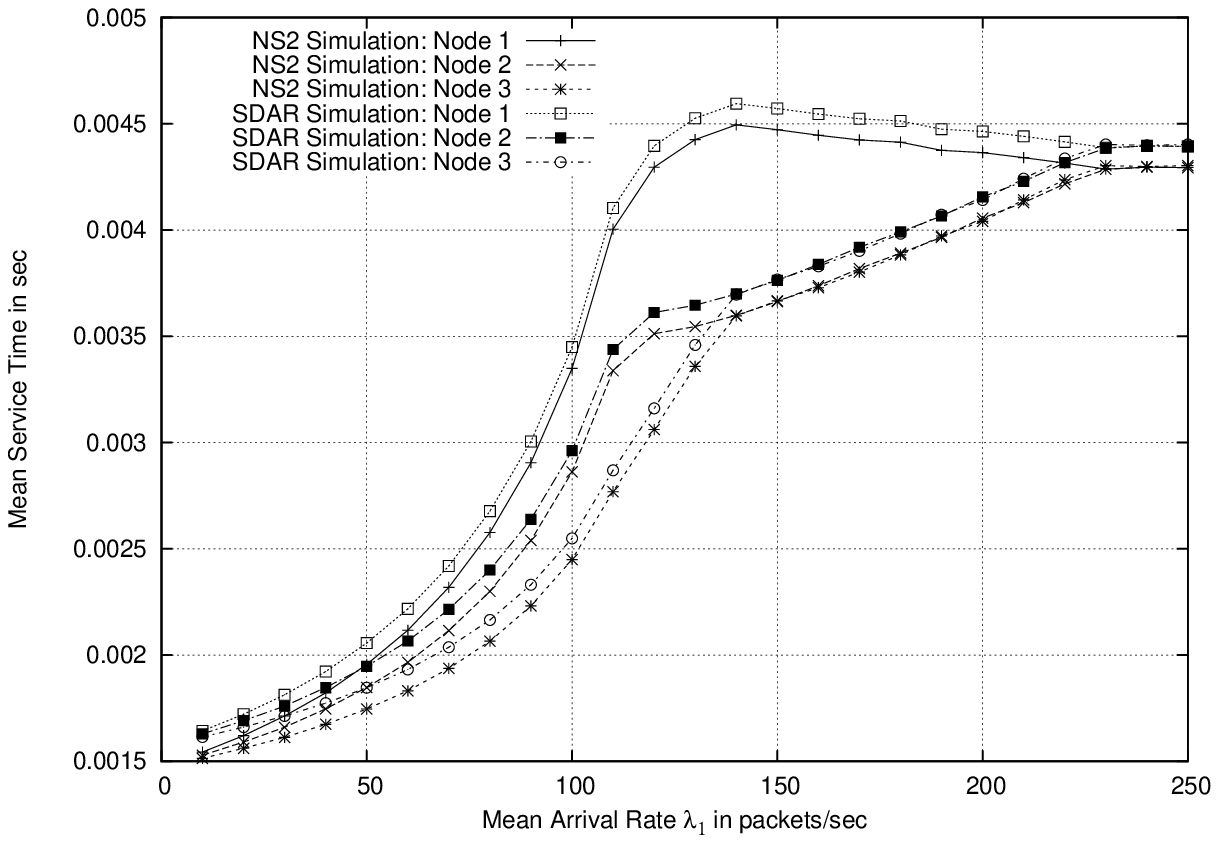}
      \caption{Comparison of mean service time with $M = 3$ nodes,
        infinite buffer size and unequal arrival
        rates. \label{fig:comparemuUnequalThreeNodes}} 
    \end{center}
  \end{minipage}
\hfill
  \begin{minipage}{8cm}
    \begin{center}
      \includegraphics[height=7.5cm,width=8cm]{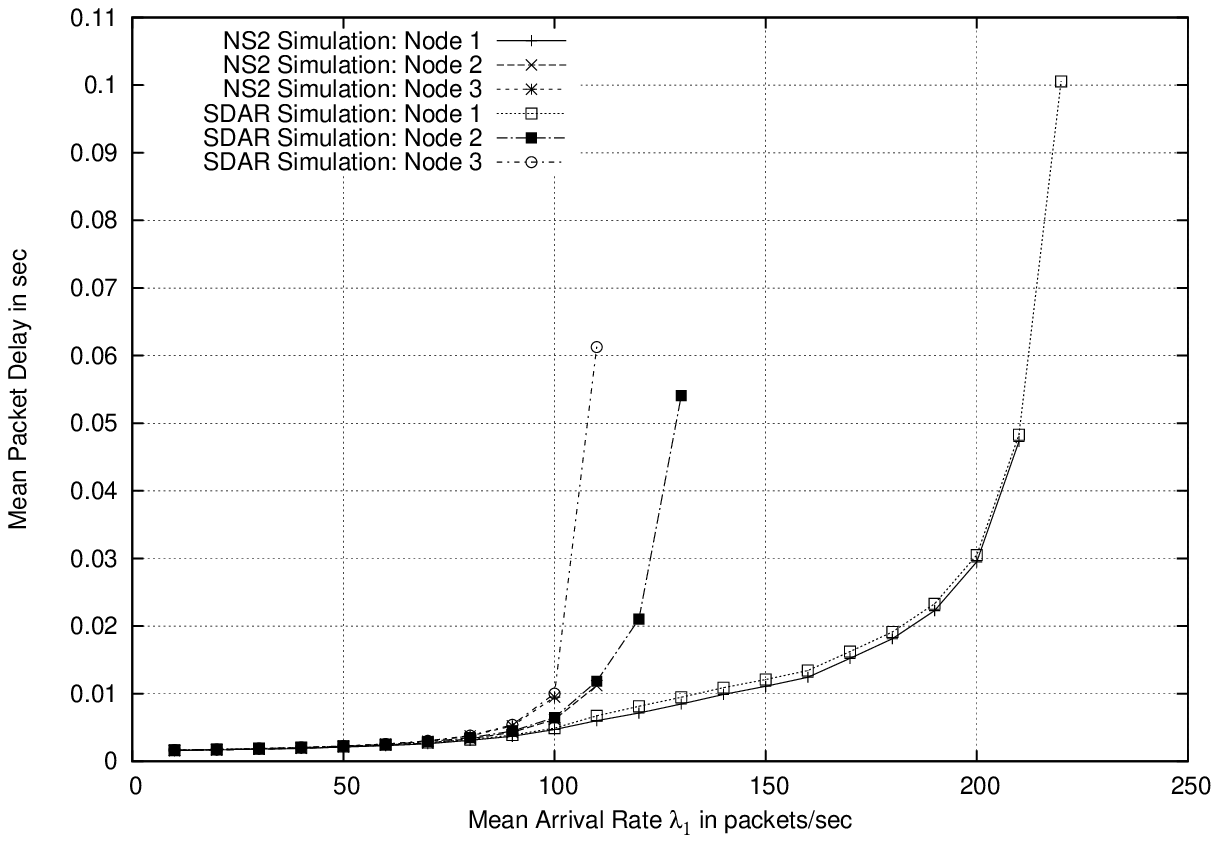}
      \caption{Comparison of mean packet delay $\bar{W}$ with $M = 3$
       nodes, infinite buffer size and unequal arrival
       rates. \label{fig:comparedelayUnequalThreeNodes}}
    \end{center}
  \end{minipage}
\end{figure*}

We now turn to some simulation results with unequal arrival rates. Figures~\ref{fig:comparegammaUnequalTwoNodes}-\ref{fig:comparedelayUnequalTwoNodes} compare the collision probability $\gamma$, the throughput, the \textit{mean service time},\footnote{Mean service time is the mean time a packet spends at the Head Of Line (HOL) position until it is dequeued. A packet is dequeued either when an ACK is received which implies a successful transmission or when a timeout occurs and the the retransmit limit has been reached.} and the mean packet delay $\bar{W}$ with $M=2$ nodes, infinite buffer size, and unequal arrival rates with ratio $\lambda_1:\lambda_2 = 1:2$. Figures~\ref{fig:comparegammaUnequalThreeNodes}-\ref{fig:comparedelayUnequalThreeNodes} compare the collision probability $\gamma$, the throughput, the mean service time, and the mean packet delay $\bar{W}$ with $M=3$ nodes, infinite buffer size, and unequal arrival rates with ratio $\lambda_1:\lambda_2:\lambda_3 = 1:2:3$. Notice that in Figures~\ref{fig:comparegammaUnequalTwoNodes}-\ref{fig:comparedelayUnequalThreeNodes} we plot the results against the arrival rate $\lambda_1$ into Node-1. The corresponding arrival rate into Node-2 (resp.~Node-3) is given by $\lambda_2 = 2 \lambda_1$ (resp.~$\lambda_3 = 3 \lambda_1$). It can be seen in Figures~\ref{fig:comparegammaUnequalTwoNodes}-\ref{fig:comparedelayUnequalThreeNodes} that results from the SDAR approximation in NS-2 match extremely well with the unmodified NS-2. This validates the SDAR approximation also for the case when the arrival rates are unequal and differ significantly from each other. 


Notice in Figure~\ref{fig:comparegammaUnequalTwoNodes} that $\gamma$ for Node-1 is larger than that for Node-2 even though $\lambda_1 < \lambda_2 = 2 \lambda_1$. This can be explained as follows.  recall that $\gamma$ is the conditional collision probability, i.e., $\gamma$ is the probability with which an attempted transmission collides. The probability that ``Node-2 is non-empty given that Node-1 is non-empty'' is larger than the probability that ``Node-1 is non-empty given that Node-2 is non-empty'' since $\lambda_2 > \lambda_1$. Hence, the probability that Node-1's attempts collide is larger than the probability that Node-2's attempts collide. The fact that $\gamma$ for Node-1 is larger than that for Node-2 explains why the mean service time of Node-1's packets is larger than the mean service time of Node-2's packets (see Figure~\ref{fig:comparemuUnequalTwoNodes}); a larger $\gamma$ implies that a larger number of attempts is need for successfully transmitting any packet. However, since $\lambda_2 > \lambda_1$, the ``queueing delay'' for Node-2 is larger than that for Node-1. This explains why the mean packet delay $\bar{W}$, which includes the queueing delay, is larger for Node-2 than that for Node-1 (see Figure~\ref{fig:comparedelayUnequalTwoNodes}). Similar explanations hold for the three-node case as well.

\section{Conclusion}
\label{sec:conclusion}

In this paper, we applied the SDAR approximation to model the attempt processes of the nodes in a single cell WLAN. Applying the SDAR approximation and considering Poisson arrivals, we developed a Markov model and reduced the problem of modeling a single cell WLAN with non-saturated nodes to the problem of analyzing a coupled queue system. We provided a sufficient condition under which the joint queue length Markov chain is positive recurrent. Assuming equal arrival rates, we proposed a technique to reduce the state space of the coupled queue system. For the case of equal arrival rates, and finite and equal buffers, we proposed an iterative method to obtain the stationary distribution of the reduced state process and obtained accurate numerical predictions for important performance measures. Our iterative method was shown to be better in terms of computational complexity than an earlier work with same level of accuracy. The reduction in complexity was achieved precisely due to the SDAR approximation. We demonstrated via simulations that the SDAR model of contention provides an accurate model for the CSMA/CA protocol (for equal or unequal arrival rates and infinite or finite buffers). In addition, when applied to the NS-2 simulator it improves the simulation speed without affecting the accuracy of results. However, a theoretical justification for the SDAR approximation is lacking at this point. 



\bibliographystyle{model1b-num-names}
\bibliography{manojbibfiles}  

\begin{thebibliography}{32}
\expandafter\ifx\csname natexlab\endcsname\relax\def\natexlab#1{#1}\fi
\providecommand{\bibinfo}[2]{#2}
\ifx\xfnm\relax \def\xfnm[#1]{\unskip,\space#1}\fi
\bibitem[{wan(1999)}]{wanet.IEEE8021199standard}
\bibinfo{title}{Wireless {LAN} {M}edium {A}ccess {C}ontrol ({MAC}) and
  {P}hysical {L}ayer ({PHY}) {S}pecifications, {IEEE} {S}td 802.11, 1999
  {E}dition}, \bibinfo{year}{1999}.
\bibitem[{wan(2007)}]{wanet.IEEE802dot11standard2007}
\bibinfo{title}{Wireless {LAN} {M}edium {A}ccess {C}ontrol ({MAC}) and
  {P}hysical {L}ayer ({PHY}) {S}pecifications, {IEEE} {S}td 802.11-2007},
  \bibinfo{year}{2007}.
\bibitem[{Baccelli and Foss(1995)}]{queueing.baccelli-foss95saturation}
\bibinfo{author}{F.~Baccelli}, \bibinfo{author}{S.~Foss}, \bibinfo{title}{On
  the {S}aturation {R}ule for the {S}tability of {Q}ueues},
  \bibinfo{journal}{Journal of Applied Probability} \bibinfo{volume}{32}
  (\bibinfo{year}{1995}) \bibinfo{pages}{494--507}.
\bibitem[{Bae et~al.(2008)Bae, Kim, Moon and
  Choi}]{wanet.choi_etal08matrix-analytic}
\bibinfo{author}{Y.H. Bae}, \bibinfo{author}{K.J. Kim}, \bibinfo{author}{M.N.
  Moon}, \bibinfo{author}{B.D. Choi}, \bibinfo{title}{{A}nalysis of {IEEE}
  802.11 {N}on-{S}aturated {DCF} by {M}atrix {A}nalytic {M}ethods},
  \bibinfo{journal}{Annals of Operations Research} \bibinfo{volume}{162}
  (\bibinfo{year}{2008}) \bibinfo{pages}{3--18}.
\bibitem[{Bianchi(2000)}]{wanet.bianchi00performance}
\bibinfo{author}{G.~Bianchi}, \bibinfo{title}{Performance {A}nalysis of the
  {IEEE} 802.11 {D}istributed {C}oordination {F}unction},
  \bibinfo{journal}{IEEE Journal on Selected Areas in Communications}
  \bibinfo{volume}{18} (\bibinfo{year}{2000}) \bibinfo{pages}{535--547}.
\bibitem[{Bruno et~al.(2006)Bruno, Conti and
  Gregori}]{wanet.bruno-etal06tcp-over-dot11}
\bibinfo{author}{R.~Bruno}, \bibinfo{author}{M.~Conti},
  \bibinfo{author}{E.~Gregori}, \bibinfo{title}{Performance {M}odeling and
  {M}easurements of {TCP} {T}ransfer {T}hroughput in 802.11-based {WLAN}s}, in:
  \bibinfo{booktitle}{MSWiM}, \bibinfo{address}{Torremolinos, Malaga, Spain},
  \bibinfo{year}{2006}.
\bibitem[{Cali et~al.(2000{\natexlab{a}})Cali, Conti and
  Gregori}]{wanet.cali-etal00throughput-limit}
\bibinfo{author}{F.~Cali}, \bibinfo{author}{M.~Conti},
  \bibinfo{author}{E.~Gregori}, \bibinfo{title}{Dynamic {T}uning of the {IEEE}
  802.11 {P}rotocol to {A}chieve a {T}heoretical {T}hroughput {L}imit},
  \bibinfo{journal}{IEEE/ACM Transactions on Networking} \bibinfo{volume}{8}
  (\bibinfo{year}{2000}{\natexlab{a}}) \bibinfo{pages}{785--799}.
\bibitem[{Cali et~al.(2000{\natexlab{b}})Cali, Conti and
  Gregori}]{wanet.cali-etal00adaptive-backoff}
\bibinfo{author}{F.~Cali}, \bibinfo{author}{M.~Conti},
  \bibinfo{author}{E.~Gregori}, \bibinfo{title}{{IEEE} 802.11 protocol:
  {D}esign and performance evaluation of an adaptive backoff mechanism},
  \bibinfo{journal}{IEEE Journal on Selected Areas in Communications}
  \bibinfo{volume}{18} (\bibinfo{year}{2000}{\natexlab{b}})
  \bibinfo{pages}{1774--1780}.
\bibitem[{Cantieni et~al.(2005)Cantieni, Ni, Barakat and
  Turletti}]{wanet.Cantieni_etal05finite-load-and-multirate}
\bibinfo{author}{G.R. Cantieni}, \bibinfo{author}{Q.~Ni},
  \bibinfo{author}{C.~Barakat}, \bibinfo{author}{T.~Turletti},
  \bibinfo{title}{Performance {A}nalysis under {F}inite {L}oad and
  {I}mprovements for {M}ultirate 802.11b}, \bibinfo{journal}{Elsevier Computer
  Communication Journal}  (\bibinfo{year}{2005}).
\bibitem[{Duffy and Ganesh(2007)}]{wanet.ganesh-duffy-07-nonsat-commletter}
\bibinfo{author}{K.~Duffy}, \bibinfo{author}{A.J. Ganesh},
  \bibinfo{title}{Modeling the {I}mpact of {B}uffering on 802.11},
  \bibinfo{journal}{IEEE Communication Letters} \bibinfo{volume}{11}
  (\bibinfo{year}{2007}) \bibinfo{pages}{219--221}.
\bibitem[{Fayolle et~al.(1995)Fayolle, Malyshev and
  Menshikov}]{theory.fayolle-etal95markovchains}
\bibinfo{author}{G.~Fayolle}, \bibinfo{author}{V.A. Malyshev},
  \bibinfo{author}{M.V. Menshikov}, \bibinfo{title}{Topics in the
  {C}onstructive {T}heory of {M}arkov {C}hains}, \bibinfo{publisher}{Cambridge
  University Press}, \bibinfo{year}{1995}.
\bibitem[{Feller(1971)}]{theory.feller71volII}
\bibinfo{author}{W.~Feller}, \bibinfo{title}{An {I}ntroduction to {P}robability
  {T}heory and {I}ts {A}pplications}, volume~\bibinfo{volume}{II},
  \bibinfo{publisher}{John Wiley \& Sons, Inc.}, \bibinfo{year}{1971}.
\bibitem[{Foh and Zukerman(2002)}]{wanet.foh-zukerman02SDSR}
\bibinfo{author}{C.H. Foh}, \bibinfo{author}{M.~Zukerman},
  \bibinfo{title}{{P}erformance {A}nalysis of the {IEEE} 802.11 {MAC}
  {P}rotocol}, in: \bibinfo{booktitle}{European Wireless Conference},
  \bibinfo{address}{Florence, Italy}, pp. \bibinfo{pages}{184--190},
  \bibinfo{year}{2002}.
\bibitem[{Garetto and
  Chiasserini(2005)}]{wanet.garetto-chiasserini05802.11-MAC-sporadic-traffic}
\bibinfo{author}{M.~Garetto}, \bibinfo{author}{C.F. Chiasserini},
  \bibinfo{title}{Performance {A}nalysis of the 802.11 {D}istributed
  {C}oordination {F}unction under {S}poradic {T}raffic}, in:
  \bibinfo{booktitle}{Networking}, \bibinfo{address}{Waterloo, Ontario,
  Canada}, \bibinfo{year}{2005}.
\bibitem[{Hegde et~al.(2005)Hegde, Proutiere and
  Roberts}]{wanet.hegde_etal05voice-capacity}
\bibinfo{author}{N.~Hegde}, \bibinfo{author}{A.~Proutiere},
  \bibinfo{author}{J.~Roberts}, \bibinfo{title}{{E}valuating the {V}oice
  {C}apacity of 802.11 {WLAN} under {D}istributed {C}ontrol}, in:
  \bibinfo{booktitle}{IEEE LANMAN}, \bibinfo{address}{Chania, Greece},
  \bibinfo{year}{2005}.
\bibitem[{Huang and Duffy(2009)}]{wanet.huang-duffy09buffering-hypothesis}
\bibinfo{author}{K.D. Huang}, \bibinfo{author}{K.R. Duffy},
  \bibinfo{title}{{O}n a {B}uffering {H}ypothesis in 802.11 {A}nalytic
  {M}odels}, \bibinfo{journal}{IEEE Communication Letters} \bibinfo{volume}{13}
  (\bibinfo{year}{2009}) \bibinfo{pages}{312--314}.
\bibitem[{Huang et~al.(2008)Huang, Duffy and
  Malone}]{wanet.huang_etal08MAC-modeling-hypotheses}
\bibinfo{author}{K.D. Huang}, \bibinfo{author}{K.R. Duffy},
  \bibinfo{author}{D.~Malone}, \bibinfo{title}{{O}n the {V}alidity of {IEEE}
  802.11 {MAC} {M}odeling {H}ypotheses}, \bibinfo{type}{Technical Report},
  Hamilton Institute, NUI Maynooth, \bibinfo{year}{2008}.
\bibitem[{Kobayashi(1978)}]{theory.kobayashi78modeling}
\bibinfo{author}{H.~Kobayashi}, \bibinfo{title}{Modeling and Analysis},
  \bibinfo{publisher}{Addison Wesley}, \bibinfo{year}{1978}.
\bibitem[{Kulkarni(1995)}]{theory.kulkarni95modeling-analysis-stochastic-syste%
ms}
\bibinfo{author}{V.G. Kulkarni}, \bibinfo{title}{Modeling and Analysis of
  Stochastic Systems}, \bibinfo{publisher}{Chapman \& Hall},
  \bibinfo{year}{1995}.
\bibitem[{Kumar et~al.(2007)Kumar, Altman, Miorandi and
  Goyal}]{wanet.kumar_etal07new_insights}
\bibinfo{author}{A.~Kumar}, \bibinfo{author}{E.~Altman},
  \bibinfo{author}{D.~Miorandi}, \bibinfo{author}{M.~Goyal},
  \bibinfo{title}{New insights from a fixed point analysis of single cell
  {IEEE}~802.11 {WLAN}s}, \bibinfo{journal}{IEEE/ACM Transactions on
  Networking} \bibinfo{volume}{15} (\bibinfo{year}{2007})
  \bibinfo{pages}{588--601}. \bibinfo{note}{Also appeared in INFOCOM 2005}.
\bibitem[{Kumar et~al.(2004)Kumar, Manjunath and
  Kuri}]{theory.kumar_etal04communication-networking}
\bibinfo{author}{A.~Kumar}, \bibinfo{author}{D.~Manjunath},
  \bibinfo{author}{J.~Kuri}, \bibinfo{title}{Communication Networking: An
  Analytical Approach}, \bibinfo{publisher}{Morgan Kaufmann Publishers, An
  Imprint of Elsevier}, \bibinfo{year}{2004}.
\bibitem[{Kumar and Patil(1997)}]{wanet.kumar-patil97cdma-aloha}
\bibinfo{author}{A.~Kumar}, \bibinfo{author}{D.~Patil},
  \bibinfo{title}{Stability and throughput analysis of unslotted {CDMA-ALOHA}
  with finite numer of users and code sharing},
  \bibinfo{journal}{Telecommunication Systems} \bibinfo{volume}{8}
  (\bibinfo{year}{1997}) \bibinfo{pages}{257--275}.
\bibitem[{Kuriakose et~al.(2009)Kuriakose, Harsha, Kumar and
  Sharma}]{wanet.harsha07WiNet}
\bibinfo{author}{G.~Kuriakose}, \bibinfo{author}{S.~Harsha},
  \bibinfo{author}{A.~Kumar}, \bibinfo{author}{V.~Sharma},
  \bibinfo{title}{Analytical {M}odels for {C}apacity {E}stimation of {IEEE}
  802.11 {WLAN}s using {DCF} for {I}nternet {A}pplication},
  \bibinfo{journal}{Wireless Networks (Springer)} \bibinfo{volume}{15}
  (\bibinfo{year}{2009}) \bibinfo{pages}{259--277}.
\bibitem[{Litjens et~al.(2003)Litjens, Roijers, van~den Berg, Boucherie and
  Fleuren}]{wanet.litjens_etalITC03integrated_packet_flow}
\bibinfo{author}{R.~Litjens}, \bibinfo{author}{F.~Roijers},
  \bibinfo{author}{J.L. van~den Berg}, \bibinfo{author}{R.J. Boucherie},
  \bibinfo{author}{M.~Fleuren}, \bibinfo{title}{{P}erformance {A}nalysis of
  {W}ireless {LAN}s: {A}n {I}ntegrated {P}acket/{F}low {L}evel {A}pproach}, in:
  \bibinfo{booktitle}{ITC 18}, \bibinfo{address}{Berlin, Germany},
  \bibinfo{year}{2003}.
\bibitem[{Malone et~al.(2007)Malone, Duffy and
  Leith}]{wanet.malone_etal07finite-load-heterogeneous}
\bibinfo{author}{D.~Malone}, \bibinfo{author}{K.~Duffy},
  \bibinfo{author}{D.~Leith}, \bibinfo{title}{Modeling the 802.11 {D}istributed
  {C}oordination {F}unction in {N}on-saturated {H}eterogeneous {C}onditions},
  \bibinfo{journal}{IEEE/ACM Transactions on Networking} \bibinfo{volume}{15}
  (\bibinfo{year}{2007}) \bibinfo{pages}{159--172}.
\bibitem[{McCanne and Floyd(????)}]{wanet.ns2}
\bibinfo{author}{S.~McCanne}, \bibinfo{author}{S.~Floyd}, \bibinfo{title}{The
  ns {N}etwork {S}imulator}, ???? \bibinfo{note}{Available for free download at
  {\tt http://www.isi.edu/nsnam/ns/}}.
\bibitem[{Miorandi et~al.(2006)Miorandi, Kherani and
  Altman}]{wanet.miorandi_etal06http_over_wlans}
\bibinfo{author}{D.~Miorandi}, \bibinfo{author}{A.A. Kherani},
  \bibinfo{author}{E.~Altman}, \bibinfo{title}{A {Q}ueueing {M}odel for {HTTP}
  {T}raffic over {IEEE} 802.11 {WLAN}s}, \bibinfo{journal}{Computer Networks
  (Elsevier)} \bibinfo{volume}{50} (\bibinfo{year}{2006})
  \bibinfo{pages}{63--79}.
\bibitem[{Neuts(1989)}]{theory.neuts89MG1}
\bibinfo{author}{M.F. Neuts}, \bibinfo{title}{Structured Stochastic Matrices of
  M/G/1 Type and their Applications}, \bibinfo{publisher}{Marcel Dekker},
  \bibinfo{year}{1989}.
\bibitem[{Panda and Kumar(2009)}]{wanet.manoj-anuragCOMSNETS09SDAR}
\bibinfo{author}{M.K. Panda}, \bibinfo{author}{A.~Kumar},
  \bibinfo{title}{{S}tate {D}ependent {A}ttempt {R}ate {M}odeling of {S}ingle
  {C}ell {IEEE} 802.11 {WLAN}s with {H}omogeneous {N}odes and {P}oisson
  {A}rrivals}, in: \bibinfo{booktitle}{COMSNETS 2009}, \bibinfo{year}{2009}.
  \bibinfo{note}{{\tt DOI: 10.1109/COMSNETS.2009.4808879}}.
\bibitem[{Sykas et~al.(1986)Sykas, Karvelas and
  Protonotarios}]{wanet.sykasetal86buffanal}
\bibinfo{author}{E.D. Sykas}, \bibinfo{author}{D.E. Karvelas},
  \bibinfo{author}{E.N. Protonotarios}, \bibinfo{title}{{Q}ueueing {A}nalysis
  of {S}ome {B}uffered {R}andom {M}ultiple {A}ccess {S}chemes},
  \bibinfo{journal}{IEEE Transactions on Communications} \bibinfo{volume}{34}
  (\bibinfo{year}{1986}) \bibinfo{pages}{790--798}.
\bibitem[{Tickoo and
  Sikdar(2004)}]{wanet.tickoo-sikdar04finite-load-queueing-model-802.11-MAC}
\bibinfo{author}{O.~Tickoo}, \bibinfo{author}{B.~Sikdar}, \bibinfo{title}{A
  {Q}ueueing {M}odel for {F}inite {L}oad {IEEE} 802.11 {R}andom {A}ccess
  {MAC}}, in: \bibinfo{booktitle}{ICC}, \bibinfo{year}{2004}.
\bibitem[{Winands et~al.(2004)Winands, Deteneer, Resing and
  Rietman}]{wanet.winands_etal04finite-source-feedback}
\bibinfo{author}{E.M.M. Winands}, \bibinfo{author}{T.J.J. Deteneer},
  \bibinfo{author}{J.A.C. Resing}, \bibinfo{author}{R.~Rietman},
  \bibinfo{title}{A {F}inite-{S}ource {F}eedback {Q}ueueing {N}etwork as a
  {M}odel for the {IEEE} 802.11 {D}istributed {C}oordination {F}unction}, in:
  \bibinfo{booktitle}{European Wireless Conference},
  \bibinfo{address}{Barcelona, Spain}, pp. \bibinfo{pages}{551--557},
  \bibinfo{year}{2004}.

\end{thebibliography}

\newpage
\appendix
\noindent \textbf{Appendix}


\section{Proof of Theorem~\ref{thm:positive-recurrence-Qt}} 
\label{app:positive-recurrence-Qt}

We apply \textit{Foster's criterion} (see page 29, chapter 2 of~\cite{theory.fayolle-etal95markovchains}) to prove Theorem~\ref{thm:positive-recurrence-Qt}. \begin{theorem}[Foster's Criterion]
\label{thm:foster} 
A time homogeneous irreducible aperiodic Markov chain $\{X(t), t \geq 0\}$ with a countable state space $\mathcal{S}$ is positive recurrent if and only if there exists a non-negative function $f(x)$, $x \in \mathcal{S}$, a number $\epsilon > 0$, and a finite set $A \subset \mathcal{S}$, such that the following conditions hold: \begin{itemize}

\item [(1)] $\EXP{f(X(t+1)) - f(X(t)) \; \big| \; X(t) = x} \leq - \epsilon,
  \forall x \in \mathcal{S}\setminus A,$ 

\item [(2)] $\EXP{f(X(t+1)) \; \big| \; X(t) = x} < \infty, \forall x \in
  A$. \hfill \qed

\end{itemize} \end{theorem} Clearly, the DTMC $\{\bmath{Q}(t), t \geq 0\}$ is time homogeneous and has a countable state space $\mathcal{S} = \mathbb{N}^M$. Irreducibility and aperiodicity can be easily shown as follows. If, $\forall i, 1 \leq i \leq M$, $\lambda_i > 0$, then any state $\bmath{k} = (k_1, k_2, \ldots, k_M) \in \mathbb{N}^M$ can be reached from the state $\bmath{0} = (0, 0, \ldots, 0) \in \mathbb{N}^M$ in one step by $k_i$, $1 \leq i \leq M$, arrivals to the $i^{th}$ queue. Also, the state $(0, 0, \ldots, 0)$ can be reached from any state $(k_1, k_2, \ldots, k_M) \in \mathbb{N}^M$ in $\sum_{i=1}^{M} k_i$ steps by $\sum_{i=1}^{M} k_i$ consecutive successes and no arrivals such that the $k_i$'s do not increase in between. Hence, the DTMC $\{\bmath{Q}(t), t \geq 0\}$ is irreducible. Since $\{\bmath{Q}(t), t \geq 0\}$ is irreducible and there exists a self loop, e.g., from the state $(0, 0, \ldots, 0)$ to itself, the DTMC $\{\bmath{Q}(t), t \geq 0\}$ is aperiodic as well. 

We define the finite set $A$ and the non-negative function $f(\cdot)$ as follows: \[A := \{\bmath{0}\} = \{(0, 0, \ldots, 0)\} \; ; \; f(k_1, k_2, \ldots, k_M) := \sum_{i=1}^{M} k_i.\] Notice that $A := \{\bmath{0}\} = \{(0, 0, \ldots, 0)\}$ consists only of the ``system-empty'' state, and for any state $\bmath{k} = (k_1, k_2, \ldots, k_M)$, the function $f(k_1, k_2, \ldots, k_M) := \sum_{i=1}^{M} k_i$ gives the total number (of packets) in the system. 


Since the transition probabilities of the DTMC $\{\bmath{Q}(t), t \geq 0\}$ in any state depend on the number of non-empty nodes in that state, we partition the state space $\mathcal{S} = \mathbb{N}^M$ as \[\mathcal{S} = \bigcup_{n=0}^M \mathcal{S}_n,\] where \[\mathcal{S}_n := \left\{\bmath{k} = (k_1, k_2, \ldots, k_M) \in \mathbb{N}^M \; : \; \sum_{i=1}^{M} \ind{k_i > 0} = n\right\}.\] For all $\bmath{k} \in \mathcal{S}_n$, we have 
\begin{eqnarray*} 
\lefteqn{\EXP{f(\bmath{Q}(t+1)) - f(\bmath{Q}(t)) \; \big| \; \bmath{Q}(t) = \bmath{k}}} \nonumber \\ 
&=& \EXP{f(\bmath{Q}(t+1)) - f(\bmath{Q}(t)) \; \big| \; \bmath{Q}(t) = \bmath{k}, N(t) = n} \nonumber \\ 
&=& P\big(L(t+1) = L_{idle} \; \big | \; N(t) = n\big) \; \EXP{f(\bmath{Q}(t+1)) - f(\bmath{Q}(t)) \; \big| \; \bmath{Q}(t) = \bmath{k}, N(t) = n, L(t+1) = L_{idle}} \nonumber \\ 
&& + \; P\big(L(t+1) = L_{coll} \; \big | \; N(t) = n\big) \; \EXP{f(\bmath{Q}(t+1)) - f(\bmath{Q}(t)) \; \big| \; \bmath{Q}(t) = \bmath{k}, N(t) = n, L(t+1) = L_{coll}} \nonumber \\ 
&& + \; P\big(L(t+1) = L_{succ} \; \big | \; N(t) = n\big) \; \EXP{f(\bmath{Q}(t+1)) - f(\bmath{Q}(t)) \; \big| \; \bmath{Q}(t) = \bmath{k}, N(t) = n, L(t+1) = L_{succ}} \nonumber \\ 
&=& p_{idle,n} \; \EXP{\displaystyle \sum_{i=1}^M \big(Q_i(t+1) - Q_i(t)\big) \; \big| \; \bmath{Q}(t) = \bmath{k}, N(t) = n, L(t+1) = L_{idle}} \nonumber \\ 
&& + \; p_{coll,n} \; \EXP{\displaystyle \sum_{i=1}^M \big(Q_i(t+1) - Q_i(t)\big) \; \big| \; \bmath{Q}(t) = \bmath{k}, N(t) = n, L(t+1) = L_{coll}} \nonumber \\ 
&& + \; p_{succ,n} \; \EXP{\displaystyle \sum_{i=1}^M \big(Q_i(t+1) - Q_i(t)\big) \; \big| \; \bmath{Q}(t) = \bmath{k}, N(t) = n, L(t+1) = L_{succ}} \nonumber \\ 
\end{eqnarray*} \begin{eqnarray}
\label{eqn:verify-condition-2}
&=& p_{idle,n} \; \EXP{\displaystyle \sum_{i=1}^M A_i(t+1) \; \big| \; L(t+1) = L_{idle}} + \; p_{coll,n} \; \EXP{\displaystyle \sum_{i=1}^M A_i(t+1) \; \big| \; L(t+1) = L_{coll}} \nonumber \\ 
&& + \; p_{succ,n} \; \EXP{\displaystyle \left(\sum_{i=1}^M A_i(t+1)\right) - 1\; \big| \; L(t+1) = L_{succ}} \nonumber \\ 
&=& p_{idle,n} \; \displaystyle \left(\sum_{i=1}^M \EXP{A_i(t+1) \; \big| \; L(t+1) = L_{idle}}\right) + \; p_{coll,n} \; \displaystyle \left(\sum_{i=1}^M \EXP{A_i(t+1) \; \big| \; L(t+1) = L_{coll}}\right)
\nonumber \\ 
&& + \; p_{succ,n} \; \left( \displaystyle \sum_{i=1}^M \EXP{A_i(t+1) \; \big| \; L(t+1) = L_{succ}} \right) - p_{succ,n} \nonumber \\ 
 &=& p_{idle,n} \; \displaystyle \left(\sum_{i=1}^M \lambda_i \sigma \right) + \; p_{coll,n} \; \displaystyle \left(\sum_{i=1}^M \lambda_i \left(\sigma + T_c\right) \right) + \; p_{succ,n} \; \displaystyle \left(\sum_{i=1}^M \lambda_i \left(\sigma + T_s\right) \right) - p_{succ,n} \nonumber \\ 
&=& \displaystyle \left(\sum_{i=1}^M \lambda_i \right) \left(\sigma + p_{coll,n} T_c + p_{succ,n} T_s \right) - p_{succ,n}, 
\end{eqnarray} where we have used the following facts: (i) $\left(\bmath{Q}(t) = \bmath{k} \in \mathcal{S}_n\right) \Rightarrow \left(N(t) = n\right)$, (ii) given $N(t) = n$, the type of $(t+1)^{th}$ channel slot does not depend on $\bmath{Q}(t)$, and (iii) given the type of channel slot, the quantity $\sum_{i=1}^M \left(Q_i(t+1) - Q_i(t)\right)$, which represents ``the increase in the total number (of packets) in the system in the $(t+1)^{th}$ channel slot,'' depends neither on $\bmath{Q}(t)$ nor on $N(t)$. In particular, (a) if the $(t+1)^{th}$ channel slot is an idle or a collision channel slot, then $\sum_{i=1}^M \left(Q_i(t+1) - Q_i(t)\right)$ is equal to $\sum_{i=1}^M A_i(t+1)$ which represents the total number of arrivals in the $(t+1)^{th}$ channel slot, and (b) if the $(t+1)^{th}$ channel slot is a success channel slot, then $\sum_{i=1}^M \left(Q_i(t+1) - Q_i(t)\right)$ is equal to $\left(\sum_{i=1}^M A_i(t+1)\right) - 1$ since there is exactly one departure in the $(t+1)^{th}$ channel slot. 

Noticing that the finite set $A = \mathcal{S}_0$, and that $p_{idle,0} = 1, p_{coll,0} = p_{succ,0} = 0$, condition \textit{(2)} of Theorem~\ref{thm:foster} is satisfied for the DTMC $\{\bmath{Q}(t), t \geq 0\}$ if $\left(\sum_{i=1}^M \lambda_i \right) \sigma < \infty$, i.e., if, $\forall i, 1 \leq i \leq M$, $\lambda_i < \infty$. Condition \textit{(1)} of Theorem~\ref{thm:foster} is satisfied for the DTMC $\{\bmath{Q}(t), t \geq 0\}$ if, there exists $\epsilon > 0$ such that, $\forall n, 1 \leq n \leq M$, we have \[\displaystyle \left(\sum_{i=1}^M \lambda_i \right) \left(\sigma + p_{coll,n} T_c + p_{succ,n} T_s \right) - p_{succ,n} \leq - \epsilon.\] 

Since $\epsilon$ can be made arbitrarily small, condition \textit{(1)} of Theorem~\ref{thm:foster} is satisfied for the DTMC $\{\bmath{Q}(t), t \geq 0\}$ if, $\forall n, 1 \leq n \leq M$, we have \[\left(\sum_{i=1}^M \lambda_i \right) < \frac{p_{succ,n}}{\left(\sigma + p_{coll,n} T_c + p_{succ,n} T_s\right)}.\] We define $L_{sat,n} := \left(\sigma + p_{coll,n} T_c + p_{succ,n} T_s\right)$ and $\Theta_{sat,n} := \displaystyle \frac{p_{succ,n}}{L_{sat,n}}$. Note that, $L_{sat,n}$ and $\Theta_{sat,n}$ represent the mean channel slot duration in seconds and the aggregate throughput in packets/sec, respectively, in a single cell consisting of $n$ homogeneous and saturated nodes~\cite{wanet.kumar_etal07new_insights}. Condition \textit{(1)} of Theorem~\ref{thm:foster} is satisfied for the DTMC $\{\bmath{Q}(t), t \geq 0\}$ if, $\forall n, 1 \leq n \leq M$, we have \[\left(\sum_{i=1}^M \lambda_i \right) < \Theta_{sat,n}.\] Since the condition \textit{(2)} of Theorem~\ref{thm:foster} holds at all finite arrival rates, the DTMC $\{\bmath{Q}(t), t \geq 0\}$ is positive recurrent if \[\left(\sum_{i=1}^M \lambda_i \right) < \min_{1 \leq n \leq M} \Theta_{sat,n} \] and Theorem~\ref{thm:positive-recurrence-Qt} is proved. \hfill \qed


\section{Proof of Lemma~\ref{lem:conditional-probability-simplification-exact}}
\label{app:proof-lemma-conditional-probability-simplification-exact}

$\forall j \geq 1$, $\forall l, 2 \leq l \leq M$, $\forall i, i = 0, 1, 2, \ldots$, $\forall n, 0 \leq n \leq M-1$, we have \begin{eqnarray} 
\label{eqn:conditional-probability-simplification-exact-reverse}
\lefteqn{P\big(L(t+1) = L_{succ}, D_l(t+1) = 1 \; \big| \; Q_1(t) = i, \mathcal{M}(t) = n, Q_l(t) > 0, Q_l(t) = j\big)} \nonumber \\ 
&=& P\big(L(t+1) = L_{succ} \; \big| \; Q_1(t) = i, \mathcal{M}(t) = n, Q_l(t) > 0\big) \nonumber \\ 
&& \; \; \; \; \; \cdot \; P\big(D_l(t+1) = 1 \; \big| \; Q_1(t) = i, \mathcal{M}(t) = n, Q_l(t) > 0, L(t+1) = L_{succ}\big) \nonumber \\ 
&=& P\big(L(t+1) = L_{succ}, D_l(t+1) = 1 \; \big| \; Q_1(t) = i, \mathcal{M}(t) = n, Q_l(t) > 0\big), 
\end{eqnarray} since (i) given that $Q_1(t) = i$ and $\mathcal{M}(t) = n$, we know that $N(t) = \ind{i > 0} + n$, (ii) given $N(t)$, the probability that $L(t+1) = L_{succ}$, does not depend on the exact value $j$ of $Q_l(t)$ and is equal to $p_{succ,N(t)}$, and (iii) given $N(t)$, $Q_l(t) > 0$ and $L(t+1) = L_{succ}$, the probability that $D_l(t+1) = 1$, does not depend on the exact value $j$ of $Q_l(t)$ and is given by $\frac{1}{N(t)}$ (see Equation~\eqref{eqn:departure-equallylikely}). Notice that, the condition $Q_l(t) > 0$ cannot be eliminated from the second factor in the first step (on the right hand side of Equation~\eqref{eqn:conditional-probability-simplification-exact-reverse}), since $Q_l(t) = 0$ would imply $D_l(t+1) = 0$. In fact, given $N(t)$, the probability that $L(t+1) = L_{succ}$ does not depend on anything else, and the condition $Q_l(t) > 0$ could have been eliminated from the first factor in the first step. However, we keep the condition $Q_l(t) > 0$ since it is required in the second factor in the first step, and hence, is required to obtain the final expression. 

Define now the events \[E_A := \{L(t+1) = L_{succ}, D_l(t+1) = 1\},\] \[E_B := \{Q_1(t) = i, \mathcal{M}(t) = n, Q_l(t) > 0\},\] \[E_C := \{Q_l(t) = j\}, j \geq 1. \; \; \; \; \; \; \; \; \; \; \; \; \; \; \; \; \; \; \; \; \; \; \; \; \] Equation~\eqref{eqn:conditional-probability-simplification-exact-reverse} says that, $P\big(E_A \; \big| \; E_B, E_C\big) = P\big(E_A \; \big| \; E_B\big)$. Lemma~\ref{lem:conditional-probability-simplification-exact} follows from \[\; \; \; \; \; \; \; \; \; \; \; \Big(P\big(E_A \; \big| \; E_B, E_C\big) = P\big(E_A \; \big| \; E_B\big)\Big) \Leftrightarrow \Big(P\big(E_C \; \big| \; E_B, E_A\big) = P\big(E_C \; \big| \; E_B\big)\Big). \; \; \; \; \; \; \; \qed\]

\section{Transition Probabilities of the Process $\{\mathcal{X}(t)\}$} 
\label{app:balance-eqn-Xt}

In this appendix we derive the one-step transition probabilities of the process $\{\mathcal{X}(t)\}$ from a generic state $(i,n)$ to a generic state $(j,k)$, $i,j = 0, 1, 2, \ldots$, $0 \leq n,k \leq M-1$.

\subsection{When the Tagged Queue is Empty in the Initial State}
\label{subsubsec:tagged-queue-empty}

Consider the transition probability $P\big(Q_1(t+1) = j, \mathcal{M}(t+1) = k \; \big| \; Q_1(t) = 0, \mathcal{M}(t) = n\big)$, which corresponds to the case when the tagged node is empty at the channel slot boundary $t$ and can be expanded according to whether the $(t+1)^{th}$ channel slot is an idle, a collision and a success channel slot as follows: \begin{eqnarray}
\label{eqn:pi-jk-tplus1-first-part}
\lefteqn{P\big(Q_1(t+1) = j, \mathcal{M}(t+1) = k \; \big| \; Q_1(t) = 0, \mathcal{M}(t) = n\big)} \nonumber \\ 
&=& P\big(Q_1(t+1) = j, \mathcal{M}(t+1) = k, L(t+1) = L_{idle} \; \big| \; Q_1(t) = 0, \mathcal{M}(t) = n\big) \nonumber \\ 
&& + \; P\big(Q_1(t+1) = j, \mathcal{M}(t+1) = k, L(t+1) = L_{coll} \; \big| \; Q_1(t) = 0, \mathcal{M}(t) = n\big) \nonumber \\ 
&& + \; P\big(Q_1(t+1) = j, \mathcal{M}(t+1) = k, L(t+1) = L_{succ} \; \big| \; Q_1(t) = 0, \mathcal{M}(t) = n\big). 
\end{eqnarray}

\subsubsection{The Case of Idle Channel Slot} 

The first term on the right hand side of Equation~\eqref{eqn:pi-jk-tplus1-first-part} corresponds to the case when the tagged queue is empty at the channel slot boundary $t$ and the $(t+1)^{th}$ channel slot is an idle channel slot, and can be expanded as follows: \begin{eqnarray*}
\lefteqn{P\big(Q_1(t+1) = j, \mathcal{M}(t+1) = k, L(t+1) = L_{idle} \; \big| \; Q_1(t) = 0, \mathcal{M}(t) = n\big)} \nonumber \\ 
&=& P\big(L(t+1) = L_{idle} \; \big| \; Q_1(t) = 0, \mathcal{M}(t) = n \big) \nonumber \\ 
&& \; \; \; \; \cdot \; P\big(Q_1(t+1) = j, \mathcal{M}(t+1) = k \; \big| \; Q_1(t) = 0, \mathcal{M}(t) = n, L(t+1) = L_{idle} \big) \nonumber \\ 
&=& P\big(L(t+1) = L_{idle} \; \big| \; Q_1(t) = 0, \mathcal{M}(t) = n\big) \nonumber \\ 
&& \; \; \; \; \cdot \; P\big(A_1(t+1) = j, \mathcal{M}(t+1) = k \; \big| \; Q_1(t) = 0, \mathcal{M}(t) = n, L(t+1) = L_{idle} \big) \nonumber \\ 
\end{eqnarray*} 

\begin{eqnarray}
\label{eqn:pi-jk-tplus1-first-part-idle-case-partial}
&=& P\big(L(t+1) = L_{idle} \; \big| \; N(t) = n\big) \cdot \; P\big(\mathcal{M}(t+1) = k \; \big| \; Q_1(t) = 0, \mathcal{M}(t) = n, L(t+1) = L_{idle} \big) \nonumber \\ 
&& \; \; \; \; \cdot \; P\big(A_1(t+1) = j \; \big| \; L(t+1) = L_{idle} \big) \nonumber \\ 
&=& p_{idle,n} \cdot d(j) \cdot P\big(\mathcal{M}(t+1) = k \; \big| \; Q_1(t) = 0, \mathcal{M}(t) = n, L(t+1) = L_{idle} \big), 
\end{eqnarray} where we have applied Equation~\eqref{eqn:channel-slot-probabilities} to write $P\big(L(t+1) = L_{idle} \; \big| \; N(t) = n\big) = p_{idle,n}$. We have also used the following facts: (a) $\left(L(t+1) = L_{idle}\right) \Rightarrow \left(D_m(t+1) = 0, 1 \leq m \leq M\right)$, (b) $\left(Q_1(t) = 0, D_1(t+1) = 0\right) \Rightarrow \left(Q_1(t+1) = A_1(t+1)\right)$ which implies that $Q_1(t+1) = j$, if and only if $A_1(t+1) = j$, (c) given that $L(t+1) = L_{idle}$, the probability that $A_1(t+1) = j$, does not depend on anything else and is equal to $d(j)$. (See Equation~\eqref{eqn:definition-dj} and recall that the arrival rates are equal). 

The conditional probability $P\big(\mathcal{M}(t+1) = k \; \big| \; Q_1(t) = 0, \mathcal{M}(t) = n, L(t+1) = L_{idle} \big)$ in Equation~\eqref{eqn:pi-jk-tplus1-first-part-idle-case-partial} can be expanded as follows: \begin{eqnarray}
\label{eqn:pi-jk-tplus1-first-part-idle-case-partial-ratio}
\lefteqn{P\big(\mathcal{M}(t+1) = k \; \big| \; Q_1(t) = 0, \mathcal{M}(t) = n, L(t+1) = L_{idle} \big)} \nonumber \\
&=& \frac{P\big(\mathcal{M}(t) = n, \mathcal{M}(t+1) = k \; \big| \; Q_1(t) = 0, L(t+1) = L_{idle} \big)}{P\big(\mathcal{M}(t) = n \; \big| \; Q_1(t) = 0, L(t+1) = L_{idle} \big)} 
\end{eqnarray}

Let $I_M := \{2, \ldots, M\}$ denote the set of non-tagged queues. Then, the numerator on the right hand side of Equation~\eqref{eqn:pi-jk-tplus1-first-part-idle-case-partial-ratio} can be expanded as follows: \begin{eqnarray} 
\label{eqn:pi-jk-tplus1-first-part-idle-case-partial-ratio-num}
\lefteqn{P\big(\mathcal{M}(t) = n, \mathcal{M}(t+1) = k \; \big| \; Q_1(t) = 0, L(t+1) = L_{idle} \big)} \nonumber \\
&=& \displaystyle \sum_{I \subset I_M \; : \; |I| = n} P\big(Q_m(t) > 0, \forall m \in I, Q_r(t) = 0, \forall r \in I_M \setminus I, \mathcal{M}(t+1) = k \; \big| \; Q_1(t) = 0, L(t+1) = L_{idle} \big) \nonumber \\ 
&=& \displaystyle \sum_{I \subset I_M \; : \; |I| = n} P\big(Q_m(t) > 0, \forall m \in I, Q_r(t) = 0, \forall r \in I_M \setminus I \; \big| \; Q_1(t) = 0, L(t+1) = L_{idle} \big) \nonumber \\ 
&& \; \; \; \; \; \; \; \; \; \; \; \; \; \; \; \; \cdot \; P\big(\mathcal{M}(t+1) = k \; \big| \; Q_m(t) > 0, \forall m \in I, Q_r(t) = 0, \forall r \in I_M \setminus I, Q_1(t) = 0, L(t+1) = L_{idle} \big). \nonumber \\ 
\end{eqnarray}

The second conditional probability inside the summation in the second step of Equation~\eqref{eqn:pi-jk-tplus1-first-part-idle-case-partial-ratio-num} can be simplified as follows: (Remember that $I \subset I_M$ and $|I| = n$.) \begin{eqnarray}
\label{eqn:pi-jk-tplus1-first-part-idle-case-partial-ratio-num-simplified}
\lefteqn{P\big(\mathcal{M}(t+1) = k \; \big| \; Q_m(t) > 0, \forall m \in I, Q_r(t) = 0, \forall r \in I_M \setminus I, Q_1(t) = 0, L(t+1) = L_{idle} \big)} \nonumber \\
&=& P\big(\mathcal{M}(t+1) = k \; \big| \; Q_m(t) > 0, \forall m \in I, Q_r(t) = 0, \forall r \in I_M \setminus I, Q_1(t) = 0, L(t+1) = L_{idle}, \nonumber \\
&& \; \; \; \; \; \; \; \; \; \; \; \; \; \; \; \; \; \; \; \; \; \; \; \; \; \; \; \; \; \; \; \; \; \; \; \; \; \; \; \; \; \; \; \; \; \; \; \; \; \; \; \; \; \; \; \; \; \; \; \; \; \; \; \; \; \; \; \; \; \; \; \; \; \; \; \; \; \; \; \; \; \; \; \; \; \; \; \; \; \; \; \; \; \; \; \; \; \; \; \; \; \; \; \; \; \; \; \; \; \; \; D_u(t+1) = 0, \forall u \in I_M \big) \nonumber \\ 
&=& P\big(\left(\sum_{m \in I} \ind{\left(Q_{m}(t)+A_{m}(t+1)\right) > 0} + \sum_{r \in I_M \setminus I} \ind{A_{r}(t+1) > 0}\right) = k \; \big| \; Q_{m}(t) > 0, \forall m \in I, L(t+1) = L_{idle}\big) \nonumber \\ 
&=& P\big(\left(\sum_{r \in I_M \setminus I} \ind{A_{r}(t+1) > 0}\right) = k-n \; \big| \; L(t+1) = L_{idle}\big) \nonumber \\ 
&=& {{M-n-1 \choose k-n}} (1-d(0))^{k-n} d(0)^{M-k-1}. 
\end{eqnarray}

In the first step we have used the fact that $\left(L(t+1) = L_{idle}\right) \Rightarrow \left(D_{u}(t+1) = 0, \forall u \in I_M\right)$. In the second step we first applied Equations~\eqref{eqn:definition-Mt} and~\eqref{eqn:queue-evolution} to write \[\mathcal{M}(t+1) = \sum_{u=2}^M \ind{Q_{u}(t+1) > 0} = \sum_{u=2}^M \ind{\left(Q_{u}(t)-D_{u}(t+1)+A_{u}(t+1)\right) > 0},\] and then substituted $D_{u}(t+1) = 0$, $2 \leq u \leq M$, and $Q_{r}(t) = 0$, $\forall r \in I_M \setminus I$. In the third step we have used the fact that $Q_{m}(t) > 0$, $\forall m \in I$, implies that $\sum_{m \in I} \ind{\left(Q_{m}(t)+A_{m}(t+1)\right) > 0} = |I| = n$. In the last step we observe that, $\sum_{r \in I_M \setminus I} \ind{A_{r}(t+1) > 0} = k-n$, if and only if arrivals occur to exactly $k-n$ of the $|I_M \setminus I|= M-n-1$ non-tagged queues that are empty at $t$, and, given that $L(t+1) = L_{idle}$, the probability that arrivals occur (resp.~do not occur) to a queue is given by $1-d(0)$ (resp.~$d(0)$). 

We recall that by definition, $\displaystyle {{a \choose b}} > 0$, only if $a \geq b \geq 0$, and $\displaystyle {{a \choose b}} = 0$, otherwise. 

Substituting Equation~\eqref{eqn:pi-jk-tplus1-first-part-idle-case-partial-ratio-num-simplified} in Equation~\eqref{eqn:pi-jk-tplus1-first-part-idle-case-partial-ratio-num}, we obtain \begin{eqnarray}
\label{eqn:pi-jk-tplus1-first-part-idle-case-partial-ratio-num-final}
\lefteqn{P\big(\mathcal{M}(t) = n, \mathcal{M}(t+1) = k \; \big| \; Q_1(t) = 0, L(t+1) = L_{idle} \big)} \nonumber \\
&=& \displaystyle \sum_{I \subset I_M \; : \; |I| = n} P\big(Q_m(t) > 0, \forall m \in I, Q_r(t) = 0, \forall r \in I_M \setminus I \; \big| \; Q_1(t) = 0, L(t+1) = L_{idle} \big) \nonumber \\ 
&& \; \; \; \; \; \; \; \; \; \; \; \; \; \; \; \; \; \; \; \; \; \; \; \; \; \; \; \; \; \; \; \; \; \; \; \; \; \; \; \; \; \; \; \; \; \; \; \; \; \; \; \; \; \; \; \; \; \; \; \; \cdot \; {{M-n-1 \choose k-n}} (1-d(0))^{k-n} d(0)^{M-k-1} \nonumber \\ 
&=& P\big(\mathcal{M}(t) = n \; \big| \; Q_1(t) = 0, L(t+1) = L_{idle} \big) \cdot \; {{M-n-1 \choose k-n}} (1-d(0))^{k-n} d(0)^{M-k-1}, 
\end{eqnarray} where, in the second step we first take the common factor ${{M-n-1 \choose k-n}} (1-d(0))^{k-n} d(0)^{M-k-1}$ outside the summation, and then observe that the summation is equal to the probability $P\big(\mathcal{M}(t) = n \; \big| \; Q_1(t) = 0, L(t+1) = L_{idle}\big)$. 

Substituting Equation~\eqref{eqn:pi-jk-tplus1-first-part-idle-case-partial-ratio-num-final} in Equation~\eqref{eqn:pi-jk-tplus1-first-part-idle-case-partial-ratio}, we obtain \begin{eqnarray} 
\label{eqn:pi-jk-tplus1-first-part-idle-case-partial-completing}
P\big(\mathcal{M}(t+1) = k \; \big| \; Q_1(t) = 0, \mathcal{M}(t) = n, L(t+1) = L_{idle} \big) = {{M-n-1 \choose k-n}} (1-d(0))^{k-n} d(0)^{M-k-1},
\end{eqnarray} where we observe that the probability $P\big(\mathcal{M}(t) = n \; \big| \; Q_1(t) = 0, L(t+1) = L_{idle}\big)$ gets canceled from both the numerator and the denominator. 

Substituting Equation~\eqref{eqn:pi-jk-tplus1-first-part-idle-case-partial-completing} in Equation~\eqref{eqn:pi-jk-tplus1-first-part-idle-case-partial} we obtain \begin{eqnarray} 
\label{eqn:pi-jk-tplus1-first-part-idle-case}
\lefteqn{P\big(Q_1(t+1) = j, \mathcal{M}(t+1) = k, L(t+1) = L_{idle} \; \big| \; Q_1(t) = 0, \mathcal{M}(t) = n\big)} \nonumber \\ 
&=& p_{idle,n} \cdot d(j) \cdot {{M-n-1 \choose k-n}} (1-d(0))^{k-n} d(0)^{M-k-1}.
\end{eqnarray}

\subsubsection{The Case of Collision Channel Slot} 

The second term on the right hand side of Equation~\eqref{eqn:pi-jk-tplus1-first-part} corresponds to the case when the tagged queue is empty at the channel slot boundary $t$ and the $(t+1)^{th}$ channel slot is a collision channel slot. By similar arguments, we obtain \begin{eqnarray} 
\label{eqn:pi-jk-tplus1-first-part-coll-case}
\lefteqn{P\big(Q_1(t+1) = j, \mathcal{M}(t+1) = k, L(t+1) = L_{coll} \; \big| \; Q_1(t) = 0, \mathcal{M}(t) = n\big)} \nonumber \\ 
&=& p_{coll,n} \cdot c(j) \cdot {{M-n-1 \choose k-n}} (1-c(0))^{k-n} c(0)^{M-k-1}. 
\end{eqnarray}

\subsubsection{The Case of Success Channel Slot} 

The third term on the right hand side of Equation~\eqref{eqn:pi-jk-tplus1-first-part} corresponds to the case when the tagged queue is empty at the channel slot boundary $t$ and the $(t+1)^{th}$ channel slot is a success channel slot. For this case, we have \begin{eqnarray} 
\label{eqn:pi-jk-tplus1-first-part-succ-case}
\lefteqn{P\big(Q_1(t+1) = j, \mathcal{M}(t+1) = k, L(t+1) = L_{succ} \; \big| \; Q_1(t) = 0, \mathcal{M}(t) = n\big)} \nonumber \\ 
&=& P\big(L(t+1) = L_{succ} \; \big| \; N(t) = n\big) \cdot P\big(\mathcal{M}(t+1) = k \; \big| \; Q_1(t) = 0, \mathcal{M}(t) = n, L(t+1) = L_{succ} \big) \nonumber \\ 
&& \; \; \; \; \cdot \; P\big(A_1(t+1) = j \; \big| \; L(t+1) = L_{succ} \big) \nonumber \\ 
&=& p_{succ,n} \cdot s(j) \cdot P\big(\mathcal{M}(t+1) = k \; \big| \; Q_1(t) = 0, \mathcal{M}(t) = n, L(t+1) = L_{succ} \big). 
\end{eqnarray}

Note that, since the tagged queue is empty at the channel slot boundary $t$, the departure in the success channel slot must occur from a non-tagged queue. Thus, we have $Q_1(t+1) = j$, if and only if $A_1(t+1) = j$. Furthermore, given that $L(t+1) = L_{succ}$, the probability that $A_1(t+1) = j$, does not depend on anything else and is equal to $s(j)$. (See Equation~\eqref{eqn:definition-sj} and recall that the arrival rates are equal). 

The non-tagged queue from which the departure occurs is empty at the channel slot boundary $t+1$ if it has exactly 1 packet at the channel slot boundary $t$ and it does not receive any packets in the $(t+1)^{th}$ channel slot. Thus, we must condition on the number of packets at the channel slot boundary $t$ in the non-tagged queue from which the departure occurs at the channel slot boundary $t+1$. To that end, we first expand the conditional probability in the second step in Equation~\eqref{eqn:pi-jk-tplus1-first-part-succ-case} as follows: \begin{eqnarray} 
\label{eqn:pi-jk-tplus1-first-part-succ-case-partial-ratio}
\lefteqn{P\big(\mathcal{M}(t+1) = k \; \big| \; Q_1(t) = 0, \mathcal{M}(t) = n, L(t+1) = L_{succ} \big)} \nonumber \\
&=& \frac{P\big(\mathcal{M}(t) = n, \mathcal{M}(t+1) = k \; \big| \; Q_1(t) = 0, L(t+1) = L_{succ} \big)}{P\big(\mathcal{M}(t) = n \; \big| \; Q_1(t) = 0, L(t+1) = L_{succ} \big)}. 
\end{eqnarray}

As before, let $I_M := \{2, \ldots, M\}$ denote the set of non-tagged queues. Then, the numerator on the right hand side of Equation~\eqref{eqn:pi-jk-tplus1-first-part-succ-case-partial-ratio} can be expanded as follows: \begin{eqnarray} 
\label{eqn:pi-jk-tplus1-first-part-succ-case-partial-ratio-num}
\lefteqn{P\big(\mathcal{M}(t) = n, \mathcal{M}(t+1) = k \; \big| \; Q_1(t) = 0, L(t+1) = L_{succ} \big)} \nonumber \\
&=& \displaystyle \sum_{I \subset I_M \; : \; |I| = n} P\big(Q_m(t) > 0, \forall m \in I, Q_r(t) = 0, \forall r \in I_M \setminus I, \mathcal{M}(t+1) = k \; \big| \; Q_1(t) = 0, L(t+1) = L_{succ} \big) \nonumber \\ 
&=& \displaystyle \sum_{I \subset I_M \; : \; |I| = n} P\big(Q_m(t) > 0, \forall m \in I, Q_r(t) = 0, \forall r \in I_M \setminus I \; \big| \; Q_1(t) = 0, L(t+1) = L_{succ} \big) \nonumber \\ 
&& \; \; \; \; \; \; \; \; \; \; \; \; \; \; \cdot \; P\big(\mathcal{M}(t+1) = k \; \big| \; Q_m(t) > 0, \forall m \in I, Q_r(t) = 0, \forall r \in I_M \setminus I, Q_1(t) = 0, L(t+1) = L_{succ} \big). \nonumber \\ 
\end{eqnarray}

The second conditional probability inside the summation in the second step in Equation~\eqref{eqn:pi-jk-tplus1-first-part-succ-case-partial-ratio-num} can be expanded as follows: (Remember that $I \subset I_M$ and $|I| = n$.) \begin{eqnarray} 
\label{eqn:pi-jk-tplus1-first-part-succ-case-partial-ratio-num-1}
\lefteqn{P\big(\mathcal{M}(t+1) = k \; \big| \; Q_m(t) > 0, \forall m \in I, Q_r(t) = 0, \forall r \in I_M \setminus I, Q_1(t) = 0, L(t+1) = L_{succ} \big)} \nonumber \\
&=& \sum_{l \in I} P\big(\mathcal{M}(t+1) = k, D_{l}(t+1) = 1 \; \big| \; Q_m(t) > 0, \forall m \in I, Q_r(t) = 0, \forall r \in I_M \setminus I, \nonumber \\ 
&& \; \; \; \; \; \; \; \; \; \; \; \; \; \; \; \; \; \; \; \; \; \; \; \; \; \; \; \; \; \; \; \; \; \; \; \; \; \; \; \; \; \; \; \; \; \; \; \; \; \; \; \; \; \; \; \; \; \; \; \; \; \; \; \; \; \; \; \; \; \; \; \; \; \; \; \; \; \; \; \; \; \; \; \; \; \; \; \; \; Q_1(t) = 0, L(t+1) = L_{succ} \big) \nonumber \\ 
&=& \sum_{l \in I} P\big(D_{l}(t+1) = 1 \; \big| \; N(t) = n, Q_{l}(t) > 0, L(t+1) = L_{succ} \big) \nonumber \\ 
&& \; \; \; \; \; \; \; \; \; \; \; \; \; \; \; \; \cdot \; P\big(\mathcal{M}(t+1) = k \; \big| \; Q_m(t) > 0, \forall m \in I, Q_r(t) = 0, \forall r \in I_M \setminus I, \nonumber \\ 
&& \; \; \; \; \; \; \; \; \; \; \; \; \; \; \; \; \; \; \; \; \; \; \; \; \; \; \; \; \; \; \; \; \; \; \; \; \; \; \; \; \; \; \; \; \; \; \; \; \; \; \; \; \; \; \; \; \; \; \; \; \; \; \; \; \; \; \; \; \; \; \; \; \; \; \; \; \; \; \; Q_1(t) = 0, L(t+1) = L_{succ}, D_{l}(t+1) = 1 \big) \nonumber \\ 
&=& \sum_{l \in I} \left(\frac{1}{n}\right) \; \cdot P\big(\mathcal{M}(t+1) = k \; \big| \; Q_m(t) > 0, \forall m \in I, Q_r(t) = 0, \forall r \in I_M \setminus I, \nonumber \\ 
&& \; \; \; \; \; \; \; \; \; \; \; \; \; \; \; \; \; \; \; \; \; \; \; \; \; \; \; \; \; \; \; \; \; \; \; \; \; \; \; \; \; \; \; \; \; \; \; \; \; \; \; \; \; \; \; \; \; \; \; \; \; \; \; \; \; \; \; \; \; \; \; \; \; \; \; \; \; \; \; Q_1(t) = 0, L(t+1) = L_{succ}, D_{l}(t+1) = 1 \big) \nonumber \\ 
&=& \sum_{l \in I} \left(\frac{1}{n}\right) \; \cdot P\big(\mathcal{M}(t+1) = k, Q_l(t) = 1 \; \big| \; Q_m(t) > 0, \forall m \in I, Q_r(t) = 0, \forall r \in I_M \setminus I, \nonumber \\ 
&& \; \; \; \; \; \; \; \; \; \; \; \; \; \; \; \; \; \; \; \; \; \; \; \; \; \; \; \; \; \; \; \; \; \; \; \; \; \; \; \; \; \; \; \; \; \; \; \; \; \; \; \; \; \; \; \; \; \; \; \; \; \; \; \; \; \; \; \; \; \; \; \; \; \; \; \; \; \; \; Q_1(t) = 0, L(t+1) = L_{succ}, D_{l}(t+1) = 1 \big) \nonumber \\ 
&& + \sum_{l \in I} \left(\frac{1}{n}\right) \; \cdot P\big(\mathcal{M}(t+1) = k, Q_l(t) > 1 \; \big| \; Q_m(t) > 0, \forall m \in I, Q_r(t) = 0, \forall r \in I_M \setminus I, \nonumber \\ 
&& \; \; \; \; \; \; \; \; \; \; \; \; \; \; \; \; \; \; \; \; \; \; \; \; \; \; \; \; \; \; \; \; \; \; \; \; \; \; \; \; \; \; \; \; \; \; \; \; \; \; \; \; \; \; \; \; \; \; \; \; \; \; \; \; \; \; \; \; \; \; \; \; \; \; \; \; \; \; \; Q_1(t) = 0, L(t+1) = L_{succ}, D_{l}(t+1) = 1 \big). \; \; \; \; \; \; \;
\end{eqnarray}

In the first step we condition on the queue from which the departure occurs at $t+1$. Note that a departure can occur from a queue at $t+1$, only if the queue is non-empty at $t$. Thus, the departure in the success channel slot can occur from one of the queues in $I$. In the second step we factor out the probability that the departure occurs from the (specific) non-empty non-tagged queue at $t+1$ given that there are $N(t) = n$ non-empty nodes in the system at $t$. In the third step we apply Equation~\eqref{eqn:departure-equallylikely}. In the fourth step we condition on whether the non-empty non-tagged queue from which the departure occurs at $t+1$ contains exactly 1 packet at $t$. 

The conditional probability inside the first summation in the last step in Equation~\eqref{eqn:pi-jk-tplus1-first-part-succ-case-partial-ratio-num-1} can be expanded as follows: \begin{eqnarray} 
\label{eqn:pi-jk-tplus1-first-part-succ-case-partial-ratio-num-1-1}
\lefteqn{P\big(\mathcal{M}(t+1) = k, Q_l(t) = 1 \; \big| \; Q_m(t) > 0, \forall m \in I, Q_r(t) = 0, \forall r \in I_M \setminus I,} \nonumber \\ 
&& \; \; \; \; \; \; \; \; \; \; \; \; \; \; \; \; \; \; \; \; \; \; \; \; \; \; \; \; \; \; \; \; \; \; \; \; \; \; \; \; \; \; \; \; \; \; \; \; \; \; \; \; \; \; \; \; \; \; \; \; \; \; \; \; \; \; \; \; \; \; \; \; \; \; \; \; \; \; \; Q_1(t) = 0, L(t+1) = L_{succ}, D_{l}(t+1) = 1 \big) \nonumber \\ 
&=& P\big(Q_l(t) = 1 \; \big| \; Q_m(t) > 0, \forall m \in I, Q_r(t) = 0, \forall r \in I_M \setminus I, \nonumber \\ 
&& \; \; \; \; \; \; \; \; \; \; \; \; \; \; \; \; \; \; \; \; \; \; \; \; \; \; \; \; \; \; \; \; \; \; \; \; \; \; \; \; \; \; \; \; \; \; \; \; \; \; \; \; \; \; \; \; \; \; \; \; \; \; \; \; \; \; \; \; \; \; \; \; \; \; \; \; \; \; \; Q_1(t) = 0, L(t+1) = L_{succ}, D_{l}(t+1) = 1 \big) \nonumber \\ 
&& \; \cdot \; P\big(\mathcal{M}(t+1) = k \; \big| \; Q_m(t) > 0, \forall m \in I \setminus \{l\}, Q_l(t) = 1, Q_r(t) = 0, \forall r \in I_M \setminus I, \nonumber \\ 
&& \; \; \; \; \; \; \; \; \; \; \; \; \; \; \; \; \; \; \; \; \; \; \; \; \; \; \; \; \; \; \; \; \; \; \; \; \; \; \; \; \; \; \; \; \; \; \; \; \; \; \; \; \; \; \; \; \; \; \; \; \; \; \; \; \; \; \; \; \; \; \; \; \; \; \; \; \; \; \; Q_1(t) = 0, L(t+1) = L_{succ}, D_{l}(t+1) = 1 \big). \; \; \; \; \; \; \; \; \;
\end{eqnarray}

The second conditional probability on the right hand side in Equation~\eqref{eqn:pi-jk-tplus1-first-part-succ-case-partial-ratio-num-1-1} can be simplified as follows: (Remember from Equation~\eqref{eqn:pi-jk-tplus1-first-part-succ-case-partial-ratio-num-1} that $l$, $l \in I$, denotes the index of the non-tagged queue from which the departure occurs at $t+1$.) \begin{eqnarray} 
\label{eqn:pi-jk-tplus1-first-part-succ-case-partial-ratio-num-1-1-simple}
\lefteqn{P\big(\mathcal{M}(t+1) = k \; \big| \; Q_m(t) > 0, \forall m \in I \setminus \{l\}, Q_l(t) = 1, Q_r(t) = 0, \forall r \in I_M \setminus I,} \nonumber \\ 
&& \; \; \; \; \; \; \; \; \; \; \; \; \; \; \; \; \; \; \; \; \; \; \; \; \; \; \; \; \; \; \; \; \; \; \; \; \; \; \; \; \; \; \; \; \; \; \; \; \; \; \; \; \; \; \; \; \; \; \; \; \; \; \; \; \; \; \; \; \; \; \; \; \; \; \; \; \; \; \; Q_1(t) = 0, L(t+1) = L_{succ}, D_{l}(t+1) = 1 \big) \nonumber \\ 
&=& P\big(\mathcal{M}(t+1) = k \; \big| \; Q_m(t) > 0, \forall m \in I \setminus \{l\}, Q_l(t) = 1, Q_r(t) = 0, \forall r \in I_M \setminus I, \nonumber \\ 
&& \; \; \; \; \; \; \; \; \; \; \; \; \; \; \; \; \; \; \; \; \; \; \; \; \; \; \; \; \; \; \; \; \; Q_1(t) = 0, L(t+1) = L_{succ}, D_{l}(t+1) = 1, D_u(t+1) = 0, \forall u \in I_M \setminus \{l\} \big) \nonumber \\ 
&=& P\big(\left(\sum_{m \in I \setminus \{l\}} \ind{\left(Q_{m}(t) + A_{m}(t+1)\right) > 0} + \ind{A_{l}(t+1) > 0} + \sum_{r \in I_M \setminus I} \ind{A_{r}(t+1) > 0}\right) = k\; \big| \; Q_{m}(t) > 0, \nonumber \\ 
&& \; \; \; \; \; \; \; \; \; \; \; \; \; \; \; \; \; \; \; \; \; \; \; \; \; \; \; \; \; \; \; \; \; \; \; \; \; \; \; \; \; \; \; \; \; \; \; \; \; \; \; \; \; \; \; \; \; \; \; \; \; \; \; \; \; \; \; \; \; \; \; \; \; \; \; \; \; \; \; \; \; \; \; \; \; \; \; \; \; \; \; \; \; \; \; \; \; \; \; \forall m \in I \setminus \{l\}, L(t+1) = L_{succ}\big) \nonumber\\ 
&=& P\big(\left(\ind{A_{l}(t+1) > 0} + \sum_{r \in I_M \setminus I} \ind{A_{r}(t+1) > 0}\right) = k-n+1 \; \big| \; L(t+1) = L_{succ}\big) \nonumber \\ 
&=& {{M-n \choose k-n+1}} (1-s(0))^{k-n+1} s(0)^{M-k-1}. 
\end{eqnarray}

In the first step we have used the fact that, given that $D_{l}(t+1) = 1$, we have $D_{u}(t+1) = 0$, $\forall u \in I_M \setminus \{l\}$, since at most one departure can occur in a channel slot. In the second step we applied Equations~\eqref{eqn:definition-Mt} and~\eqref{eqn:queue-evolution}, and the conditioning queue length information of the first step. Note that, $D_l(t+1) = 1$ and $Q_l(t) = 1$ cancel each other in the indicator variable corresponding to $l$. Thus, the queue from which the departure occurs at $t+1$ can be empty at $t+1$ if it does not receive any packets in the $(t+1)^{th}$ channel slot. In the third step we have used the fact that $Q_m(t) > 0$, $\forall m \in I \setminus \{l\}$, implies that $\sum_{m \in I \setminus \{l\}} \ind{\left(Q_{m}(t) + A_{m}(t+1)\right) > 0} = |I \setminus \{l\}| = n-1$. The fourth step follows by observing that the sum of the $M-n$ indicator variables in the third step is equal to $k-n+1$ if and only if arrivals occur to exactly $k-n+1$ of the corresponding queues, and, given that $L(t+1) = L_{succ}$, the probability that arrivals occur (resp.~do not occur) to a queue is given by $1-s(0)$ (resp.~$s(0)$).

The conditional probability inside the second summation in the last step in Equation~\eqref{eqn:pi-jk-tplus1-first-part-succ-case-partial-ratio-num-1} can be expanded as follows: 

\begin{eqnarray} 
\label{eqn:pi-jk-tplus1-first-part-succ-case-partial-ratio-num-1-2}
\lefteqn{P\big(\mathcal{M}(t+1) = k, Q_l(t) > 1 \; \big| \; Q_m(t) > 0, \forall m \in I, Q_r(t) = 0, \forall r \in I_M \setminus I,} \nonumber \\ 
&& \; \; \; \; \; \; \; \; \; \; \; \; \; \; \; \; \; \; \; \; \; \; \; \; \; \; \; \; \; \; \; \; \; \; \; \; \; \; \; \; \; \; \; \; \; \; \; \; \; \; \; \; \; \; \; \; \; \; \; \; \; \; \; \; \; \; \; \; \; \; \; \; \; \; \; \; \; \; \; Q_1(t) = 0, L(t+1) = L_{succ}, D_{l}(t+1) = 1 \big) \nonumber \\ 
&=& P\big(Q_l(t) > 1 \; \big| \; Q_m(t) > 0, \forall m \in I, Q_r(t) = 0, \forall r \in I_M \setminus I, \nonumber \\ 
&& \; \; \; \; \; \; \; \; \; \; \; \; \; \; \; \; \; \; \; \; \; \; \; \; \; \; \; \; \; \; \; \; \; \; \; \; \; \; \; \; \; \; \; \; \; \; \; \; \; \; \; \; \; \; \; \; \; \; \; \; \; \; \; \; \; \; \; \; \; \; \; \; \; \; \; \; \; \; \; Q_1(t) = 0, L(t+1) = L_{succ}, D_{l}(t+1) = 1 \big) \nonumber \\ 
&& \; \cdot \; P\big(\mathcal{M}(t+1) = k \; \big| \; Q_m(t) > 0, \forall m \in I \setminus \{l\}, Q_l(t) > 1, Q_r(t) = 0, \forall r \in I_M \setminus I, \nonumber \\ 
&& \; \; \; \; \; \; \; \; \; \; \; \; \; \; \; \; \; \; \; \; \; \; \; \; \; \; \; \; \; \; \; \; \; \; \; \; \; \; \; \; \; \; \; \; \; \; \; \; \; \; \; \; \; \; \; \; \; \; \; \; \; \; \; \; \; \; \; \; \; \; \; \; \; \; \; \; \; \; \; Q_1(t) = 0, L(t+1) = L_{succ}, D_{l}(t+1) = 1 \big). \; \; \; \; \; \; \; \;
\end{eqnarray}

The second conditional probability on the right hand side in Equation~\eqref{eqn:pi-jk-tplus1-first-part-succ-case-partial-ratio-num-1-2} can be simplified by similar arguments as follows: \begin{eqnarray} 
\label{eqn:pi-jk-tplus1-first-part-succ-case-partial-ratio-num-1-2-simple}
\lefteqn{P\big(\mathcal{M}(t+1) = k \; \big| \; Q_m(t) > 0, \forall m \in I \setminus \{l\}, Q_l(t) > 1, Q_r(t) = 0, \forall r \in I_M \setminus I,} \nonumber \\ 
&& \; \; \; \; \; \; \; \; \; \; \; \; \; \; \; \; \; \; \; \; \; \; \; \; \; \; \; \; \; \; \; \; \; \; \; \; \; \; \; \; \; \; \; \; \; \; \; \; \; \; \; \; \; \; \; \; \; \; \; \; \; \; \; \; \; \; \; \; \; \; \; \; \; \; \; \; \; \; \; Q_1(t) = 0, L(t+1) = L_{succ}, D_{l}(t+1) = 1 \big) \nonumber \\ 
&=& P\big(\mathcal{M}(t+1) = k \; \big| \; Q_m(t) > 0, \forall m \in I \setminus \{l\}, Q_l(t) > 1, Q_r(t) = 0, \forall r \in I_M \setminus I, \nonumber \\ 
&& \; \; \; \; \; \; \; \; \; \; \; \; \; \; \; \; \; \; \; \; \; \; \; \; \; \; \; \; \; \; \; \; \; Q_1(t) = 0, L(t+1) = L_{succ}, D_{l}(t+1) = 1, D_u(t+1) = 0, \forall u \in I_M \setminus \{l\} \big) \nonumber \\ 
&=& P\big(\left(\sum_{m \in I \setminus \{l\}} \ind{\left(Q_{m}(t) + A_{m}(t+1)\right) > 0} + \ind{Q_l(t) - 1 + A_{l}(t+1) > 0} + \sum_{r \in I_M \setminus I} \ind{A_{r}(t+1) > 0}\right) = k\; \big| \; Q_{m}(t) > 0, \nonumber \\ 
&& \; \; \; \; \; \; \; \; \; \; \; \; \; \; \; \; \; \; \; \; \; \; \; \; \; \; \; \; \; \; \; \; \; \; \; \; \; \; \; \; \; \; \; \; \; \; \; \; \; \; \; \; \; \; \; \; \; \; \; \; \; \; \; \; \; \; \; \; \; \; \; \; \; \; \; \; \forall m \in I \setminus \{l\}, Q_l(t) > 1, L(t+1) = L_{succ}\big) \nonumber\\ 
&=& P\big(\sum_{r \in I_M \setminus I} \ind{A_{r}(t+1) > 0} = k-n \; \big| \; L(t+1) = L_{succ}\big) \nonumber \\ 
&=& {{M-n-1 \choose k-n}} (1-s(0))^{k-n} s(0)^{M-k-1}, 
\end{eqnarray} where, in the second step, we observe that $(Q_l(t) - 1) > 0$, and hence, the indicator variable corresponding to $l$ is equal to 1. 

We emphasize that, $\forall I \subset I_M$, $\forall l \in I$, the conditional probability \[P\big(Q_l(t) = 1 \; \big| \; Q_m(t) > 0, \forall m \in I, Q_r(t) = 0, \forall r \in I_M \setminus I, Q_1(t) = 0, L(t+1) = L_{succ}, D_{l}(t+1) = 1 \big),\] which appears in Equation~\eqref{eqn:pi-jk-tplus1-first-part-succ-case-partial-ratio-num-1-1}, and the conditional probability \[P\big(Q_l(t) > 1 \; \big| \; Q_m(t) > 0, \forall m \in I, Q_r(t) = 0, \forall r \in I_M \setminus I, Q_1(t) = 0, L(t+1) = L_{succ}, D_{l}(t+1) = 1 \big),\] which appears in Equation~\eqref{eqn:pi-jk-tplus1-first-part-succ-case-partial-ratio-num-1-2} are not known, precisely because the queue lengths of the non-tagged nodes are not tracked by the process $\{\mathcal{X}(t)\}$. Moreover, the above unknown probabilities cannot be obtained from the known quantities, namely, the arrival rate $\lambda$ and the state-dependent attempt probabilities $\beta_n$'s. All the other probabilities, namely, $p_{idle,n}$, $p_{coll,n}$, $p_{succ,n}$, $0 \leq n \leq M$, $d(j)$, $c(j)$, $s(j)$, $j = 0, 1, 2, \ldots$, can be obtained from $\lambda$ and the $\beta_n$'s. Later in this appendix we shall derive the transition probabilities out of a state in which the tagged node is non-empty, and it will turn out that the transition probabilities then involve unknown probabilities of the form \[P\big(Q_l(t) = 1 \; \big| \; Q_m(t) > 0, \forall m \in I, Q_r(t) = 0, \forall r \in I_M \setminus I, Q_1(t) = i > 0, L(t+1) = L_{succ}, D_{l}(t+1) = 1 \big),\] and \[P\big(Q_l(t) > 1 \; \big| \; Q_m(t) > 0, \forall m \in I, Q_r(t) = 0, \forall r \in I_M \setminus I, Q_1(t) = i > 0, L(t+1) = L_{succ}, D_{l}(t+1) = 1 \big).\]

Recall that $l$, $l \in I$, denotes the index of the non-tagged queue from which a departure occurs at $t+1$, and $I$ denotes the set of non-tagged queues that are non-empty at $t$. At any $t$, $I$ is a specific subset of $I_M$, and $l$ takes some specific value between $2$ and $M$. However, due to exchangeability, once we fix the \textit{number} of non-tagged queues that are non-empty at $t$, i.e., once we fix $|I|$, the values of the above unknown conditional probabilities do not depend on the specific set $I$ and the specific queue $l$, $l \in I$, from which the departure occurs at $t+1$. Thus, given that $\mathcal{M}(t) = |I|$, we do not have to retain the empty/non-empty status of the non-tagged queues, except that of the queue from which the departure occurs at $t+1$. We must, retain the fact that $Q_l(t) > 0$; otherwise, $Q_l(t) = 0$ would imply $D_l(t+1) = 0$, which is a contradiction. 

Applying the above exchangeability argument, we can write, $\forall t \geq 0$, $\forall i \geq 0$, $\forall I \subset I_M$, $\forall l \in I$, \begin{eqnarray} 
\label{eqn:simplification-by-exchangeability}
\lefteqn{P\big(Q_l(t) = 1 \; \big| \; Q_m(t) > 0, \forall m \in I, Q_r(t) = 0, \forall r \in I_M \setminus I, Q_1(t) = i, L(t+1) = L_{succ}, D_{l}(t+1) = 1 \big)} \nonumber \\ 
&=& P\big(Q_{l}(t) = 1 \; \big| \; Q_1(t) = i, \mathcal{M}(t) = |I|, Q_{l}(t) > 0, D_{l}(t+1) = 1, L(t+1) = L_{succ}\big). \; \; \; \; \; \; \; \; \; \; \; \; \; \; \; \; \; \; \; \; \; \; \;  
\end{eqnarray}

To simplify notations, we define, $\forall t \geq 0$, $\forall i \geq 0$, $\forall n$, $0 \leq n \leq M-1$, $\forall l$, $2 \leq l \leq M$, \begin{equation} 
\label{eqn:qint-definition-one}
q(i,n,t) := P\big(Q_{l}(t) = 1 \; \big| \; Q_1(t) = i, \mathcal{M}(t) = n, Q_{l}(t) > 0, D_{l}(t+1) = 1, L(t+1) = L_{succ}\big).
\end{equation} Note that, $q(i,n,t)$ on the left hand side of Equation~\eqref{eqn:qint-definition-one} does not involve $l$ because, due to exchangeability, the value of the conditional probability on the right hand side of Equation~\eqref{eqn:qint-definition-one} does not depend on the specific value of $l$. Notice that, $\forall t \geq 0$, $\forall i \geq 0$, $\forall n$, $0 \leq n \leq M-1$, $\forall l$, $2 \leq l \leq M$, we have 
\begin{eqnarray} 
\label{eqn:complementary-eqn1}
\lefteqn{P\big(Q_{l}(t) > 1 \; \big| \; Q_1(t) = i, \mathcal{M}(t) = n, Q_{l}(t) > 0, D_{l}(t+1) = 1, L(t+1) = L_{succ}\big)} \nonumber \\ 
&=& 1 - P\big(Q_{l}(t) = 1 \; \big| \; Q_1(t) = 0, \mathcal{M}(t) = n, Q_{l}(t) > 0, D_{l}(t+1) = 1, L(t+1) = L_{succ}\big) \nonumber \\ 
&=& 1-q(i,n,t). 
\end{eqnarray}

Substituting Equation~\eqref{eqn:pi-jk-tplus1-first-part-succ-case-partial-ratio-num-1-1-simple} in Equation~\eqref{eqn:pi-jk-tplus1-first-part-succ-case-partial-ratio-num-1-1}, and applying the notation given by Equation~\eqref{eqn:qint-definition-one}, we obtain \begin{eqnarray} 
\label{eqn:pi-jk-tplus1-first-part-succ-case-partial-ratio-num-1-1-final}
\lefteqn{P\big(\mathcal{M}(t+1) = k, Q_l(t) = 1 \; \big| \; Q_m(t) > 0, \forall m \in I, Q_r(t) = 0, \forall r \in I_M \setminus I,} \nonumber \\ 
&& \; \; \; \; \; \; \; \; \; \; \; \; \; \; \; \; \; \; \; \; \; \; \; \; \; \; \; \; \; \; \; \; \; \; \; \; \; \; \; \; \; \; \; \; \; \; \; \; \; \; \; \; \; \; \; \; \; \; \; \; \; \; \; \; \; \; \; \; \; \; \; \; \; \; \; \; \; \; \; Q_1(t) = 0, L(t+1) = L_{succ}, D_{l}(t+1) = 1 \big) \nonumber \\ 
&=& q(0,n,t) \; \cdot \; {{M-n \choose k-n+1}} (1-s(0))^{k-n+1} s(0)^{M-k-1}. 
\end{eqnarray}

Substituting Equation~\eqref{eqn:pi-jk-tplus1-first-part-succ-case-partial-ratio-num-1-2-simple} in Equation~\eqref{eqn:pi-jk-tplus1-first-part-succ-case-partial-ratio-num-1-2}, and applying Equation~\eqref{eqn:complementary-eqn1}, we obtain \begin{eqnarray} 
\label{eqn:pi-jk-tplus1-first-part-succ-case-partial-ratio-num-1-2-final}
\lefteqn{P\big(\mathcal{M}(t+1) = k, Q_l(t) > 1 \; \big| \; Q_m(t) > 0, \forall m \in I, Q_r(t) = 0, \forall r \in I_M \setminus I,} \nonumber \\ 
&& \; \; \; \; \; \; \; \; \; \; \; \; \; \; \; \; \; \; \; \; \; \; \; \; \; \; \; \; \; \; \; \; \; \; \; \; \; \; \; \; \; \; \; \; \; \; \; \; \; \; \; \; \; \; \; \; \; \; \; \; \; \; \; \; \; \; \; \; \; \; \; \; \; \; \; \; \; \; \; Q_1(t) = 0, L(t+1) = L_{succ}, D_{l}(t+1) = 1 \big) \nonumber \\ 
&=& \left(1- q(0,n,t)\right) \; \cdot \; {{M-n-1 \choose k-n}} (1-s(0))^{k-n} s(0)^{M-k-1}. 
\end{eqnarray}

Substituting Equations~\eqref{eqn:pi-jk-tplus1-first-part-succ-case-partial-ratio-num-1-1-final} and~\eqref{eqn:pi-jk-tplus1-first-part-succ-case-partial-ratio-num-1-2-final} in Equation~\eqref{eqn:pi-jk-tplus1-first-part-succ-case-partial-ratio-num-1}, we obtain \begin{eqnarray} 
\label{eqn:pi-jk-tplus1-first-part-succ-case-partial-ratio-num-1-final}
\lefteqn{P\big(\mathcal{M}(t+1) = k \; \big| \; Q_m(t) > 0, \forall m \in I, Q_r(t) = 0, \forall r \in I_M \setminus I, Q_1(t) = 0, L(t+1) = L_{succ} \big)} \nonumber \\
&=& \sum_{l \in I} \left(\frac{1}{n}\right) \; \cdot q(0,n,t) \; \cdot \; {{M-n \choose k-n+1}} (1-s(0))^{k-n+1} s(0)^{M-k-1} \nonumber \\
&& + \sum_{l \in I} \left(\frac{1}{n}\right) \; \cdot \left(1- q(0,n,t)\right) \; \cdot \; {{M-n-1 \choose k-n}} (1-s(0))^{k-n} s(0)^{M-k-1} \nonumber \\
&=& q(0,n,t) \; \cdot \; {{M-n \choose k-n+1}} (1-s(0))^{k-n+1} s(0)^{M-k-1} \nonumber \\ 
&& + \left(1- q(0,n,t)\right) \; \cdot \; {{M-n-1 \choose k-n}} (1-s(0))^{k-n} s(0)^{M-k-1}. \; \; \; \; \; \; \; \; \; \; \; \; \; \; \; \; \; \; \; \; \; \; \; \; \; \; \; \; \; \; \; \; \; \;  
\end{eqnarray}

Substituting Equation~\eqref{eqn:pi-jk-tplus1-first-part-succ-case-partial-ratio-num-1-final} in Equation~\eqref{eqn:pi-jk-tplus1-first-part-succ-case-partial-ratio-num}, we obtain \begin{eqnarray} 
\label{eqn:pi-jk-tplus1-first-part-succ-case-partial-ratio-num-final}
\lefteqn{P\big(\mathcal{M}(t) = n, \mathcal{M}(t+1) = k \; \big| \; Q_1(t) = 0, L(t+1) = L_{succ} \big)} \nonumber \\
&=& \displaystyle \sum_{I \subset I_M \; : \; |I| = n} P\big(Q_m(t) > 0, \forall m \in I, Q_r(t) = 0, \forall r \in I_M \setminus I \; \big| \; Q_1(t) = 0, L(t+1) = L_{succ} \big) \nonumber \\ 
&& \; \; \; \; \; \; \; \; \; \; \; \; \; \; \; \; \; \; \; \; \; \; \cdot \; \left[q(0,n,t) \; \cdot \; {{M-n \choose k-n+1}} (1-s(0))^{k-n+1} s(0)^{M-k-1} \right. \nonumber \\ 
&& \; \; \; \; \; \; \; \; \; \; \; \; \; \; \; \; \; \; \; \; \; \; \; \; \; \; \; \; \left. + \left(1- q(0,n,t)\right) \; \cdot \; {{M-n-1 \choose k-n}} (1-s(0))^{k-n} s(0)^{M-k-1} \right] \nonumber \\
&=& P\big(\mathcal{M}(t) = n, \; \big| \; Q_1(t) = 0, L(t+1) = L_{succ} \big) \nonumber \\
&& \; \; \; \; \; \; \; \; \; \; \; \; \; \; \; \; \; \; \; \; \; \; \cdot \; \left[q(0,n,t) \; \cdot \; {{M-n \choose k-n+1}} (1-s(0))^{k-n+1} s(0)^{M-k-1} \right. \nonumber \\ 
&& \; \; \; \; \; \; \; \; \; \; \; \; \; \; \; \; \; \; \; \; \; \; \; \; \; \; \; \; \left. + \left(1- q(0,n,t)\right) \; \cdot \; {{M-n-1 \choose k-n}} (1-s(0))^{k-n} s(0)^{M-k-1} \right], 
\end{eqnarray} where in the second step we take the common factor within square brackets outside the summation, and then observe that the summation is equal to the conditional probability $P\big(\mathcal{M}(t) = n \; \big| \; Q_1(t) = 0, L(t+1) = L_{succ} \big)$. 

Substituting Equation~\eqref{eqn:pi-jk-tplus1-first-part-succ-case-partial-ratio-num-final} in Equation~\eqref{eqn:pi-jk-tplus1-first-part-succ-case-partial-ratio}, we obtain \begin{eqnarray} 
\label{eqn:pi-jk-tplus1-first-part-succ-case-partial-ratio-final}
\lefteqn{P\big(\mathcal{M}(t+1) = k \; \big| \; Q_1(t) = 0, \mathcal{M}(t) = n, L(t+1) = L_{succ} \big)} \nonumber \\ 
&=& q(0,n,t) \; \cdot \; {{M-n \choose k-n+1}} (1-s(0))^{k-n+1} s(0)^{M-k-1} \nonumber \\ 
&& + \left(1- q(0,n,t)\right) \; \cdot \; {{M-n-1 \choose k-n}} (1-s(0))^{k-n} s(0)^{M-k-1} \nonumber \\
&=& {{M-n-1 \choose k-n}} (1-s(0))^{k-n} s(0)^{M-k-1} \nonumber \\ 
&& + \; q(0,n,t) \cdot (1-s(0))^{k-n} s(0)^{M-k-1} \cdot \left({{M-n \choose k-n+1}} (1-s(0)) - {{M-n-1 \choose k-n}} \right), \; \; \; \; 
\end{eqnarray} where in the first step we observe that $P\big(\mathcal{M}(t) = n \; \big| \; Q_1(t) = 0, L(t+1) = L_{succ} \big)$ gets canceled from both the numerator and the denominator on the right hand side in Equation~\eqref{eqn:pi-jk-tplus1-first-part-succ-case-partial-ratio}. 

Substituting Equation~\eqref{eqn:pi-jk-tplus1-first-part-succ-case-partial-ratio-final} in Equation~\eqref{eqn:pi-jk-tplus1-first-part-succ-case} we obtain \begin{eqnarray} 
\label{eqn:pi-jk-tplus1-first-part-succ-case-final}
\lefteqn{P\big(Q_1(t+1) = j, \mathcal{M}(t+1) = k, L(t+1) = L_{succ} \; \big| \; Q_1(t) = 0, \mathcal{M}(t) = n\big)} \nonumber \\ 
&=& p_{succ,n} \cdot s(j) \cdot {{M-n-1 \choose k-n}} (1-s(0))^{k-n} s(0)^{M-k-1} \nonumber \\ 
&& + \; p_{succ,n} \cdot s(j) \cdot q(0,n,t) \cdot (1-s(0))^{k-n} s(0)^{M-k-1} \cdot \left( {{M-n \choose k-n+1}} (1-s(0)) - {{M-n-1 \choose k-n}} \right). \; \; \; \; \; \; \; \; \; 
\end{eqnarray}

Substituting Equations~\eqref{eqn:pi-jk-tplus1-first-part-idle-case},~\eqref{eqn:pi-jk-tplus1-first-part-coll-case} and~\eqref{eqn:pi-jk-tplus1-first-part-succ-case-final} in Equation~\eqref{eqn:pi-jk-tplus1-first-part}, we obtain \begin{eqnarray} 
\label{eqn:pi-jk-tplus1-first-part-final}
\lefteqn{P\big(Q_1(t+1) = j, \mathcal{M}(t+1) = k \; \big| \; Q_1(t) = 0, \mathcal{M}(t) = n\big)} \nonumber \\ 
&=& {{M-n-1 \choose k-n}} \left( p_{idle,n} \cdot d(j) (1-d(0))^{k-n} d(0)^{M-k-1} + p_{coll,n} \cdot c(j) (1-c(0))^{k-n} c(0)^{M-k-1} \right. \nonumber \\
&& \; \; \; \; \; \; \; \; \; \; \; \; \; \; \; \; \; \; \; \; \; \; \; \; \; \; \; \; \; \; \; \; \; \; \; \; \; \; \; \; \; \; \; \; \; \; \; \; \; \; \; \; \; \; \; \; \; \; \; \; \; \; \; \; \; \; \; \; \; \; \; \; \; \; \; \; \; \; \; \; \; \; \; \; \; \; \; \; + \left. p_{succ,n} \cdot s(j) (1-s(0))^{k-n} s(0)^{M-k-1} \right) \nonumber \\ 
&& + \; p_{succ,n} \cdot s(j) \cdot q(0,n,t) \cdot (1-s(0))^{k-n} s(0)^{M-k-1} \cdot \; \left( {{M-n \choose k- +1}} (1-s(0)) - {{M-n-1 \choose k-n}} \right), \; \; \; \; \; \; \; \; \; \; 
\end{eqnarray} which is the transition probability from a state $(0,n)$, $0 \leq n \leq M-1$, to a state $(j,k)$, $j = 0, 1, 2, \ldots$, $0 \leq k \leq M-1$. Next, we derive the transition probabilities out of a state in which the tagged node is non-empty.

\subsection{When the Tagged Queue is Non-Empty in the Initial State}
\label{subsubsec:tagged-queue-non-empty}

The transition probability $P\big(Q_1(t+1) = j, \mathcal{M}(t+1) = k \; \big| \; Q_1(t) = i > 0, \mathcal{M}(t) = n\big)$ corresponds to the case when the tagged node is non-empty at the channel slot boundary $t$ and can be expanded according to whether the $(t+1)^{th}$ channel slot is an idle, a collision and a success channel slot as follows: \begin{eqnarray} 
\label{eqn:pi-jk-tplus1-second-part}
\lefteqn{P\big(Q_1(t+1) = j, \mathcal{M}(t+1) = k \; \big| \; Q_1(t) = i > 0, \mathcal{M}(t) = n\big)} \nonumber \\ 
&=& P\big(Q_1(t+1) = j, \mathcal{M}(t+1) = k, L(t+1) = L_{idle} \; \big| \; Q_1(t) = i > 0, \mathcal{M}(t) = n\big) \nonumber \\ 
&& + \; P\big(Q_1(t+1) = j, \mathcal{M}(t+1) = k, L(t+1) = L_{coll} \; \big| \; Q_1(t) = i > 0, \mathcal{M}(t) = n\big) \nonumber \\  
&& + \; P\big(Q_1(t+1) = j, \mathcal{M}(t+1) = k, L(t+1) = L_{succ} \; \big| \; Q_1(t) = i > 0, \mathcal{M}(t) = n\big). 
\end{eqnarray}

\subsubsection{The Case of Idle Channel Slot} 

The first term on the right hand side of Equation~\eqref{eqn:pi-jk-tplus1-second-part} can be simplified as follows: \begin{eqnarray} 
\label{eqn:pi-jk-tplus1-second-part-idle-case}
\lefteqn{P\big(Q_1(t+1) = j, \mathcal{M}(t+1) = k, L(t+1) = L_{idle} \; \big| \; Q_1(t) = i > 0, \mathcal{M}(t) = n\big)} \nonumber \\ 
&=& P\big(L(t+1) = L_{idle} \; \big| \; Q_1(t) = i > 0, \mathcal{M}(t) = n\big) \nonumber \\
&& \; \; \; \; \; \; \; \; \; \; \; \; \; \; \cdot \; P\big(A_1(t+1) = j-i, \mathcal{M}(t+1) = k \; \big| \; Q_1(t) = i > 0, \mathcal{M}(t) = n, L(t+1) = L_{idle} \big) \nonumber \\ 
&=& P\big(L(t+1) = L_{idle} \; \big| \; N(t) = n+1 \big) \cdot P\big(\mathcal{M}(t+1) = k \; \big| \; Q_1(t) = i > 0, \mathcal{M}(t) = n, L(t+1) = L_{idle} \big) \nonumber \\ 
&& \; \; \; \; \; \; \; \; \; \; \; \; \; \; \; \; \; \; \; \; \; \; \; \cdot P\big(A_1(t+1) = j-i \; \big| L(t+1) = L_{idle} \big) \nonumber \\ 
&=& p_{idle,n+1} \cdot d(j-i) \cdot {{M-n-1 \choose k-n}} (1-d(0))^{k-n} d(0)^{M-k-1}, 
\end{eqnarray} where we have used the following facts: (a) $\left(L(t+1) = L_{idle}\right) \Rightarrow \left(D_m(t+1) = 0, 1 \leq m \leq M\right)$, (b) $Q_1(t) = i$, $i > 0$, and $D_1(t+1) = 0$ imply that $Q_1(t+1) = i + A_1(t+1)$, and hence, $Q_1(t+1) = j$, if and only if $A_1(t+1) = j-i$, (c) given that $L(t+1) = L_{idle}$, the probability that $A_1(t+1) = j-i$, does not depend on anything else and is equal to $d(j-i)$, and (d) $D_m(t+1) = 0$, $2 \leq m \leq M$, implies that the set of non-empty non-tagged queues can only increase, and, given that $\mathcal{M}(t) = n$, we have $\mathcal{M}(t+1) = k$, if and only if arrivals occur to exactly $k-n$ of the $M-n-1$ empty non-tagged queues in the idle channel slot, which happens with probability ${{M-n-1 \choose k-n}} (1-d(0))^{k-n} d(0)^{M-k-1}$.

\vspace{1mm}

\subsubsection{The Case of Collision Channel Slot} 

The second term on the right hand side of Equation~\eqref{eqn:pi-jk-tplus1-second-part} can be simplified by similar arguments, and we obtain \begin{eqnarray} 
\label{eqn:pi-jk-tplus1-second-part-coll-case}
\lefteqn{P\big(Q_1(t+1) = j, \mathcal{M}(t+1) = k, L(t+1) = L_{coll} \; \big| \; Q_1(t) = i > 0, \mathcal{M}(t) = n\big)} \nonumber \\ 
&=& p_{coll,n+1} \cdot c(j-i) \cdot {{M-n-1 \choose k-n}} (1-c(0))^{k-n} c(0)^{M-k-1}. \; \; \; \; \; \; \; \; \; \; \; \; \; \; \; \; \; \; \; \; \; \; 
\end{eqnarray}

\vspace{-1mm}

\subsubsection{The Case of Success Channel Slot} 

The third term on the right hand side of Equation~\eqref{eqn:pi-jk-tplus1-second-part} can be expanded as follows: \begin{eqnarray} 
\label{eqn:pi-jk-tplus1-second-part-succ-case}
\lefteqn{P\big(Q_1(t+1) = j, \mathcal{M}(t+1) = k, L(t+1) = L_{succ} \; \big| \; Q_1(t) = i > 0, \mathcal{M}(t) = n\big)} \nonumber \\ 
&=& P\big(L(t+1) = L_{succ} \; \big| \; Q_1(t) = i > 0, \mathcal{M}(t) = n\big) \nonumber \\
&& \; \; \; \; \; \; \; \cdot \; P\big(Q_1(t+1) = j, \mathcal{M}(t+1) = k \; \big| \; Q_1(t) = i > 0, \mathcal{M}(t) = n, L(t+1) = L_{succ} \big) \nonumber \\ 
&=& p_{succ,n+1} \cdot P\big(Q_1(t+1) = j, \mathcal{M}(t+1) = k \; \big| \; Q_1(t) = i > 0, \mathcal{M}(t) = n, L(t+1) = L_{succ} \big) \nonumber \\ 
&=& p_{succ,n+1} \cdot P\big(Q_1(t+1) = j, \mathcal{M}(t+1) = k, D_1(t+1) = 1 \; \big| \; Q_1(t) = i > 0, \mathcal{M}(t) = n, L(t+1) = L_{succ} \big) \nonumber \\ 
&+& p_{succ,n+1} \cdot P\big(Q_1(t+1) = j, \mathcal{M}(t+1) = k, D_1(t+1) = 0 \; \big| \; Q_1(t) = i > 0, \mathcal{M}(t) = n, L(t+1) = L_{succ} \big), \nonumber \\ 
\end{eqnarray} where in the last step we condition on whether the departure occurs from the tagged queue or from a non-tagged queue. 

When the departure occurs from the tagged queue, we have \begin{eqnarray} 
\label{eqn:pi-jk-tplus1-second-part-succ-case-tagged}
\lefteqn{P\big(Q_1(t+1) = j, \mathcal{M}(t+1) = k, D_1(t+1) = 1 \; \big| \; Q_1(t) = i > 0, \mathcal{M}(t) = n, L(t+1) = L_{succ} \big)} \nonumber \\ 
&=& P\big(D_1(t+1) = 1 \; \big| \; Q_1(t) = i > 0, \mathcal{M}(t) = n, L(t+1) = L_{succ} \big) \nonumber \\
&& \cdot \; P\big(A_1(t+1) = j-i+1, \mathcal{M}(t+1) = k \; \big| \; Q_1(t) = i > 0, \mathcal{M}(t) = n, L(t+1) = L_{succ}, D_1(t+1) = 1 \big) \nonumber \\ 
&=& P\big(D_1(t+1) = 1 \; \big| \; Q_1(t) = i > 0, N(t) = n+1, L(t+1) = L_{succ} \big) \nonumber \\ 
&& \; \; \; \; \; \; \; \cdot \; P\big(\mathcal{M}(t+1) = k \; \big| \; Q_1(t) = i > 0, \mathcal{M}(t) = n, L(t+1) = L_{succ}, D_1(t+1) = 1 \big) \nonumber \\ 
&& \; \; \; \; \; \; \; \; \; \; \; \; \cdot \; P\big(A_1(t+1) = j-i+1 \; \big| \; L(t+1) = L_{idle} \big) \nonumber \\ 
&=& \frac{1}{n+1} \cdot s(j-i+1) \cdot {{M-n-1 \choose k-n}} (1-s(0))^{k-n} s(0)^{M-k-1}, 
\end{eqnarray} where we have used the following facts: (a) given that the tagged queue is non-empty and $n$ non- tagged nodes are non-empty at the channel slot boundary $t$, the departure occurs from the tagged queue with probability $\frac{1}{n+1}$, (b) given that $D_1(t+1) = 1$ and $Q_1(t) = i > 0$, we have $Q_1(t+1) = j$ if and only if $A_1(t+1) = j-i+1$, (c) given that $L(t+1) = L_{succ}$, the probability that $A_1(t+1) = j-i+1$ does not depend on anything else, and is equal to $s(j-i+1)$, and (d) $\left(D_1(t+1) = 1\right) \Rightarrow \left(D_m(t+1) = 0, 2 \leq m \leq M\right)$, which implies that the set of non-empty non-tagged nodes can only increase, and, given that $\mathcal{M} t) = n$, we have $\mathcal{M}(t+1) = k$, if and only if arrivals occur to exactly $k-n$ of the $M-n-1$ empty non- tagged nodes in the success channel slot, which happens with probability ${{M-n-1 \choose k-n}} (1-s(0))^{k-n} s(0)^{M-k-1}$.

When the departure occurs from a non-tagged queue, we have \begin{eqnarray} 
\label{eqn:pi-jk-tplus1-second-part-succ-case-non-tagged}
\lefteqn{P\big(Q_1(t+1) = j, \mathcal{M}(t+1) = k, D_1(t+1) = 0 \; \Big| \; Q_1(t) = i > 0, \mathcal{M}(t) = n, L(t+1) = L_{succ} \big)} \nonumber \\ 
&=& P\big(D_1(t+1) = 0 \; \big| \; Q_1(t) = i > 0, \mathcal{M}(t) = n, L(t+1) = L_{succ} \big) \nonumber \\ 
&& \; \; \; \; \cdot \; P\big(A_1(t+1) = j-i, \mathcal{M}(t+1) = k \; \big| \; Q_1(t) = i > 0, \mathcal{M}(t) = n, L(t+1) = L_{succ}, D_1(t+1) = 0 \big) \nonumber \\ 
&=& \left(1-P\big(D_1(t+1) = 1 \; \big| \; Q_1(t) = i > 0, N(t) = n+1, L(t+1) = L_{succ} \big)\right) \nonumber \\ 
&& \; \; \; \; \cdot \; P\big(\mathcal{M}(t+1) = k \; \Big| \; Q_1(t) = i > 0, \mathcal{M}(t) = n, L(t+1) = L_{succ}, D_1(t+1) = 0 \big) \nonumber \\ 
&& \; \; \; \; \; \; \cdot \; P\big(A_1(t+1) = j-i \; \Big| \; L(t+1) = L_{idle} \big) \nonumber \\ 
&=& \frac{n}{n+1} \cdot s(j-i) \cdot P\big(\mathcal{M}(t+1) = k \; \big| \; Q_1(t) = i > 0, \mathcal{M}(t) = n, L(t+1) = L_{succ}, D_1(t+1) = 0 \big).  
\end{eqnarray}

The conditional probability in the last step in Equation~\eqref{eqn:pi-jk-tplus1-second-part-succ-case-non-tagged} can be simplified by similar arguments as before, and we obtain \begin{eqnarray} 
\label{eqn:pi-jk-tplus1-second-part-succ-case-non-tagged-detailed}
\lefteqn{P\big(\mathcal{M}(t+1) = k \; \big| \; Q_1(t) = i > 0, \mathcal{M}(t) = n, L(t+1) = L_{succ}, D_1(t+1) = 0 \big)} \nonumber \\ 
&=& {{M-n-1 \choose k-n}} (1-s(0))^{k-n} s(0)^{M-k-1} \nonumber \\ 
&& + \; q(i,n,t) \cdot (1-s(0))^{k-n} s(0)^{M-k-1} \cdot \left({{M-n \choose k-n+1}} (1-s(0)) - {{M-n-1 \choose k-n}} \right), \; \; \; \; 
\end{eqnarray} where notice the similarity with the last step in Equation~\eqref{eqn:pi-jk-tplus1-first-part-succ-case-partial-ratio-final}. However, now we have the unknown probability $q(i,n,t)$ instead of $q(0,n,t)$ since the tagged node contains $i > 0$ packets at $t$. 

Substituting Equation~\eqref{eqn:pi-jk-tplus1-second-part-succ-case-non-tagged-detailed} in Equation~\eqref{eqn:pi-jk-tplus1-second-part-succ-case-non-tagged}, we obtain \begin{eqnarray} 
\label{eqn:pi-jk-tplus1-second-part-succ-case-non-tagged-final}
\lefteqn{P\big(Q_1(t+1) = j, \mathcal{M}(t+1) = k, D_1(t+1) = 0 \; \Big| \; Q_1(t) = i > 0, \mathcal{M}(t) = n, L(t+1) = L_{succ} \big)} \nonumber \\ 
&=& \frac{n}{n+1} \cdot s(j-i) \cdot {{M-n-1 \choose k-n}} (1-s(0))^{k-n} s(0)^{M-k-1} \nonumber \\ 
&& + \; \frac{n}{n+1} \cdot s(j-i) \cdot q(i,n,t) \cdot (1-s(0))^{k-n} s(0)^{M-k-1} \cdot \left({{M-n \choose k-n+1}} (1-s(0)) - {{M-n-1 \choose k-n}}\right). \nonumber \\
\end{eqnarray}

Substituting Equations~\eqref{eqn:pi-jk-tplus1-second-part-succ-case-tagged} and~\eqref{eqn:pi-jk-tplus1-second-part-succ-case-non-tagged-final} in Equation~\eqref{eqn:pi-jk-tplus1-second-part-succ-case} and rearranging, we obtain \begin{eqnarray} 
\label{eqn:pi-jk-tplus1-second-part-succ-case-final}
\lefteqn{P\big(Q_1(t+1) = j, \mathcal{M}(t+1) = k, L(t+1) = L_{succ} \; \big| \; Q_1(t) = i > 0, \mathcal{M}(t) = n\big)} \nonumber \\ 
&=& p_{succ,n+1} \cdot s(j-i) \cdot {{M-n-1 \choose k-n}} (1-s(0))^{k-n} s(0)^{M-k-1} \nonumber \\ 
&& + \; p_{succ,n+1} \left(\frac{1}{n+1}\right) \cdot {{M-n-1 \choose k-n}} (1-s(0))^{k-n} s(0)^{M-k-1} \left(s(j-i+1) - s(j-i)\right) \nonumber \\ 
&& + \; p_{succ,n+1} \left(\frac{n}{n+1}\right) \cdot s(j-i) \cdot q(i,n,t) \cdot (1-s(0))^{k-n} s(0)^{M-k-1} \nonumber \\ 
&& \; \; \; \; \; \; \; \; \; \; \; \; \; \; \; \; \; \; \; \; \; \; \; \; \; \; \; \; \; \; \; \; \; \; \; \; \; \; \; \; \; \; \; \; \; \cdot \; \left( {{M-n \choose k-n+1}} (1-s(0)) - {{M-n-1 \choose k-n}} \right). \; \; \; \; \; \; \; \; 
\end{eqnarray}

Substituting Equations~\eqref{eqn:pi-jk-tplus1-second-part-idle-case},~\eqref{eqn:pi-jk-tplus1-second-part-coll-case} and~\eqref{eqn:pi-jk-tplus1-second-part-succ-case-final} in Equation~\eqref{eqn:pi-jk-tplus1-second-part}, we obtain \begin{eqnarray} 
\label{eqn:pi-jk-tplus1-second-part-final}
\lefteqn{P\big(Q_1(t+1) = j, \mathcal{M}(t+1) = k \; \big| \; Q_1(t) = i > 0, \mathcal{M}(t) = n\big)} \nonumber \\ 
&=& {{M-n-1 \choose k-n}} \left(p_{idle,n+1} \cdot d(j-i) (1-d(0))^{k-n} d(0)^{M-k-1} \right. \nonumber \\ 
&& \; \; \; \; \; \; \; \; \; \; \; \; \; \; \; \; \; \; \; \; \; \; \; \; \; \; \; \; \; \; \; \; \; \; \; \; \; \; \; \; \; \; \; \; + p_{coll,n+1} \cdot c(j-i) (1-c(0))^{k-n} c(0)^{M-k-1} \nonumber \\ 
&& \; \; \; \; \; \; \; \; \; \; \; \; \; \; \; \; \; \; \; \; \; \; \; \; \; \; \; \; \; \; \; \; \; \; \; \; \; \; \; \; \; \; \; \; \; \; \; \; \; \; \; \; \; \; \; \; \; \; \; \; \; \; \; \; \; \; \; \; \; \left. + p_{succ,n+1} \cdot s(j-i) (1-s(0))^{k-n} s(0)^{M-k-1} \right) \nonumber \\ 
&& + \; p_{succ,n+1} \left(\frac{1}{n+1}\right) \cdot {{M-n-1 \choose k-n}} (1-s(0))^{k-n} s(0)^{M-k-1} \left(s(j-i+1) - s(j-i)\right) \nonumber \\ 
&& + \; p_{succ,n+1} \left(\frac{n}{n+1}\right) \cdot s(j-i) \cdot q(i,n,t) \cdot (1-s(0))^{k-n} s(0)^{M-k-1} \nonumber \\ 
&& \; \; \; \; \; \; \; \; \; \; \; \; \; \; \; \; \; \; \; \; \; \; \; \; \; \; \; \; \; \; \; \; \; \; \; \; \; \; \; \; \; \; \; \; \; \; \; \; \; \; \; \cdot \left( {{M-n \choose k-n+1}} (1-s(0)) - {{M-n-1 \choose k-n}} \right), \; \; \; \; \; \; \; \; \; \; 
\end{eqnarray} which is the transition probability from a state $(i,n)$ to a state $(j,k)$, $i = 1, 2, \ldots$, $j = 0, 1, 2, \ldots$, $0 \leq n,k \leq M-1$. 


It is important to emphasize again that the transition probabilities of the process $\{\mathcal{X}(t)\}$ involve the unknown probabilities $q(i,n,t)$'s which depend on $t$. Certain transition probabilities can be zero, since $\displaystyle {{a \choose b}} = 0$ when $b < 0$ or $a < b$, and the probabilities $d(j)$, $c(j)$ and $s(j)$ are zero for $j < 0$. 



\section{Transition Probabilities of the Process $\{\tilde{\mathcal{X}}(t)\}$} 
\label{app:balance-eqn-tilde-Xt}

The transition probabilities of the process $\{\tilde{\mathcal{X}}(t)\}$ can be derived from the transition probabilities of the process $\{\mathcal{X}(t)\}$ as follows. First we apply Lemma~\ref{lem:conditional-probability-simplification-exact} and redefine $q(i,n,t)$, $\forall t \geq 0$, $\forall i \geq 0$, $\forall n$, $0 \leq n \leq M-1$, $\forall l$, $2 \leq l \leq M$, as given by Equation~\eqref{eqn:qint-definition-two}. Then, we apply Approximation~\ref{approx:independence-exact-number} and Equation~\eqref{eqn:relation-qint-qnt} to write \begin{eqnarray*} 
q(i,n,t) = \left\{ \begin{array}{ll} q(n,t) & \mbox{if} \; i = 0, \\ q(n+1,t) & \mbox{if} \; i > 0. \end{array} \right.
\end{eqnarray*} Finally, we replace the $q(n,t)$'s by the constant time-independent probabilities $\tilde{q}(n)$'s and obtain the transition probabilities of the process $\{\tilde{\mathcal{X}}(t)\}$ in terms of the $\tilde{q}(n)$'s as follows. \begin{eqnarray} 
\label{eqn:pi-jk-tplus1-first-part-final-tilde}
\lefteqn{P\big(\tilde{Q}_1(t+1) = j, \tilde{\mathcal{M}}(t+1) = k \; \big| \; \tilde{Q}_1(t) = 0, \tilde{\mathcal{M}}(t) = n\big)} \nonumber \\ 
&=& {{M-n-1 \choose k-n}} \left( p_{idle,n} \cdot d(j) (1-d(0))^{k-n} d(0)^{M-k-1} + p_{coll,n} \cdot c(j) (1-c(0))^{k-n} c(0)^{M-k-1} \right. \nonumber \\
&& \; \; \; \; \; \; \; \; \; \; \; \; \; \; \; \; \; \; \; \; \; \; \; \; \; \; \; \; \; \; \; \; \; \; \; \; \; \; \; \; \; \; \; \; \; \; \; \; \; \; \; \; \; \; \; \; \; \; \; \; \; \; \; \; \; \; \; \; \; \; \; \; \; \; \; \; \; \; \; \; \; \; \; \; \; \; \; \; + \left. p_{succ,n} \cdot s(j) (1-s(0))^{k-n} s(0)^{M-k-1} \right) \nonumber \\ 
&& + \; p_{succ,n} \cdot s(j) \cdot \tilde{q}(n) \cdot (1-s(0))^{k-n} s(0)^{M-k-1} \cdot \; \left( {{M-n \choose k-n+1}} (1-s(0)) - {{M-n-1 \choose k-n}} \right), \; \; \; \; \; \; 
\end{eqnarray} \begin{eqnarray} 
\label{eqn:pi-jk-tplus1-second-part-final-tilde}
\lefteqn{P\big(\tilde{Q}_1(t+1) = j, \tilde{\mathcal{M}}(t+1) = k \; \big| \; \tilde{Q}_1(t) = i > 0, \tilde{\mathcal{M}}(t) = n\big)} \nonumber \\ 
&=& {{M-n-1 \choose k-n}} \left(p_{idle,n+1} \cdot d(j-i) (1-d(0))^{k-n} d(0)^{M-k-1} \right. \nonumber \\ 
&& \; \; \; \; \; \; \; \; \; \; \; \; \; \; \; \; \; \; \; \; \; \; \; \; \; \; \; \; \; \; \; \; \; \; \; \; \; \; \; \; \; \; \; \; + p_{coll,n+1} \cdot c(j-i) (1-c(0))^{k-n} c(0)^{M-k-1} \nonumber \\ 
&& \; \; \; \; \; \; \; \; \; \; \; \; \; \; \; \; \; \; \; \; \; \; \; \; \; \; \; \; \; \; \; \; \; \; \; \; \; \; \; \; \; \; \; \; \; \; \; \; \; \; \; \; \; \; \; \; \; \; \; \; \; \; \; \; \; \; \; \; \; \left. + p_{succ,n+1} \cdot s(j-i) (1-s(0))^{k-n} s(0)^{M-k-1} \right) \nonumber \\ 
&& + \; p_{succ,n+1} \left(\frac{1}{n+1}\right) \cdot {{M-n-1 \choose k-n}} (1-s(0))^{k-n} s(0)^{M-k-1} \left(s(j-i+1) - s(j-i)\right) \nonumber \\ 
&& + \; p_{succ,n+1} \left(\frac{n}{n+1}\right) \cdot s(j-i) \cdot \tilde{q}(n+1) \cdot (1-s(0))^{k-n} s(0)^{M-k-1} \nonumber \\ 
&& \; \; \; \; \; \; \; \; \; \; \; \; \; \; \; \; \; \; \; \; \; \; \; \; \; \; \; \; \; \; \; \; \; \; \; \; \; \; \; \; \; \; \; \; \; \; \; \; \; \; \; \cdot \left( {{M-n \choose k-n+1}} (1-s(0)) - {{M-n-1 \choose k-n}} \right). \; \; \; \; \; \; \; \; \; \; 
\end{eqnarray}

Due to the constant time-independent probabilities $\tilde{q}(n)$'s, the process $\{\tilde{\mathcal{X}}(t)\}$ is a \textit{time-homogeneous} DTMC embedded at the channel slot boundaries. However, the constant time-independent probabilities $\tilde{q}(n)$'s are not known, and they must be treated as unknown parameters of the process $\{\tilde{\mathcal{X}}(t)\}$.


\section{Derivation of Equation~\eqref{eqn:pidj0toK-2}}
\label{app:derivation-eqn-pidj0toK-2}

Recall that the DTMC $\{\bmath{Q}^{(K)}(t), t \geq 0\}$ is stationary and ergodic. We apply the discrete time version of \textit{Birkoff's strong ergodic theorem} (see page 800 of~\cite{theory.kumar_etal04communication-networking}) to obtain $p^{(d)}(j)$, $0 \leq j \leq K-2$. (Note that, the time argument $t$ in Theorem~\ref{thm:birkoff-strong-ergodic-theorem} is discrete.)

\begin{theorem}[Birkoff's Strong Ergodic Theorem]
\label{thm:birkoff-strong-ergodic-theorem} 
Let $\{X(t), t \geq 0\}$, be a stationary and ergodic process, and let $f(\cdot)$ be a function that maps realizations of the process (i.e., $\{X(t,\omega), t \geq 0\}$) to the set of real numbers $\mathbb{R}$, such that $\EXP{|f(\{X(t), t \geq 0\})|} < \infty$. Then, with probability 1, \[\; \; \; \; \; \; \; \; \; \; \; \; \; \; \; \; \; \; \; \; \; \; \; \; \; \; \; \; \; \; \; \; \; \; \; \; \; \; \; \; \displaystyle \lim_{\tau \rightarrow \infty} \frac{1}{\tau} \sum_{t' = 0}^{\tau-1} f(\{X(t+t'), t \geq 0\}) = \EXP{f(\{X(t), t \geq 0\})}. \; \; \; \; \; \; \; \; \; \; \; \; \; \; \; \; \; \; \; \; \; \; \; \; \; \; \; \; \; \; \; \; \; \; \; \; \; \; \; \; \hfill \qed\] 

\end{theorem}


Noting that departures can occur only at the end of channel slots, we write \begin{eqnarray*} 
p^{(d)}(j) &=& \displaystyle \lim_{\tau \rightarrow \infty} \displaystyle \frac{\displaystyle \sum_{t'=0}^{\tau-1}   \ind{D_1^{(K)}(t_0+1+t') = 1} \ind{Q_1^{(K)}(t_0+1+t') = j}}{\displaystyle \sum_{t'=0}^{\tau-1} \ind{D_1^{(K)}(t_0+1+t') = 1}} = \frac{ \displaystyle \lim_{\tau \rightarrow \infty} \frac{1}{\tau} \displaystyle \sum_{t'=0}^{\tau-1} \ind{D_1^{(K)}(t_0+1+t') = 1, Q_1^{(K)}(t_0+1+t') = j}}{\displaystyle \lim_{\tau \rightarrow \infty} \frac{1}{\tau} \displaystyle \sum_{t'=0}^{\tau-1} \ind{D_1^{(K)}(t_0+1+t') = 1}}, 
\end{eqnarray*}  where, applying ergodicity of the process $\{\bmath{Q}^{(K)}(t), t \geq 0\}$, we equate the probability $p^{(d)}(j)$ with the long-term fraction of departures from the tagged queue that leave $j$ packets behind (starting from an arbitrary time $t_0+1$). Then, we can apply Theorem~\ref{thm:birkoff-strong-ergodic-theorem}, observing that expectation of the indicator function of an event is equal to the probability of the event, and hence, is always bounded between 0 and 1. Thus, we have \begin{eqnarray} 
\label{eqn:pidj-definition}
&=& \frac{\EXP{\ind{D_1^{(K)}(t_0+1) = 1, Q_1^{(K)}(t_0+1) =j}}}{\EXP{\ind{D_1^{(K)}(t_0+1) = 1}}} = \frac{P\big(D_1^{(K)}(t_0+1) = 1, Q_1^{(K)}(t_0+1) = j\big)}{P\big(D_1^{(K)}(t_0+1) = 1\big)} \nonumber \\ 
&=& \frac{\displaystyle \sum_{i=1}^{K} \sum_{n=0}^{M-1} P\big(Q_1^{(K)}(t_0) = i, \mathcal{M}^{(K)}(t_0) = n, D_1^{(K)}(t_0+1) = 1, Q_1^{(K)}(t_0+1) = j\big)}{\displaystyle \sum_{i=1}^{K} \sum_{n=0}^{M-1} P\big(Q_1^{(K)}(t_0) = i, \mathcal{M}^{(K)}(t_0) = n, D_1^{(K)}(t_0+1) = 1\big)}, \; \; \; \; \; \; \; \; \; \; \; \; \; \;  
\end{eqnarray} where, in the last step, we expand by summing over all possible states at time $t_0$. Notice that, in the last step, we do not consider the possibility $i=0$ since $\left(D_1^{(K)}(t_0+1) = 1\right) \Rightarrow \left(Q_1^{(K)}(t_0) > 0\right)$. 

For $0 \leq j \leq K-2$, Equation~\eqref{eqn:pidj-definition} can be simplified as follows: \begin{eqnarray} 
\label{eqn:pidj0toK-2-derivation}
p^{(d)}(j) &=& \frac{\displaystyle \sum_{i=1}^{K} \sum_{n=0}^{M-1}
  \left(\begin{array}{l} P\big(Q_1^{(K)}(t_0) = i,
    \mathcal{M}^{(K)}(t_0) = n\big) \\ \; \times \;
    P\big(D_1^{(K)}(t_0+1) = 1 \; \big| \; Q_1^{(K)}(t_0) = i,
    \mathcal{M}^{(K)}(t_0) = n\big) \\ \; \; \times \;
    P\big(Q_1^{(K)}(t_0+1) = j \; \big| \; Q_1^{(K)}(t_0) = i,
    \mathcal{M}^{(K)}(t_0) = n, D_1^{(K)}(t_0+1) = 1\big) \end{array}
  \right)}{\displaystyle \sum_{i=1}^{K} \sum_{n=0}^{M-1}
  \left( \begin{array}{l} P\big(Q_1^{(K)}(t_0) = i,
    \mathcal{M}^{(K)}(t_0) = n\big) \\ \; \times \;
    P\big(D_1^{(K)}(t_0+1) = 1 \; \big| \; Q_1^{(K)}(t_0) = i,
    \mathcal{M}^{(K)}(t_0) = n\big) \end{array} \right)} \nonumber \\ 
&=& \frac{\displaystyle \sum_{i=1}^{K} \sum_{n=0}^{M-1}
  \left(\begin{array}{l} P\big(Q_1^{(K)}(t_0) = i,
    \mathcal{M}^{(K)}(t_0) = n\big) \\ \; \times \;
    P\big(D_1^{(K)}(t_0+1) = 1 \; \big| \; Q_1^{(K)}(t_0) = i,
    \mathcal{M}^{(K)}(t_0) = n\big) \\ \; \; \times \;
    P\big(A_1^{(K)}(t_0+1) = j-i+1 \; \big| \; Q_1^{(K)}(t_0) = i,
    \mathcal{M}^{(K)}(t_0) = n, D_1^{(K)}(t_0+1) = 1\big) \end{array}
  \right)}{\displaystyle \sum_{i=1}^{K} \sum_{n=0}^{M-1}
  \left( \begin{array}{l} P\big(Q_1^{(K)}(t_0) = i,
    \mathcal{M}^{(K)}(t_0) = n\big) \\ \; \times \;
    P\big(D_1^{(K)}(t_0+1) = 1 \; \big| \; Q_1^{(K)}(t_0) = i,
    \mathcal{M}^{(K)}(t_0) = n\big) \end{array} \right)} \nonumber \\ 
&=& \frac{\displaystyle \sum_{i=1}^{K} \sum_{n=0}^{M-1}
  P\big(Q_1^{(K)}(t_0) = i, \mathcal{M}^{(K)}(t_0) = n\big)
  \left(\frac{p_{succ,n+1}}{n+1}\right) s(j-i+1)}{\displaystyle
  \sum_{i=1}^{K} \sum_{n=0}^{M-1} P\big(Q_1^{(K)}(t_0) = i,
  \mathcal{M}^{(K)}(t_0) = n\big)
  \left(\frac{p_{succ,n+1}}{n+1}\right)} \nonumber \\ 
&=& \displaystyle \frac{\displaystyle \sum_{i=1}^{j+1}
  \displaystyle \sum_{n=0}^{M-1} \pi^{(K)}(i,n) \displaystyle
  \left(\frac{p_{succ,n+1}}{n+1}\right) s(j-i+1)}{\displaystyle
  \sum_{i=1}^{K} \displaystyle \sum_{n=0}^{M-1} \pi^{(K)}(i,n)
  \displaystyle \left(\frac{p_{succ,n+1}}{n+1}\right)}. 
\end{eqnarray} 

In the second step, we used the fact that, given $Q_1^{(K)}(t_0) = i$ and $D_1^{(K)}(t_0+1) = 1$, we have $Q_1^{(K)}(t_0+1) = j$ if and only if $A_1^{(K)}(t_0+1) = j-i+1$. In the third step, we observe that $Q_1^{(K)}(t_0) = i > 0$ and $\mathcal{M}^{(K)}(t_0) = n$ implies that $N^{(K)}(t_0) = n+1$. Thus, the probability that the $(t_0+1)^{th}$ channel slot is a success channel slot is given by $p_{succ,n+1}$, and given that a success occurs, the departure occurs from the tagged node with probability $1/(n+1)$. In the fourth step, we let $t_0 \longrightarrow \infty$ on both sides. Since, the left hand side of Equation~\eqref{eqn:pidj0toK-2-derivation} does not depend on $t_0$, it remains unchanged. However, for the stationary and ergodic DTMC $\{\bmath{Q}^{(K)}(t), t \geq 0\}$, we have, $\lim_{t_0 \rightarrow \infty} P\big(Q_1^{(K)}(t_0) = i, \mathcal{M}^{(K)}(t_0) = n\big) = \pi^{(K)}(i,n)$ where $\pi^{(K)}(i,n)$ denotes the stationary probability that the tagged queue contains $i$ packets and $n$ non-tagged queues are non-empty. Notice that the first sum index of the numerator in the last step on the right hand side of Equation~\eqref{eqn:pidj0toK-2-derivation} runs only up to $j+1$. For $i > j+1$, the argument $j-i+1$ of $s(j-i+1)$ becomes negative so that $s(j-i+1) = 0$.

\end{document}